\let\myover=\over
\documentclass[12pt,twoside,a4paper]{article}
\usepackage{epsfig,bezier,amssymb,amsmath}
\newcommand{\mycaption}[1]{\begin{center}
                            \parbox{4.5in}{
                            \caption{\protect {\footnotesize #1}}}
                            \end{center}
                           }

\def\e{{\rm e}}

\def\be{\begin{equation}}
\def\ee{\end{equation}}
\def\d{\partial}
\def\l{\left(}
\def\r{\right)}
\def\la{\langle }
\def\ra{\rangle }
\newcommand{\tg}{\mathop{\rm tan}\nolimits}
\renewcommand{\ln}{\mathop{\rm log}\nolimits}

\newcommand{\sm}[1]{{\scriptscriptstyle #1}}
\newcommand{\Tr}{{\rm Tr}}
\newcommand{\bg}{\begin{gather}}
\newcommand{\eg}{\end{gather}}
\def\str{{\rm STr}\,m^2_{\rm mess}}

\newcommand{\eq}[1]{(\ref{#1})}
\begin{document}
\let\over=\myover
\def\half{{1 \over 2}}
\begin{titlepage}
\title{
Gauge mechanism of mediation of supersymmetry breaking
}
\author{S.~L.~Dubovsky, D.~S.~Gorbunov, and S.~V.~Troitsky \\
{\small{\em
Institute for Nuclear Research of the Russian Academy of Sciences,
}}\\
{\small{
\em
60th October Anniversary prospect, 7a, 117312, Moscow, Russia
}}\\ }
\date{}
\end{titlepage}
\maketitle

\begin{abstract} 
  \small Most phenomenologically acceptable supersymmetric models
imply breaking of supersymmetry in a separate sector of fields
introduced just for this purpose. Supersymmetry breaking is
transferred to the
Standard Model due to a certain interaction with this
additional sector. In this review we disuss one of the popular mechanisms
of such mediation with the key role played by the Standard
Model gauge interactions. We consider general principles of gauge
mediation and give the critical analysis of specific models together with their
phenomenological and cosmological implications. We present also 
basic facts concerning mechanisms of dynamical supersymmetry breaking.  
\end{abstract} 
\newpage 

\tableofcontents 
\section{Introduction}
\label{sec:intro}
In spite of the fact that no experimental evidence for supersymmetry has
been discovered so far, the possibility that nature is supersymmetric
attracts continuous attention of 
scientists working in particle 
theory. Search for supersymmetry is one of the
principal tasks for huge accelerators under construction,
particularly, for the Large Hadronic Collider (see, for instance, Ref.\
\cite{MatveevKrasnikov}). In this review we will discuss one of the approaches to
construct realistic supersymmetric theories, which underwent
intense exploration in the last few years. It might be useful,
however, to recall first why the models that yet have no experimental
evidence for, attract so great attention.

Supersymmetry \cite{GolfandLikhtman} is a symmetry between bosons and
fermions which provides a nontrivial extension of Poincare
algebra. Supersymmetric quantum field models
have been constructed \cite{VolkovAkulov,WessZum} and further
investigated in great detail
(see, for instance, textbooks Ref.\ \cite{WessBagger,West} and
reviews Ref.\ \cite{Ogiev,nilles}). These models turned out to possess a
number of features attractive from the particle physics point of
view. The main point is that the ultraviolet behaviour of these
theories is better than in non-supersymmetric models, that is,
additional cancellation of
ultraviolet divergencies occurs due
to supersymmetry. In particular, quadratically divergent radiative
corrections to the masses of scalar particles (e.g., Higgs boson) are
absent in these models. This fact is the main motivation for constructing
realistic supersymmetric theories.  This property allows one to
solve the so-called gauge hierarchy problem, which is often considered
as the main theoretical drawback of the Standard Model of particle
interactions. Namely, suppose that the Standard Model provides a good
description of subnuclear physics either up to the energies where
quantum gravitational effects become essential (Planck mass,
$M_{Pl}\sim 10^{19}$~GeV), or up to the Grand Unification scale
($M_{GUT}\sim 10^{16}$~GeV). Then the characteristic scale of the
electroweak symmetry breaking determined by $Z$ or Higgs boson masses
$\sim 10^2$~GeV $\ll M_{GUT},M_{Pl}$ gets huge radiative
corrections due to quadratically divergent diagrams. That is, the
hierarchy of mass scales is unstable with respect to quantum
effects. On the other hand, scalar masses in realistic supersymmetric
models get only logariphmic radiative corrections, so
the hierarchy of mass scales is stable. The latter fact provides
serious support to supersymmetric generalizations of the Standard
Model.

Supersymmetry, however, is very restrictive, as far as the particle
spectrum is concerned. Namely, together with known particles, their
superpartners should exist which are the particles of a different spin
but with the same quantum numbers and masses. Since scalar particles
with masses of quarks and leptons as well as fermions with masses of
gauge bosons are absent, unbroken supersymmetry in nature is ruled
out. Nevertheless it is possible to break supersymmetry without spoiling
attractive cancellation of quadratic divergencies
\cite{Gir}. This kind of breaking is called soft breaking and yields heavy
superpartners in a natural way. These arguments show that broken
supersymmetry might be relevant to nature, thus posing the problem of
how it is broken.

Softly broken supersymmetry, provided its breaking is not very
strong, has one additional advantage --- it allows to explain not only
stability but even the origin of the gauge hierarchy
\cite{witten_ier}. 
Namely, supersymmetry breaking occurs in many models  dynamically,
due to nonperturbative effects. The latter are important at scales
$\Lambda\sim\exp\left(-O(1/g^2)\right)M$, where $M\sim M_{Pl}$ or
$M_{GUT}$, and $g$ is some coupling constant. If $g$ is small enough,
then the supersymmetry breaking scale $\Lambda\ll M$. On the other hand, in most
realistic models the vacuum expectation value of Higgs boson is
generated by radiative corrections, so it is determined by
$\Lambda$ up to powers of coupling constants. This could provide the origin
of gauge hierarchy.

Among the reasons to be interested in supersymmetric particle
theories, much better (as compared to the Standard Model) unification of
gauge couplings at the scale $M_{GUT}$ is worth mentioning.  One
should remember also about purely aesthetic advantages of the theories
possessing larger symmetries. In practice, this leads to
the possibility to gain nontrivial dynamical information by making use
of
symmetry arguments (see section \ref{dsb}).

In the simplest models of supersymmetry breaking, the latter occurs
spontaneously at the tree level (O'Raifeartaigh \cite{LOR} and Fayet --
Iliopoulos \cite{FI} models). Explaining the origin ofgauge hierarchy in
these models requires fine tuning of mass parameters, that is,
introducing a new hierarchy. Moreover, there is a more
fundamental reason to exclude the possibility of direct implementation
of these mechanisms in supersymmetric generalizations of the Standard
Model. 
Namely, 
even in theories with broken
supersymmetry, 
particle spectrum cannot be arbitrary.  In
the physically interesting case, the so-called supertrace of the mass
squared matrix (sum of the masses squared for all particles in a given
supermultiplet weighted by plus sign for bosons and minus sign for
fermions) is equal to zero, ${\rm STr}\,m^2=0$, at tree level
(neglecting effects of supergravity). This
means that the Standard Model fermions should have scalar
superpartners not heavier than fermions themselves, the result being
clearly phenomenologically unacceptable. So, realistic theories should
overcome one of the conditions of this theorem, that is either to deal with 
supergravity, or to exploit loop effects in
renormalizable models. The attempts to use the second approach
yielded the gauge mediation mechanism, the subject of this
review.

General problems of supersymmetric particle theories will not be
considered here in detail since they are carefully discussed, for
instance, in the reviews  
 \cite{nilles,broad_brush,20questions}.
Dynamical supersymmetry breaking is considered also in the reviews
\cite{NelsonDSB,PopTrDSB}, 
while gauge mediation is discussed in Refs.\ 
\cite{Colda,GiudiceRattazziReview}.

In section 2, basic notations are introduced and the minimal
supersymmetric extension of the Standard Model (MSSM) is discussed in
brief. In section 3, we discuss the mechanism of gauge mediation of
supersymmetry breaking and phenomenology of the simplest, Minimal
model, which exhibits most characteristic features of gauge
mediation models. Extensions of the Minimal model and specific aspects
of their phenomenology are discussed in section 4. Section 5
is devoted to different realizations of an attractive feature of gauge
mediation models --- the possibility to construct and analyse
quantitatively a full theory incorporating both dynamical supersymmetry
breaking and its mediation to MSSM fields.  Section \ref{D3.3}
deals with cosmological implications, generic to models under
discussion, and with constraints on mechanisms of supersymmetry
breaking and its mediation coming from cosmology. In Conclusion, we
outline 
main features of gauge mediated supersymmetry breaking models and
mention a few problems interesting for future investigation.

\section{Basics of the MSSM}
\label{how-to-break}
Minimal supersymmetric standard model (MSSM) is
usually considered as a low energy theory describing phenomenology of
supersymmetric models. Standard formalism of supersymmetric models of
particle physics is discussed in detail in reviews and textbooks,
Refs.\ \cite{WessBagger,West,Ogiev,nilles},
while various aspects of MSSM are considered, for instance, in
reviews \cite{nilles,visotsky,kazakov}. 
In this section, we will briefly discuss the basic features of the
MSSM and introduce our notations.

MSSM Lagrangian is naturally separated in two parts, the first being
supersymmetric and the second containing soft supersymmetry breaking
terms. Let us consider the supersymmetric part first. A convenient formalism
to deal with supersymmetric Lagrangians is the superspace
technique. In this formalism, functions depend not
only on usual space-time coordinates $x^{\mu}$ but also on
anticommuting coordinates $\theta_{\alpha}$ and their conjugates,
$\bar{\theta}_{\dot{\alpha}}$, which are two-component spinors of the
Poincare algebra. These functions are called superfields; usual
fields appear as coefficients of expansion of superfields in series in
$\theta_{\alpha}$, $\bar{\theta}_{\dot{\alpha}}$. Since
$\theta_{\alpha}$, $\bar{\theta}_{\dot{\alpha}}$ are anticommuting,
these series are in fact finite polynomials. Action of the fields is
an ordinary integral of the Lagrangian density with respect to
space-time coordinates and Berezin integral \cite{berezin} with
respect to supercoordinates.

To construct a realistic supersymmetric model, it is necessary to
ascribe a superfield to each field of the Standard Model,
the fact which leads to the presence of new particles in the 
spectrum. These new particles (superpartners) carry the
same global quantum numbers as the ordinary fields but 
different spin.

Gauge fields\footnote{Hereafter we consider only the simplest, $N=1$
supersymmetry.} 
are described by vector (real) superfields. In the most commonly used
Wess-Zumino gauge, a real superfield ${\cal{V}}$ consists of a usual vector field 
$A_\mu$, its superpartner (gaugino) --- fermion
$\lambda$ in the adjoint representation of the gauge group --- and an
auxiliary scalar field $D$, the latter having no kinetic term so it may be
integrated out by means of equations of motion.
In superfield formulation, the action reads 
\begin{equation}
\label{gauge_part}
S_g=\mbox{Im}\left[\tau\int  
d^2\theta d^4x\rm{Tr}({\cal W}^{\alpha} {\cal W}^{\alpha})\right] 
\;, 
\end{equation}
where ${\cal W}_\alpha$ is a superfield containing the gauge stress
tensor $F_{\mu\nu}$ and
$\tau=\frac{\Theta}{2\pi}+i\frac{4\pi}{g^2}$ is a combination of
the gauge coupling $g$ and the $CP$-violating parameter 
$\Theta$. Written in terms of components, this is the usual
gauge-invariant action of gauge bosons and massless gauginos 
with $\Theta$-term included. In the MSSM, the names of superpartners
of $SU(3)\times SU(2)\times U(1)_Y$ gauge bosons are gluino (for
$SU(3)$ gluons), wino and zino (for weak $W$ and $Z$ bosons) and bino
(for $U(1)_Y$). Photon's superpartner is photino (for the electromagnetic
gauge group $U(1)_{\rm EM}$ which remains unbroken after the electroweak symmetry
breaking).

Matter fields are described by another representation of the
superalgebra, the chiral superfields
$$
\Phi=\phi(y)+\sqrt{2}{\psi}(y)\theta+{\theta}\theta F(y)\;,
$$
where $\phi$ is a scalar component of the superfield,
${\psi}$ is a fermion and $F$ is an auxiliary scalar 
field. Supersymmetric action which describes interactions of chiral superfields
(with each other and with gauge superfields) is given
by the functional,
\begin{equation}
\label{matter_part}
S_m=\int d^2\bar{\theta} d^2\theta d^4x\Phi^{\dagger}\e^{\cal{V}}\Phi+ 
\left(\int d^2\theta W(\Phi)+h.c.\right)\;. 
\end{equation}
The first term is called Kahler potential; it
contains gauge invariant kinetic terms for $\phi$ and $\psi$ and their
Yukawa interaction with
gauginos. The second term in Eq.\ \eq{matter_part} is called superpotential
and contains self-interaction of components of the chiral
supermultiplet. The full action is a sum of
$S_g$ and 
$S_m$. 
With the auxiliary fields integrated out,
scalar potential takes the form
\begin{equation}
\label{FandD}
V=\sum_i|F_i|^2+\sum_a{1\over 2}D_a^2\;,
\end{equation}
where $F$-terms are determined from the superpotential: 
$$ 
F_i=\frac{\partial W(\phi)}{\partial\phi_i}\;, 
$$ 
and $D$-terms in the Wess-Zumino gauge are 
$$ 
D^a=\sum_i\phi^{\dagger}_iT^a\phi_i\;. 
$$ 
($T^a$ are the gauge group generators).

Quarks and leptons in the MSSM are fermionic components
of chiral superfields (their scalar superpartners are called squarks
and sleptons) while Higgs fields are the bosonic components of other
chiral superfields (their fermionic
superpartners are called higgsinos). Note that, contrary to the
Standard Model, two Higgs fields are required in the
MSSM. Electrically neutral fermions (photino, zino, and neutral
higgsino) may mix with each other, forming physically observable linear
combinations, neutralinos. Charged fermions, that is winos and charged higgsinos,
may mix also.

Superpotential of the MSSM contains terms describing Yukawa
interaction of the matter and Higgs fields,
\begin{equation}
\label{MSSM}
W_{MSSM}=H_D(L_iY^l_{ij}E_j+Q_iY^d_{ij}D_j)+H_UQ_iY^u_{ij}U_j+\mu H_UH_D\;,
\end{equation}
where $L_i$, $E_i$, $Q_i$, $D_i$, and $U_i$ are superfields containing
left- and right-handed leptons, left- and right-handed down and up quarks,
respectively, together with their superpartners;
$H_D$ and $H_U$ contain Higgs doublets (hereafter we define superfilds
by the upper-case, and their scalar components by the lower-case
letters); 
$Y^l_{ij}$, $Y^d_{ij}$ É $Y^u_{ij}$ are Yukawa mixing
matrices. Superpotential \eq{MSSM} respects all symmetries of the
MSSM, including lepton and baryon number conservation. Note that
it contains the so-called $\mu$-term, a supersymmetric mixing mass
term for the Higgs fields.

Generally, additional terms in the superpotential are allowed by
$SU(3)\times SU(2)\times U(1)_Y$ gauge symmetry. These terms lead to
violation of lepton and baryon numbers. To satisfy experimental
constraints coming, for instance, from proton decay, coefficients of
these terms should be very small. The usual assumption is that these
terms are forbidden by an additional $U(1)_R$ global symmetry. The
symmetry is broken down to a discrete $Z_2$ symmetry called $R$-parity
when supersymmetry is broken. All Standard Model particles have charge
$+1$ under this symmetry, while their superpartners carry charge
$-1$. $R$ parity conservation leads to stability of the lightest
superpartner (LSP) of the Standard Model particles.

Besides the MSSM fields, an important role in the phenomenology of
gauge mediated supersymmetry breaking is played by the superpartner of
graviton --- gravitino, a particle of spin $3/2$ which is massless in
the limit of unbroken supersymmetry.

Soft supersymmetry breaking terms are added to the MSSM Lagrangian
explicitly. These terms contain mass terms for gauginos, $(\half
M_{\lambda_i}\lambda_i\lambda_i+{\rm h.c.})$, and for scalar fields,
Higgses \footnote{It can be shown that a strong constraint on the
mass of the lightest Higgs boson arises in the MSSM. At tree level, it
implies that $m_h<M_Z$. Though it is somewhat relaxed by loop
corrections, most realistic models predict the lightest Higgs boson
mass not larger than 150~GeV \cite{boer}. This distinction from
the Standard Model allows the MSSM to be ruled out if the 
Higgs particle is not discovered at future 
accelerators.} for example,
\begin{equation}
m^2_{h_U}|h_U|^2+
m^2_{h_D}|h_D|^2+ \left(B_\mu h_U h_D +{\rm h.c.}\right),
\label{8**}
\end{equation}
as well as trilinear interactions of the scalar fields,
\begin{equation}
h_D(l_iy^l_{ij}e_j+q_iy^d_{ij}d_j)+h_Uq_iy^u_{ij}u_j+ {\rm h.c.},
\label{8*}
\end{equation}
where $y_{ij}$ are new matrices of coupling constants.
So, about a hundred new parameters are introduced in addition to those
of the Standard Model.

Generally, such amount of parameters might lead to nontrivial flavour
physics. Absence of flavour-changing neutral currents (FCNC),
processes with CP- and lepton flavour violation poses
severe constraints on these parameters.

Usually, these constraints come from consideration of the first
two generations. A characteristic example is provided by
the system of neutral kaons where absence of FCNC implies
\cite{FCNC}
\begin{equation}
\varkappa\l\frac{1~\mbox{TeV}^2}{m^2_{\sm{\tilde{Q}}}}\r\l
\frac{\delta m^2_{\sm{\tilde{Q}}}}{m^2_{\sm{\tilde{Q}}}}\r^2\lesssim
5\times 10^{-3},
\label{FCNC-rate}
\end{equation} 
where $\delta m_{\sm{\tilde{Q}}}$ is a mass difference between $u$
and $s$ squarks, $m_{\sm{\tilde{Q}}}$ denotes their average mass squared, and
$\varkappa$ is a product of elements of the rotation matrix between mass
eigenbases of quarks and squarks. Similar bounds appear from other
flavour violating processes.

One can see from Eq.\ \eq{FCNC-rate} that there are three possibilities
to satisfy experimental bounds
on FCNC contribution. Namely, one may:
\begin{enumerate}
\item
Consider models with heavy squarks,
$m_{\sm{\tilde{Q}}}\gtrsim 10$~TeV. 
\item
Enforce $\varkappa$ to be small by requiring that mass eigenbases for
scalar and fermions are parallel. 
\item
Provide squark mass degeneracy, $\delta
m^2_{\sm{\tilde{Q}}} \ll m^2_{\sm{\tilde{Q}}}$ (this approach is most
commonly used).
\end{enumerate}
To choose one of the three possibilities in a natural way, one should
understand the origin of the soft terms.

An observation important for model building is that supersymmetry
cannot be broken in a phenomenologically acceptable way in a framework
of a theory which contains only MSSM fields. Indeed, breaking at the
tree level in the MSSM is excluded because of the mentioned above
property ${\rm  STr}\,m^2=0$, while dynamical breaking by means of
nonperturbative effects is possible only at the energy scale where QCD
is strongly coupled. This scale, however, is a few orders of magnitude
less than one required to explain the gauge hierarchy and absence of
observed superpartners. So, supersymmetry breaking should occur in
a new sector which is introduced exclusively for this purpose and
contains either O'Raifeartaigh model or a gauge theory where
supersymmetry is broken due to Fayet-Iliopoulos mechanism or
dynamically. As it has already been pointed out, to explain not only
stability but also the origin of the gauge hierarchy, dynamical
supersymmetry breaking is prefered. So, it is usually assumed that
there exists some gauge group in addition to the Standard Model.
There are some matter fields carrying quantum numbers of this group
and dynamics of this new sector provides
supersymmetry breaking at
the required energy scale. Since no evidence for this sector has been
found in the experiments, it is natural to suppose that interaction of
MSSM fields with that sector is rather weak and becomes essential at
energies not available to current experiments. It is this interaction
which is responsible for generating the soft terms. Depending on the
nature of this interaction, one distinguishes gravitational
mediation at scales of order the Planck scale and gauge
mediation. Effects of supersymmetry breaking are transferred to the MSSM
at much lower scales in the latter case, so entire model may
be described in terms of field theory. This possibility is an
advantage of the gauge mediation mechanism since no reliable theory of
Planck scale physics is known. In the low energy mediation models,
the supersymmetry breaking sector is often called
\cite{vacuum_relaxation} a {\em secluded}
sector --- contrary to a {\em hidden} sector of gravitational models.

The vacuum energy is zero in the case ofunbroken supersymmetry, as a
consequence of the supersymmetry algebra. So, Eq.\ (\ref{FandD})
relates supersymmetry breaking to appearance of
a nonzero vacuum expectation value of some $F$-- or $D$--
component. In realistic models, supersymmetry is often broken with a
nonzero vacuum expectation value $F_{\sm{DSB}}$ of an auxiliary
component of some field in the additional sector. In the gravity
mediation case, the corresponding soft masses are of order
$\frac{F_{\sm{DSB}}}{M_{Pl}}$. To make these terms of order the
electroweak scale, one needs $F_{\sm{DSB}}\sim M_{Pl}m_{\sm Z}$.  

Difficulties with flavour violation are usually overcome in the
supergravity approach
by conjecturing the universality of the soft terms \cite{DimH}. This
conjecture states that, at the gravitational scale, all scalar
superpartners of the Standard Model fermi\-ons get equal masses $m_0$
while all gauginos get equal masses $m_{1/2}$. The additional trilinear
scalar interaction, Eq.\ \eq{8*}, is assumed to be equal, up to some
coefficient $A_0$ with dimensionality of a mass, to Yukawa part of the
superpotential,
$$
y_{i,j}^{l,d,u}=A_0 Y_{i,j}^{l,d,u}.
$$
So, at high energies new sources of flavour violation additional to
the Standard Model do not appear --- scalar masses are degenerate, and
GIM mechanism is operating. Nevertheless, in the low energy theory
flavour violation occurs not only due to Cabibbo-Kobayashi-Maskawa
(CKM) matrix, but also because of mixing in the squark sector. The
latter effect originates from renormalization group scaling with
nondegenerate Yukawa couplings. These additional contributions are,
however, small enough not to contradict current experimental bounds in
a wide range of parameters. The same mechanism may help to solve the
problem of $CP$ violation in the system of neutral kaons, but generally results
in too large electric dipole moments of leptons and
quarks. In particular, for the electric dipole moment of neutron in
theories with universal squark masses $\tilde m$ one has \cite{dipole}
\begin{equation}
d_{\sm N}\simeq 2\l\frac{100\mbox{GeV}}{\tilde{m}}\r^2\sin\phi\cdot 
10^{-23}e\cdot\mbox{cm}\;,
\label{dip_moment}
\end{equation}   
while the experimental limit is $d_{\sm N}<1.1\cdot
10^{-25}e\cdot$cm~\cite{dipole_exp}.  Here we denote $\phi={\rm arg}\l
A_0^*M_{gluino}\r-{\rm arg}\l B_\mu^*\mu M_{gluino}\r$. So, 
some 
additional mechanism is required to suppress the CP violating
parameter, 
$\sin\phi\ll 1$ unless squark masses (in the case of universality,
this is true for
slepton masses too) are large enough,
$\tilde m\gtrsim 1$~TeV. This is the so-called supersymmetric CP-problem
(see Ref.\ \cite{dipole_review} for details).

With the universality assumption, MSSM has only five additional parameters 
as compared to the Standard Model: besides
$m_0$, $m_{1/2}$ and $A_0$, one has supersymmetric
$\mu$ and soft $B_\mu$ masses of Higgs fields. The number of
parameters may be increased still satisfying the experimental
constraints on flavour physics if one replaces the universality condition by
the so-called horizontal universality, when different
mass parameters $m^i_0$ for different generations are introduced
\cite{DimH}.

Another approach exploits the so-called low energy dynamical
supersymmetry breaking mediated by gauge interactions. Though its
basics have been formulated some time ago
\cite{spectrum_old,parents}, these theories underwent intense
development in the last few years
\cite{DN,DNS,DNNS}. The main reason for this fact is the 
considerable progress in understanding nonperturbative effects which
break supersymmetry (see section \ref{dsb} of this review). Mediation
itself occurs by means of usual gauge interactions of the visible
sector and provides effective superpartner mass scale
of order the electroweak scale.

The key advantadge of this class of models is a possibility to {\em
calculate} (by usual tools of quantum field theory) supersymmetry breaking
parameters of the MSSM. Resulting values of these parameters are such that
automatic suppression of flavour violating processes occurs and a
small value of electric dipole moment of neutron appears without fine
tuning. Yet the theory has very few free parameters, thus it is highly
predictive. Electroweak symmetry breaking occurs due to radiative
corrections, and to construct unified theories with supersymmetry
breaking one has not to invoke nonrenormalizable theories
(gravity). 

One of shortcomings of models with gauge mediation is their rather
complicated structure. Among other problems one might mention the
following: 1)~in the specific realizations of the mechanism,
parameters of the Higgs sector often remain free, and sometimes
require fine tuning; 2)~in a wide class of models, supersymmetry
breaking vacuum is metastable, though its lifetime is much larger than
the age of the Universe for a natural choice of parameters; 3)~in
models with unbroken R-parity, light gravitino is stable leading to
cosmological problems; 4)~though flavour violation due to squarks
and sleptons is naturally suppressed in minimal models, the most
general superpotential which respects all MSSM symmetries may contain
terms leading to tree level violation of corresponding quantum
numbers.  We will discuss these questions, among others, in the
following sections.

\section[The Minimal model]{Phenomenology of the Minimal gauge
mediation model} 
\label{singlet} 
\subsection{Mass spectrum and soft terms}
\label{spectrum}
Let us begin with the most general, model-independent features of
gauge mediation of supersymmetry breaking. The key ingredient of the
mediation mechanism is a set of new fields, messengers, absent in the
MSSM but charged under its gauge group. Interaction with a secluded
sector results in their supersymmetry breaking spectrum, namely, in mass
splittings in the messenger supermultiplets. Consequently, soft terms
are generated in the low energy Lagrangian by raditive corrections
due to the MSSM gauge interactions.

Messengers are massive charged fields in vector-like
representations of
$SU(3)\times SU(2)\times U(1)$. As a rule, they are taken to compose
complete multiplets of a unified gauge group, $SU(5)$ for example, in
order not to spoil the gauge coupling unification in the MSSM.

Let the messengers
$Q$ and $\bar{Q}$ transform as fundamental and antifundamental
representations of 
$SU(5)$, respectively. The simplest way to obtain mass splitting in
messenger supermultiplet is to introduce 
the following 
Yukawa interaction with a chiral superfield $S$ in the superpotential, 
\begin{equation}
\label{scalar}
W_{ms} = \lambda SQ \bar{Q}.
\end{equation}
$S$ is MSSM singlet which
gets, due to some (not specified yet) dynamics in the secluded sector,
nonzero vacuum expectation values 
$\la s\ra $ and $\la F_s\ra $ of its scalar and auxiliary components,
$$\la S\ra  = \la s\ra +\la F_s\ra \theta^2\;.$$
Consequently, scalar messengers get tree-level masses,
\begin{equation}
M_{\pm}^2=\frac{\Lambda^2}{x^2}(1\pm x),
\label{12*}
\end{equation}
where
\be
\label{Lambda_x}
\Lambda=\frac{\la F_s\ra }{\la s\ra }
 ~~~~~~~\makebox{and}~~~~~~~
x=\frac{\lambda \la F_s\ra }{\lambda^2\la s\ra ^2}.
\end{equation} 
In the physically interesting case\footnote{Otherwise some messenger
would have negative mass squared, see Eq.\ \eq{12*}, which would result in
breaking of electromagnetic and colour gauge groups.} $0<x<1$. 
Masses of all fermionic components of the messengers are equal to
$$M=\frac{\Lambda}{x}\;.$$

Interaction with messenger fields results in the soft terms in the
MSSM sector. Gauginos get masses at the messenger scale\footnote{To
obtain the physical mass spectrum from Eqs.\ \eq{gaugin},
\eq{scalmas}, one has to take into account renormalization group
corrections.} $M$ through
interaction with messengers at one-loop level \cite{spectrum_old},
see Fig.\ \ref{gc},
\begin{equation}
\label{gaugin}
M_{\lambda_{\sm{i}}} = c_i\frac{\alpha_i}{4\pi}\Lambda f_1(x).
\end{equation}
Here $\alpha_1=\alpha/\cos^2\theta_W$, $\alpha_2$, $\alpha_3$ are
coupling constants of weak and strong interactions;
$c_i$ is the sum of Dynkin indices of messenger fields running in the
loop, $c_1=5/3$, $c_2=c_3=1$. Masses depend weakly on $x$ :
$f_1(x)$ is a non-decreasing function whose values are close to $1$ in
the most part of the domain of definition
\cite{Martin}: 
$$
f_1(x)=\frac{1}{x^2}\left[(1+x)\ln (1+x)+(1-x)\ln (1-x)\right].
$$
Two--loop diagrams like one presented in Fig.~\ref{gc} 
\begin{figure}[htb]
\begin{picture}(0,0)%
\epsfig{file=rr.pstex}%
\end{picture}%
\setlength{\unitlength}{3947sp}%
\begingroup\makeatletter\ifx\SetFigFont\undefined%
\gdef\SetFigFont#1#2#3#4#5{%
  \reset@font\fontsize{#1}{#2pt}%
  \fontfamily{#3}\fontseries{#4}\fontshape{#5}%
  \selectfont}%
\fi\endgroup%
\begin{picture}(6924,1690)(514,-1532)
\put(2601,-586){\makebox(0,0)[lb]{\smash{\SetFigFont{12}{14.4}{\rmdefault}
{\mddefault}{\updefault}gaugino}}}
\put(500,-586){\makebox(0,0)[lb]{\smash{\SetFigFont{12}{14.4}{\rmdefault}
{\mddefault}{\updefault}gaugino}}}
\put(4501,-586){\makebox(0,0)[lb]{\smash{\SetFigFont{12}{14.4}{\rmdefault}
{\mddefault}{\updefault}scalar}}}
\put(6676,-586){\makebox(0,0)[lb]{\smash{\SetFigFont{12}{14.4}{\rmdefault}
{\mddefault}{\updefault}scalar}}}
\put(5501,-61){\makebox(0,0)[lb]{\smash{\SetFigFont{12}{14.4}{\rmdefault}
{\mddefault}{\updefault}fermion}}}
\put(4726, 14){\makebox(0,0)[lb]{\smash{\SetFigFont{12}{14.4}{\rmdefault}
{\mddefault}{\updefault}b)}}}
\put(2162,-28){\makebox(0,0)[lb]{\smash{\SetFigFont{12}{14.4}{\rmdefault}
{\mddefault}{\updefault}$\la F_s\ra$}}}
\put(976, 14){\makebox(0,0)[lb]{\smash{\SetFigFont{12}{14.4}{\rmdefault}
{\mddefault}{\updefault}a)}}}
\put(2101,-1505){\makebox(0,0)[lb]{\smash{\SetFigFont{12}{14.4}{\rmdefault}
{\mddefault}{\updefault}$\la s\ra$}}}
\put(5451,-986){\makebox(0,0)[lb]{\smash{\SetFigFont{12}{14.4}{\rmdefault}
{\mddefault}{\updefault}gaugino}}}
\end{picture}
\mycaption{a) Gauginos get masses via diagrams with loops of messenger fields.
b) A typical diagram contributing to soft masses of MSSM scalars; a cross
denotes the insertion of the loop presented in the figure a).
\label{gc}
}
\end{figure}
contribute to masses squared of scalar fields --- Higgs bosons,
squarks and sleptons
\cite{spectrum_old}:
\begin{equation}
\label{scalmas}
\tilde{m}^2=2\Lambda ^2\Biggl[ C_3\Big( \frac{\alpha_3}
{4\pi}\Big)^2+
C_2\Big( \frac{\alpha_2}{4\pi}\Big)^2+
\frac{5}{3}\Big( \frac{Y}{2}\Big)^2\Big( \frac{\alpha_1}{4\pi}\Big)
^2\Biggr]f_2
(x)=
2\Lambda^2\sum_{i=1}^3C_ic_i\l\frac{\alpha_i}{4\pi}\r^2 f_2(x).
\end{equation}
In Eq.\ \eq{scalmas}, $C_i$ are eigenvalues of the quadratic Casimir
operator for squark, slepton, or Higgs representation,
$C_3=4/3$ for colour triplets (zero for singlets), 
$C_2=3/4$ for weak doublets (zero for singlets),
$C_1=\l\frac{Y}{2}\r^2$, where $Y$ is the weak hypercharge.
Again, $f_2(x)$ is a smooth function close to $1$ at the most part of
its domain of definition
\cite{messenger's_cosmology}: 
$$
f_2(x)=
\frac{1+x}{x^2}\left[\ln(1+x)-2 \makebox{Li}_2\!\left(\frac{x}{1+x}\right)+
\frac{1}{2}\makebox{Li}_2\!\left(\frac{2x}{1+x}\right)\right]+(x \to -x).
$$

Let us note that to obtain a phenomenologically acceptable spectrum, one
needs a mechanism which generates two nonzero vacuum expectation values,
$\la F_s\ra$ and $\la s\ra$, simultaneously. 
It is $\la F_s\ra$ which results in supersymmetry breaking in the
visible sector. The nonzero $\la s\ra$ is required to provide (at $x<1$)
positive signs of masses squared of the scalar messengers and to induce
chirality breaking insertions in diagrams 
Fig.~\ref{gc}, thus resulting in nonsupersymmetric contributions to
the superpartner masses. 

Fermions of the MSSM chiral superfields do not get masses from
interaction with messengers because of nonrenormalization theorems
\cite{renorm}
(masses of matter fermions, not being soft terms \cite{Gir}, 
are not generated when supersymmetry is broken spontaneously).
Trilinear coupling, $A_0$, appears only at two loop level (see, for
instance, Refs.\ \cite{Dimopoulos_Thomas_Wells,kolda}), and its
effects are negligible compared to mass terms since it is
suppressed by an additional power of
$\alpha_i/4\pi$,
$$
A_0\sim
\frac{\alpha_i}{4\pi}M_{\lambda_i}\ln\frac{M_\pm}{M_{\lambda_i}}. 
$$

In the scheme presented above, which we hereafter refer to as the
Minimal Model (MM), superparicle spectrum is given by Eqs.\ 
(\ref{gaugin}), (\ref{scalmas}). Since
$f_1(x)$, $f_2(x)$ are almost constant, all soft masses are determined
by a single parameter additional to the Standard Model, that is,
supersymmetry breaking scale
$\Lambda$.
One requires 
  $\Lambda \gtrsim 72$~TeV in the Minimal
Model \cite{borzumati} in order to satisfy
the limits on slepton masses from LEP data. 

Since soft masses of particles are proportional to their coupling
constants, the heaviest superparticles are squarks
($m_{sq}\sim 1$~TeV at
$\Lambda\sim100$~TeV).  
The lightest superparticle (LSP) is gravitino, 
the least interacting superpartner.
This kind of LSP is a charachteristic
feature of models with gauge mediation. Below, we will discuss in detail
possible values of gravitino mass and supersymmetry
breaking scale which are closely related. The next-to-lightest
superpartner (NLSP) in the Minimal model is a combination
of
$\tilde{\tau}_{{\sm R}}$ and $\tilde{\tau}_{{\sm L}}$~\cite{borzumati},  
superpartners of the right- and left-handed $\tau$-leptons. This fact
is due to mass splitting caused by a mixing term,
$\mu m_{\tau}\tan{\beta}\tilde{\tau}_{{\sm R}}\tilde{\tau}_{{\sm L}}$, 
which appears in the MSSM Lagrangian after electroweak symmetry is
broken. We use the standard notation
$\tan\beta={\la h_U\ra \over \la h_D \ra}$ for the ratio of the vacuum
expectation values of Higgs particles;
$m_{\tau}$ is  
$\tau$-lepton mass.
In some extensions of the Minimal model, NLSP may turn out to be
neutralino.

Let us emphasize that scalar and gaugino masses are of the same order,
both numerically and parametrically, in corresponding couplings
$\alpha_i$. 

\subsection{Radiative electroweak symmetry breaking in the Minimal
model} 
  
Consider some characteristics of the electroweak symmetry breaking in
the MM. Supersymmetry preserving masses of Higgs particles are given
by the 
$\mu$ parameter which was not important in the above consideration. While
the 
soft mass of $h_{\sm D}$ scalar is given by the general formula, Eq.\ 
(\ref{scalmas}), and is equal to 
$$
m_{h_{\sm D}}^2=\frac{3}{2}\l\frac{\alpha_2}{4\pi}\r^2\Lambda^2\;,
$$
mass squared of another Higgs boson,
$h_{\sm U}$, gets a significant negative contribution due to 
interaction with 
$\tilde{t}$ squark, resulting in the leading order in
\begin{equation}
m_{h_{\sm U}}^2=m_{h_{\sm D}}^2\l
1-\frac{4Y_t^2}{3\pi^2}\l\frac{\alpha_3}{\alpha_2}\r^2
\ln\frac{M}{m_{\tilde t}}\r\;,
\label{13-*}
\end{equation}
where 
$Y_t$ is $t$ quark Yukawa constant. In this expression, three-loop
contribution is proportional to 
$\alpha_3^2 $ and is not suppressed compared to two-loop one. It is
the three-loop contribution which gives rise to negative
$ m_{h_{\sm U}}^2$ resulting in the electroweak symmetry breaking.

To minimize Higgs potential of the MSSM, one should solve the
following two equations,
\begin{gather}
\mu^2=\frac{m^2_{h_{\sm D}}-m^2_{h_{\sm U}}\tan^2\beta}
{\tan^2\beta}-\frac{1}{2}m_{\sm Z}^2,
\label{mequations}\\
\sin 2\beta=-\frac{2B_\mu}{m^2_{h_{\sm D}}+m^2_{h_{\sm U}}+2\mu^2}\nonumber\;,
\end{gather}
where
$B_\mu$ is the soft mixing term in the mass matrix of Higgs bosons,
Eq.\ \eq{8**}. 
In the MM, this term appears only in the third order of perturbation
theory and is suppressed even further due to accidental cancellations.
Small mixing in the Higgs sector results in large values of
$\tan{\beta}$. Indeed, at zero mixing, 
$\tan{\beta}={\la h_u\ra \over \la   h_d \ra}$ would be infinitely
large since only
$ h_u$ gets the vacuum expectation value.
Note that at large $\tan\beta$, other corrections to Higgs masses
become important (for instance, contribution of $b$ quark to $h_D$
mass analogous to Eq.\ \eq{13-*} should be taken into account). 
Accurate calculations \cite{kolda} give
$\tan\beta\sim 50$ in the Minimal model at
$\Lambda\sim 100$~TeV.

The value of $\mu$ parameter as a function of supersymmetry breaking
scale $\Lambda$ may be extracted from Eqs.\ 
(\ref{mequations}), 
with given mass of $Z$ boson. In the MM at
$\Lambda\sim 100$~TeV, the value of $\mu$ is about 500~GeV. However, the
$\mu$ parameter is an independent parameter of MSSM Lagrangian which
is not a priori related to 
parameters of the
secluded sector. It means that though one succeeds in generating the
small scale
$\Lambda\ll M_{Pl}$ dynamically, one more free parameter determining the
electroweak scale is still present. This fact is the essence of
the so-called $\mu$-problem. A few approaches to solve this problem
are discussed in section
\ref{tg-beta}. From the point of view of phenomenology, their net
effect is that
$B_\mu$ is one more free parameter of the low-energy theory.

\subsection{Constraints on the parameters of a secluded sector}
\label{D1.3}
Let us consider in more detail parameters of the theory which are
related to the secluded sector. The main of them is the supersymmetry breaking
scale $\Lambda$.  
As it has been already pointed out, 
$\Lambda$ is experimentally bounded from below since it is related
to the superparticle masses. The fact that sleptons have not been
observed in collider experiments (conservative bound from $Z$ boson
width, $\tilde{m}_{sl}>45$~GeV) together with Eq.\ \eq{scalmas} mean
that $\Lambda\sqrt{n}>30$~TeV (we take into account the possibility to
have $n$ generations of
messengers instead of one). 
To get the correct value of $Z$ boson mass
from Eq.\ (\ref{mequations}) without fine tuning of the $\mu$
parameter, soft masses of Higgs bosons should not exceed
significantly the electroweak scale. The corresponding ``aesthetical''
upper bound depends on the allowed amount of fine tuning. In any case,
one expects that
$\Lambda$ does not exceed a few hundred TeV.

Most upper bounds on the supersymmetry breaking scale deal with the
largest $F$ term in the secluded sector,
$F_{DSB}$ 
(as a rule\footnote{Exact relation between 
$F_S$ and $F_{DSB}$ depends on the way the secluded
sector is incorporated, see section \ref{sec:secluded}.}, 
$F_{DSB}>\langle F_S\rangle$), 
and with messenger masses
$M\sim \langle s \rangle$. 
These estimates result in conservative upper bounds on
$\Lambda$ since
$\Lambda< \langle s\rangle$,
$\sqrt{\langle F_S\rangle}$ (see Eq.\ \eq{Lambda_x} at $x<1$).

Bounds on
$F_{DSB}$ are mainly related to the gravitino mass
$m_{3/2}$ which is non-zero if supersymmetry is broken and which is
proportional to
$\frac{F_{\sm DSB}}{M^*_{Pl}}$ 
(hereafter 
$M^*_{Pl}={M_{Pl}}/\sqrt{8\pi}$ is the reduced Planck mass).
Namely, once supersymmetry is broken spontaneously, a massless
fermion, goldstino, emerges in the spectrum (like  massless
Nambu--Goldstone bosons emerge due to spontaneous breaking of a Lie group
of global symmetry). With gravitational effects taken into account,
goldstino is eaten, by means of super-Higgs mechanism, by a longitudinal component of a massless spin
$3/2$ particle, gravitino $\tilde{G}$ (superpartner of graviton).
Longitudinal component of gravitino,
$\tilde{G}_\mu$, is expressed in terms of goldstino 
$\psi$ as 
$\tilde{G}_\mu=\sqrt{\frac{2}{3}}{1\over m_{3/2}}\d^\mu\psi$. 
Resulting gravitino mass is
$m_{3/2}=\frac{F_{\sm DSB}}{\sqrt{3}M^*_{Pl}}$.

One of the limits on the scale at which supersymmetry breaking is
transferred to the Standard Model follows from requirement of
suppression of
flavour violating processes. In models of supergravity mediation, such
a suppression requires additional assumptions about parameters of the
model, as it has been discussed in section 
\ref{how-to-break}. Effects of gauge mediation should be much stronger
than gravitational ones, so characteristic messenger scale should be
considerably lower than Planck scale. To be specific, let soft
superparticle mass originated from interaction with messengers be
approximately 
\begin{equation}
m_{soft}\sim\sqrt{n}{\alpha\over4\pi}\frac{\la F_S \ra}{\la s \ra}
\label{16*}
\end{equation}
Supergravity contribution to the soft
masses is of order gravitino mass, $m_{3/2}\sim F_{\sm{DSB}}/
M^*_{Pl}$.  If supergravity contribution is non-universal, then Eq.\
\eq{16*} leads to a typical bound from conditions like Eq.\
\eq{FCNC-rate}, 
\begin{equation}
\la s \ra\lesssim \sqrt{n}\times 10^{15}~{\rm GeV}.
\label{mass-bound:gravity}
\end{equation}
This way of reasoning provides another ground for the fact that in models with
gauge mediation, gravitino is the LSP.

One more upper bound, the most restrictive up to now, comes from
cosmology. It is based on the estimates of influence of NLSP decays on
nucleosynthesis \cite{Dimop-1,new-nucl}.
Recall that in gauge mediation, NLSP is either neutralino or $\tau$
slepton. Its decays during nucleosynthesis may result in
too large amount of light nuclei of deuterium, helium-3,
or lithium in the Universe. If NLSP is $\tau$ slepton, nuclei are
produced due to hadronic decays of
$\tau$ lepton in the intermediate state.
If NLSP is neutralino, photons from their decay produce
quark-antiquark pairs.
To avoid
these effects, NLSP should decay before nucleosynthesis begins. One of
decay products is goldstino, which interacts with matter via
supersymmetric current. The interaction is given 
\cite{generalization_GT}
by the generalized
Goldberger--Treiman formula
of the current algebra,
\begin{equation}
\label{1}
{\cal L}=-\frac{1}{F_{\sm{DSB}}}\d_\mu\psi^\alpha j^\mu_\alpha+h.c. 
\end{equation}    

With increasing of $F_{\sm{DSB}}$, the interaction becomes less intense,
and NLSP lifetime increases. Requirement of NLSP decay before
nucleosynthesis thus set an upper bound on
$F_{\sm{DSB}}$ \cite{Dimop-1}, 
\begin{equation}
\la F_{\sm{DSB}} \ra \lesssim  10^{8}~{\rm GeV}.
\label{mass-bound:nucleos}
\end{equation}
Note that at a special choice of parameters the restriction is not so
severe \cite{new-nucl}. 
An assumption of small $R$ parity breaking may relax this
bound significantly.
Some other cosmological bounds on the supersymmetry breaking
scale related to gravitino mass will be discussed in section
\ref{D3.3}.

\subsection{Prospects of the Grand Unification} The possibility to
construct realistic supersymmetric Grand Unified Theory (GUT) is one
of the most significant arguments in favour of supersymmetric
extensions of the Standard Model. Indeed, two principal objections
against the minimal non-supersymmetric $SU(5)$ model are difficulties
with precise unification of coupling constants and unacceptably fast
proton decay. On the other hand, in supersymmetric $SU(5)$ model coupling
constants unify with much larger accuracy and proton life-time is
longer due to increase of the unification scale
$M_{GUT}$. Nevertheless, recent high precision
measurements of $\alpha_s$ indicate that the same problems are present at the
two-loop level in the simplest GUTs based on MSSM (see, e.g., Ref.\ 
\cite{kabak}). More accurate account of the threshold splitting
effects at the GUT scale may be one of the possible ways out of these
problems.

The situation in gauge mediation models is somewhat worse because of new
messenger thresholds. For example, the limits from  proton decay do
not allow one to obtain exact unification of coupling constants by
fine tuning of the triplet Higgs mass in GUTs based on MM as well as
its extensions with arbitrary value of $\tg{\beta}$ or several
messenger generations. Another problem is related to the
unification of Yukawa couplings at $M_{\sm{GUT}}$
\cite{GUT_fall_1}.  This unification is possible only provided
$\tg\beta$ is in the narrow ranges near $\tg\beta\simeq 2$ or
$\tg\beta\simeq 50\div 60$.

To invoke messengers which do not compose complete GUT
multiplets at low energies is one of the possible ways to solve
these problems~\cite{GUT_fall_3}.  A mechanism similar to one which is
responsible for doublet-triplet splitting in the Higgs sector (e.g.,
fine tuning of parameters in $SU(5)$ model) can result in this 
kind of splitting. Moreover, incomplete messenger multiplets may appear in
the framework of more complicated GUTs.  For instance, messengers may
carry only down-quark quantum numbers \cite{Faraggi}. This kind of
spectrum may
appear due to the dynamics of the underlying string
theory where non-perturbative effects lead to unusually low scale of
supersymmetry breaking\footnote{Typically,
  supersymmetry breaking scale in the string inspired models is much
  higher~\cite{string_high_scale}.} $\Lambda\sim 100$~TeV. The
resulting masses of weak gauginos are
small but experimentally acceptable~\cite{Particle_Data}, so this model can
be considered as realistic.

The limit from the nucleon decay takes the form $\tg\beta<10$ at
$\Lambda\sim 100$~TeV in the case of incomplete messenger
multiplets\footnote{$\tg\beta$ is a relevant parameter here because
the rate of $n\to K^0\bar{\nu_\mu}$ decay is proportional to
$\sin{2{\beta}}$. }. Only for three specific sets of messengers 
$\tg\beta<17$~\cite{GUT_fall_3}. In order to satisfy this limit one
should consider one of the extensions of the MM where $\tg{\beta}$ is
an additional free parameter.

To extend the model by introducing new heavy multiplets is another
possible solution of the unification problem. For instance, it was
demonstrated in Ref.~\cite{Non-Unified} that one can make the limits
from $b-\tau$ unification in $SU(5)$ GUT and from $b-t$
unification in $SO(10)$ GUT weaker provided new fields are introduced.
This happens due to the breaking of the corresponding symmetry between
Yukawa couplings at the unification scale caused by the tree-level
mixing of heavy fields with MSSM fields.

\section{Extensions of the Minimal model}
\label{sec:extensions}
\subsection{The $\mu$ problem}
\label{mu-problem}
As we have already pointed out, some extensions of the Minimal
model aim to solve the $\mu$ problem and their net effect is that
$B_{\mu}$ (or, in another formulation,
$\tan{\beta}$) is a free parameter.
In this section, we consider a few approaches to solve the
$\mu$ problem. Phenomenology of gauge mediation at different values of
$\tan{\beta}$ will be discussed in the next subsection.

Taking into account the relations Eq.\ 
(\ref{mequations}), one can identify conditions on the theory which
solves the $\mu$ problem in a natural way
\cite{vacuum_relaxation}:
\begin{itemize}
\item 
One and the same mechanism yields generation of various
parameters in the Higgs sector.
\item $\mu$ term arises in the first order of perturbation theory while
$B_\mu$ arises only in the second order 
\cite{mu_1--B_mu_2}, so
$\mu^2$ and $B_\mu$ are of equal order, like gaugino and scalar masses.
\item 
No new particles appear at the electroweak scale.
\item 
All new coupling constants are of order one.
\end{itemize}

In supergravity models, one of the most attractive solutions to the $\mu$
problem has been suggested in Ref.\ 
\cite{Giudice-Masiero}. There, a nonrenormalizable mixing
$\frac{1}{M_{Pl}}\Phi^+H_{\sm U}H_{\sm D}$ 
in the Kahler potential was introduced, where 
$\Phi^+$ is a hidden sector field with nonzero auxiliary component, 
$F_\phi\sim m_{\sm Z}M_{Pl}$. 
This approach fails in gauge mediation models since it is required
there that 
$F_\phi\ll  m_{\sm Z}M_{Pl}$ so that gauge contribution dominates over gravitational
one, see section \ref{D1.3}.

One may try to obtain a similar solution to the Higgs sector problems in
models with gauge mediation. In this case, superpotential
\begin{equation}
W=aZH_{\sm U}H_{\sm D},
\label{21}
\end{equation}
is introduced, where the vacuum expectation value of the scalar
component of $Z$ superfield results in appearance of the $\mu$ term. 
Since one singlet field, $S$, has been already involved in the model 
(see Eq.\ (\ref{scalar})), it seems natural to use it as the field 
$Z$~\cite{DNS}. However, in this case
$a\la s\ra =\mu$ and $a\la F_s\ra =B_\mu$, so to satisfy requirements
from LEP experiments one should have
$\frac{B_\mu}{\mu}=\Lambda\gg m_{\sm Z}$. This excludes the
possibility that
$\mu\sim m_{\sm Z}$ and $B_\mu\sim m_{\sm Z}^2$
simultaneously. Moreover, from the expressions for messenger masses
one gets 
$\l\lambda \la s\ra \r^2\gtrsim \lambda
\la F_s\ra \gg m_{\sm Z}^2$, so satisfying even a single requirement 
(either $\mu\sim m_{\sm Z}$ or $B_\mu\sim m_{\sm Z}^2$) results in
the unnatural condition
$\frac{a}{\lambda}\ll 1$.
It is possible, of course, to try to explain the smallness of
$\frac{a}{\lambda}$ as a consequence of some additional
symmetry. Attempts to exploit this approach are presented in Ref.\ 
\cite{Gherghetta}
where anomalous global abelian symmetries were discussed which proved
useful for explaining the fermion mass hierarchy.
In any case, the requirement
$B_\mu\sim m_{\sm Z}^2$ leads inevitably to
$\mu\ll m_{\sm Z}$, so higgsino mass is smaller than the value allowed by
experiments. The requirement
$\mu\sim m_{\sm Z}$ contradicts the second equation in 
(\ref{mequations}) because of too large values of
$B_\mu$.

It is possible to consider the case when 
$Z$ and $S$ are differet fields, and somewhat modify the interaction,
Eq.\ (\ref{21}), so that $Z$ components would have appropriate vacuum
expectation values. This may be achieved, for instance, by adding a
cubic term in Eq.\ (\ref{21})~\cite{DNS,DNNS} or by introducing an
additional interaction with  secluded sector fields ``symmetric'' to
the visible sector messengers
\cite{vacuum_relaxation}.

All these approaches face several difficulties. A general problem
arises, for instance, when global singlet fields are involved. This
problem is caused by quadratically divergent tadpoles which appear in
two-loop order (even in one loop if the Kahler potential is
nontrivial). These terms result in too large  
vacuum expectation values of the singlets and too large masses of the fields
interacting with them.

Four conditions of satisfactory solution to the $\mu$ problem
mentioned above are fulfilled in the approach with so-called dynamical
relaxation mechanism
\cite{vacuum_relaxation,Dimop-1} but specific models require many
additional fields and interactions to be introduced.

Other ways to solve the $\mu$ problem were suggested in several
papers, but most of them lead to new serious difficulties.
In this sense, elegant or minimal solution is
lacking by now.

\subsection{Phenomenology at arbitrary $\tan{\beta}$}
\label{tg-beta}
At different values of
$\tan\beta$, different spectra of superpartners may
emerge. 
Electroweak symmetry breaking results in the generation of D-terms whose
contribution to the scalar masses increases with $\tan\beta$.
However, Yukawa mixing of left- and right- handed sleptons
is proportional to 
$\tan\beta$. As a 
result,
$\tilde{\tau}_{\sm R}$ may be lighter than bino. It is
$\tilde{\tau}_{\sm R}$ which is the NLSP at large values of
$\tan\beta$
(in the Minimal model with additional free parameter
$B_\mu$, for instance, this is the case at
$\tan\beta>25$~\cite{kolda}). 

Right handed $\tau$ slepton becomes massless if
$\tan\beta$ significantly exceeds 50
\cite{kolda} which leads to the upper bound on the value of
$\tan\beta$. At so large $\tan\beta$, mass squared of the axial Higgs
boson may become negative too
\cite{weak_scale_phenomenology}. 
Bounds on
$\tan\beta$ were discussed in more detail in Ref.\ 
\cite{RS_2}. Say, at
$x=0.1$, the absence of light sleptons contributing to $Z$ boson width
implies that
$\tan\beta<55$; suppression of tunnelling rate to a vacuum where 
$U(1)_{\rm EM}$ is broken results in 
$\tan\beta<50$ while absence of such a vacuum gives
$\tan\beta<47$.
Lower bounds,
$\tan\beta>1.0\div 1.5$, follow from the smallness of $t$ quark
Yukawa coupling at 
$M_{\sm{GUT}}$, $Y_t(M_{\sm{GUT}})<3.5$
\cite{weak_scale_phenomenology} (this requirement is not necessary, of
course). The similar requirement for $b$ quark Yukawa coupling leads
to upper bounds on
$\tan\beta$ 
\cite{Sparticle-Spectroscopy}.
In unified model with proton decay mediated by triplet Higgses, one
has an additional lower bound
($\tan\beta\gtrsim 0.85$~\cite{GUT_fall_2}) 
because proton lifetime 
$\tau_{proton}\sim\sin^22\beta$.

Let us consider phenomenology of models with gauge mediated supersymmetry breaking
at different values of $\tan\beta$ in more detail.
As it has already been pointed out, the key difference from other
supersymmetric models is that the lightest superparticle is
gravitino. The latter interacts with matter not only 
gravitationally, but, once supersymmetry is broken spontaneously, by
interaction of its longuitudinal component, goldstino, with supercurrent.
The form of the latter interaction can be obtained from Eq.\ 
(\ref{1}) by substituting an
explicit expression for supersymmetric current
$j^\mu_\alpha$.  
Integrating by parts, one finally gets the following interaction
of goldstino
$\psi$ with a chiral multiplet whose scalar and fermionic components
have masses
$m_\phi$ and $m_\chi$, and with a vector multiplet (gaugino mass is
$M_\lambda$), 
\begin{equation}
\label{gold}
{\cal L}=
\frac{M_{\lambda}}{4\sqrt{2}F_{\sm{DSB}}}\bar{\psi}\sigma_{\mu\nu}\lambda
F_{\mu\nu}+
\frac{m^2_\chi-m^2_\phi}{F_{\sm{DSB}}}\bar{\psi}\chi_{\sm L}\phi^*+h.c.
\end{equation}
In models of low energy supersymmetry breaking, the inequality 
$F_{\sm{DSB}}\ll M_{Pl}^2$ is always satisfied, so the interaction Eq.\
\eq{gold} dominates over the pure gravitational one.

The lightest superpartner of the Standard Model particles, that is
NLSP, may be either neutralino or slepton
(see discussion in section
\ref{how-to-break}, Eqs.\  
(\ref{gaugin}) and (\ref{scalmas})).
Since gravitational effects are negligible, the width of NLSP decay to
gravitino is described by the equations which may be obtained from the
Lagrangian Eq.\ 
(\ref{gold}), 
\cite{gravitino_decay,MMY,Sparticle-Spectroscopy}, 
\be
\label{2}
\Gamma\l\tilde{B}\to\gamma\tilde{G}\r=
\frac{\cos^2\theta_{\sm W}}{16\pi}\frac{m^5_{\sm{\tilde{B}}}}{F^2_{\sm{DSB}}},
~~~~~~~~~~~~\Gamma\l\tilde{l}\to\tilde{G}l\r=
\frac{1}{16\pi}\frac{m^5_{\sm{\tilde{l}}}}{F^2_{\sm{DSB}}}.
\end{equation}
In models under consideration, gravitino mass is usually between 
$1$~eV and $1$~GeV. 
Since the interaction Eq.\ 
(\ref{1}) is highly suppressed by
$F_{\sm{DSB}}$, 
the most probable process leading to gravitino production is the NLSP
decay. 
Note that
$\Gamma\bigl( \it \mbox{NLSP} \to \tilde{G} + \mbox{ SM particle}$ 
is determined by
$F_{\sm{DSB}}$, see Eq.\ (\ref{2}), so by measuring this width one gains
information about supersymmetry breaking scale in the secluded sector.
Other superparticles decay to Standard Model particles and NLSP via
electroweak and strong interactions.
 
Accelerator phenomenology of models with gauge mediation has been discussed
in detail in Ref.\ 
\cite{Signatures}. Depending on the parameters of the model, the decay 
$\it \mbox{NLSP}\to\tilde{G}+\mbox{SM particle}$
may show up in experiment in different ways 
\cite{gravitino_phenomenology}.
One has three possiblities, the choice being determined by the NLSP mean
path,
$$
l\approx 
5\l\frac{100~\mbox{TeV}}{\Lambda}\r^5\l
\frac{m_{3/2}}{1~\mbox{keV}}\r^2~\mbox{m}~~~~\mbox{(NLSP is photino)}
$$
\begin{enumerate}
\item 
The width is so small that NLSP decays outside the detector, so it is
``experimentally stable''. The gravitino mass does not exceed 1 keV in
this case 
\cite{Signatures,CDF_predictions}, so that 
$\sqrt{F_{\sm{DSB}}}>2\times 10^3$~TeV.
\item 
The width is so large that the corresponding vertex is hard to distinguish
from the NLSP production vertex.
\item 
It is possible to reconstruct the NLSP trajectory from its production
to its decay. 
\end{enumerate} 
Different signatures will correspond to these possibilities, depending
on the nature of NLSP. Namely, if NLSP is neutralino, at $e^+e^-$
colliders cases 2)\ and 3)\ result in the signature
$\gamma\gamma+E_{\sm T}\hspace{-.45cm}/\hspace{.45cm}$, while the
first possibility --- in missing transverse momentum, $E_{\sm
T}\hspace{-.45cm}/\hspace{.45cm}$.  If NLSP is slepton, the case 2)\
corresponds to the signature $ll+E_{\sm
T}\hspace{-.45cm}/\hspace{.45cm}$, and cases 1)\ and 3)\ show up as a
track of a charged particle much heavier than a lepton. In the latter
case, curvature of the track will change abruptly to leptonic
one. Since the initial beams leading to NLSP production consist of the
usual particles (protons or electrons), the $R$ parity conservation
requires even number of superparticles in the final state.

If NLSP is right-handed slepton, a characteristic
process in electron--positron collisions is
$e^+e^-\to\tilde{l}^+_{\sm R}\tilde{l}^-_{\sm R}\to
l^+_{\sm R}l^-_{\sm R}+E_{\sm T}\hspace{-.45cm}/\hspace{.45cm}$. Note
that four jets production (from hadronic decays of
four tau leptons) will be competetive in this case  with
$e^+e^-\to \chi_0\chi_0\to 4b+E_{\sm T}\hspace{-.45cm}/\hspace{.45cm}$.
The latter process is characteristic to models where NLSP is higgsino
\cite{CDF_predictions,Signatures}.

Phenomenology of neutralino (bino or higgsino) decays
$\tilde{N}\to\gamma\tilde{G}$ 
has been discussed in connection with different experiments: future
linear colliders
\cite{gravitino_collider}, Tevatron
\cite{Dimopoulos_Thomas_Wells,gravitino_phenomenology},
electron-photon accelerators
\cite{ep}.  

If NLSP is bino, mass and hypercharge difference between
right-- and left-handed sleptons results in a strong dependence of the
cross section from the
polarization of electrons
\cite{Signatures},
$$
\sigma(e^+e^-_{\sm L}\to \tilde{B}\tilde{B})/
\sigma(e^+e^-_{\sm R}\to \tilde{B}\tilde{B})\simeq 0.01,
$$
Besides, cascade processes
$e^+e^-\to\tilde{l}^+_{\sm R}\tilde{l}^-_{\sm R}\to
l^+l^-\gamma\gamma+E_{\sm T}\hspace{-.45cm}/\hspace{.45cm}$ 
would be seen if selectron masses are close to mass of bino.

Proton--antiproton colliders demonstate another kind of phenomenology. If
sleptons are much heavier than bino, then the following processes are
possible: 
$$p\bar{p}\to\tilde{W}^\pm\tilde{W}^0\to 4j\gamma\gamma+ 
E_{\sm
T}\hspace{-.45cm}/\hspace{.45cm},jjl\gamma\gamma+E_{\sm
T}\hspace{-.45cm}/\hspace{.45cm},l^+l^-l'\gamma\gamma+E_{\sm
T}\hspace{-.45cm}/\hspace{.45cm}
$$
Processes
$p\bar{p}\to\tilde{B}\tilde{B}\to\gamma\gamma+E_{\sm T}
\hspace{-.45cm}/\hspace{.45cm}$
will be suppressed by large squark masses, while Drell-Yan
process will dominate,
$p\bar{p}\to l^+l^-\gamma\gamma+E_{\sm
T}\hspace{-.45cm}/\hspace{.45cm}$.
The latter occurs either directly or 
in cascade through bino, depending on the type of NLSP.
One event of this type has been discovered at Tevatron by the CDF
collaboration 
\cite{CDF} with a decay-like kinematics. The possibility to interpret
this event in the framework of models with neutralino as NLSP has been
suggested in Refs.\ 
\cite{Signatures,CDF_predictions}. 

The possibility to interpret the
$e^+e^-\gamma\gamma+E_{\sm T}\hspace{-.45cm}/\hspace{.45cm}$
event in gauge mediated supersymmetry breaking models where NLSP is either
slepton or neutralino has been discussed in detail in Ref.\ 
\cite{CDF_predictions}. To make such an interpretation possible, one
should have  
$\tan\beta\simeq 1\div 3$ in models with neutralino as NLSP while
neutralino and higgsino masses have to be in the region
$50\div 100$~GeV. In Ref.\ \cite{CDF_predictions}, fairly narrow
regions of MSSM parameters were found where this kind of
interpretation is allowed, and other processes to look for at Tevatron
and LEP were considered which might refine these regions. Given this
explanation of the CDF event, numerous other processes should
be seen. In case the evidence for these processes would not be found in current
experimental data, the interpretation of an anomalous event is
questionable \cite{Dimopoulos_Thomas_Wells,CDF_predictions}.
Note that for 
$\tilde{\tau}_{\sm R}$ NLSP, 
it is unlikely to explain the CDF event in such a way \cite{kolda}. 
The event may be explained in the framework of supergravity approach,
but this possibility is not very realistic
\cite{CDF_predictions} since it requires rather strong fine tuning
\cite{Ambrosanio_Mele}. 

 Large values of $\tan\beta>30$ are
characteristic to the Minimal Model. To distinguish the models with $\tan\beta>30$, one may
use the processes $b\to sl^+l^-$ $(l=e,\mu)$ whose probabilities are
proportional to $\tan^6\beta$. At large $\tan\beta$, the
corresponding contribution is 2 to 3 times larger than in the Standard Model
\cite{B_to_ll}, so 
experimental discovery of these processes would be an indirect argument in favour of validity of
the Minimal Model of gauge mediated supersymmetry breaking.

\subsection{The most general messenger sector}
\label{D2.3}
In the Minimal model considered above, the messengers of supersymmetry
breaking are $Q$ and $\bar{Q}$ fields in
fundamental and antifundamental multiplets of the simplest Grand
Unification group, $SU(5)$. In general case, one can have several sets
of messengers in different vector--like representations,
for instance, ${10}+{\overline{10}}$, ${15}+{\overline{15}}$, ${24}$ 
for
$SU(5)$,
${16}+{\overline{16}}$ 
for $SO(10)$ etc. Each specific case has certain distinctive
characteristics (e.g., fermionic messengers in the adjoint
representation may mix with gauginos). 

\subsubsection{Perturbative unification}
Given the requirement of weakness of gauge interactions at
$M_{GUT}$, one has an upper bound on the number of messengers
$n_r$ in representations $r$ which is related to their contribution to the $\beta$
function coefficients. If $\alpha_{GUT}\lesssim 1$, then the effective
number of messenger generations
$n\equiv \sum_{(r)}n_rc_r$,
where 
$c_r$ are Dynkin indices of the corresponding representations  
($c_r=1$ for the fundamental and
$c_r=3$ for the antisymmetric tensor representation of $SU(5)$),
should not exceed four. This means that one can use either four
fundamental generations of messengers or one antisymmetric and one
fundamental 
\cite{weak_scale_phenomenology} 
in the $SU(5)$ case, or not more than one generation of
(${16}+ \bar{16}$) in $SO(10)$.  
These bounds may be somewhat relaxed for specific relative
values of messenger thresholds.

Obviously, messengers should not necessary belong to complete GUT
multiplets. This class of models has been considered in Ref.\ 
\cite{Martin}. In that paper, the most general inequalities are presented
which are true for any mass and any representation of
messengers. Suppose that the fields completing the
messenger multiplet to a GUT multiplet have masses of order the
GUT scale. Let
$n_{\sm X}$ be the number of messengers carrying the same gauge quantum
numbers as the SM fields
($X=L,D,E,U,Q$).  
Then bounds from the smallness of the MSSM gauge couplings at
$M_{\sm{GUT}}$ result  
in the following allowed sets of multiplets 
\cite{Martin} for messenger masses
$M_\pm\lesssim 10^7$~çÜ÷,
\begin{displaymath}
\label{n_x_mes}
\begin{array}{lll}
(n_{\sm L},n_{\sm D},n_{\sm E},n_{\sm U},n_{\sm Q}) 
 & \le & ({\bf 1},{\bf 2},{\bf 2},{\bf 0},{\bf 1}) \\
 & \le & ({\bf 1},{\bf 1},{\bf 1},{\bf 1},{\bf 1}) \\
 & \le & ({\bf 1},{\bf 0},{\bf 0},{\bf 2},{\bf 1}) \\
 & \le & ({\bf 4},{\bf 4},{\bf 0},{\bf 0},{\bf 0})
\end{array}
\end{displaymath}

Generalization of the expressions for soft terms given in section
\ref{spectrum} to the case of several messenger generations and,
perhaps, several singlets
$S$, is straightforward even when mixing among generations is taken into
account. Important for phenomenology is that gaugino masses increase
as the effective number of messenger generations, 
$n$, 
while the scalar masses increase as 
$\sqrt{n}$. 
Thus slepton is the most probable candidate NLSP in models with a few
generations of messengers. Lower bound on $\Lambda$ is somewhat
relaxed in this class of models.

\subsubsection{Strong unification}
\label{strong-unif}
It is not necessary, however, to require the smallness of couplings
up to 
$M_{\rm GUT}$. In some cases, especially in direct mediation models
(see section
\ref{sec:direct}),
where 
$n$ is typically large, one may try to relax the constraints coming
from perturbative unification since controllable and
phenomenologically acceptable unification in the strongly coupled
domain may be considered. 

The possibility of gauge coupling unification in the strong coupling
regime has been considered in the framework of both the Standard Model
and its supersymmetric extensions
\cite{Maiani--Cabibbo-Farrar}. Recently, this problem attracted
interest again
\cite{nonpert-unif} after more precise measurements of the 
gauge coupling constants at $M_Z$ have been carried out. The latter results
differ from the two-loop unification predictions by more than one
standard deviation (see, e.g., Ref.\ \cite{kabak}). 

Note that running gauge couplings of the MSSM $\alpha_1$ and
$\alpha_2$ increase with energy, so $SU(2)\times U(1)$ is not
asymptotically free. These couplings, however, run relatively slow, so
Landau poles of these two groups appear at energies higher than the
unification scale. Together with the asymptotic freedom of QCD this
means that below $M_{GUT}$ all gauge couplings are small, and
perturbative analysis is valid. This picture implies the existence of
the ``desert'', i.e. absence of particles in huge region of masses
between superparticle and unification scales. When new particles, like
several multiplets of messengers, are introduced, the first
coefficients of the $\beta$ functions increase, so gauge couplings may
become large at the unification scale.

Despite the fact that unification in this case occurs at the strong
coupling, it is unexpectedly controllable from the low energy point of
view, especially in the supersymmetric case \cite{Ross}. 
Let us consider one-loop evolution of the coupling constants in an asymptotically
non-free unified theory. If
$M_G$ is the unification scale and $\alpha_G$ is the value of the unified
gauge coupling at that scale, then the renormalisation group
equations 
\begin{equation}
{d\alpha_i\over dt}=b'_i\alpha_i^2
\label{RG}
\end{equation}
have a solution
$$
\alpha_i^{-1}(Q)=\alpha_G^{-1}+b'_i t,
$$
where $t={1\over 2\pi}\ln{Q\over M_G}$ and $b'_i$ 
are the first coefficients of the beta functions of the gauge
couplings in the asymptotically non-free theory 
($b_1=33/5$, $b_2=1$, $b_3=-3$ are the corresponding coefficients of
the MSSM). 
Consider running of the {\em ratios} of pairs of the gauge couplings.
It follows from Eq.\ \eq{RG} that at one loop these ratios have
infrared fixed points,
$$
{\alpha_i\over\alpha_j}={b'_j\over b'_i}.
$$
These fixed points are reached at the energies which are model-dependent
and are determined by the renormalization group flow of ratios of couplings,
$$
{\alpha_i(Q)\over\alpha_j(Q)}={\alpha_G^{-1}+b'_j t \over
\alpha_G^{-1}+ b'_i t}.
$$
The condition that the fixed point is almost reached is
$|t|\gg\alpha_G^{-1}/b'_i$.  In the case of MSSM, one has
$\alpha_G^{-1}\sim 24$, so that the fixed point of $\alpha_2/\alpha_1$ occurs
at $|t|\gg 24$, i.e., at $Q\ll
M_G\cdot\exp(-48\pi)\sim 10^{-66} M_G$ which certainly rules out the
possibility of the fixed point analysis. However, with new matter
added, the situation changes drastically --- $b_i$ increase and
$\alpha_G^{-1}$ decreases. 

In the case of messengers in the complete vector-like representations of the
unified gauge group, it is a single number,
$n$, 
that determines the contribution of messengers to the
$\beta$ functions of all three couplings,
$$
b'_i=b_i+n.
$$
For $n\ge 5$, the unification occurs
at strong coupling. To estimate the energy scale where the ratios of
couplings get close to the fixed point values, let us take
$\alpha_G=1$. Then even at $n=5$, the ratios are almost constant at
$Q<0.04 M_G$. 

For given $n$, the threshold corresponding to messenger mass
is uniquely determined. Indeed, the low energy running of MSSM couplings is
known, and the couplings {\em should} have the ratios
equal to $b'_i/b'_j$ at the threshold. The energy where ratios of MSSM running
couplings, 
determined experimentally at $M_Z$, reach their fixed point values
corresponds to the messenger threshold. Note that at $n\ge 5$ the
corresponding thresholds are deep in the region of attraction of the fixed
points. For $n=5$, for example, the threshold is between 1
and 10~TeV, much lower than $0.04\cdot M_G$. This means that the
fixed-point approach is self-consistent. The values of thresholds can
be read out from Ref.\cite{Ross}; values of $6\le n \le 20$ are
consistent with current bounds on the messenger mass \cite{to-appear}.

The most interesting feature of this scenario is that the strong
unification constrains significantly the parameter space of
multi-messenger models of gauge mediation. Namely, the mass scale of the
messenger fields -- one of the two parameters describing the
superpartner masses -- is determined from self-consistency condition
for a given effective number of messengers $n$. 
Decreasing the number of
parameters results in stronger phenomenological limits
on the secluded sector
\cite{q98,to-appear}. 

From the low energy MSSM point of view we just have new boundary
conditions for running of the gauge couplings. Instead of requiring
the equality of couplings at
$M_G$ (as in the case of perturbative unification), one should fix their
ratios at the messenger scale. Details of evolution of the couplings
near $M_G$, where they are large, are unknown; however, they do not affect
significantly the low-energy predictions
\cite{Ross,NewRoss}.

\subsection{Messenger-matter mixing}
\label{D3.4}

Another possibility to extend  the Minimal Model is to introduce
messenger-matter mixing~\cite{dinem}. Detailed analisys of the model
with mixing can be found in Refs.~\cite{we,we_new}. Here we consider
only its most significant features.

Let us consider a theory with one messenger generation in fundamental
representation. Messenger quantum numbers may coincide with
quantum numbers either of quarks and leptons or Higgs,
depending on R-parity of messengers.  In the latter case, triplet
messenger fields might give rise to fast proton decay due to 
possible mixing with Higgs fields, unless the
corresponding Yukawa couplings are smaller than
$10^{-21}$ at messengers masses about
100~TeV~\cite{Zhang}~\footnote{It is worth noting that a
  possibility to identify messengers and Higgs fields has been
  discussed~\cite{Dvali-Shifman,zurab}. The main problem of this
approach is related to triplet Higgs
  fields. Namely, if they are light then
  it is highly non-trivial (but  still possible in some cases) to
  suppress fast proton decay while heavy triplet messengers  lead
  to very light gluinos which is hardly acceptable from
  phenomenological point of view. Light gluino problem also arises if
  one identifies messengers with Higgs particles responsible for intermediate
symmetry breaking, $SU(2)_{\sm L}\times SU(2)_{\sm
R}$ to $SU(2)_{\sm L}\times U(1)_{\sm Y}$~\cite{cc}.}.  

Let us consider in more detail the case when mixing of the messengers
with matter fields is possible. Then doublet components
of the $Q$ multiplets have the same quantum numbers as MSSM left
leptons and tree-level messenger-matter mixing is possible
\cite{dinem}. It is convenient to introduce common notations for
left leptons and doublet messengers,
\begin{displaymath}
L_{\tilde{i}} = (\tilde{l}_{\tilde{i}},l_{\tilde{i}}) =
 \left\{ \begin{array}{ll}
(\tilde{e}_{{\sm L }\tilde{i}}, e_{{\sm L }\tilde{i}}) & 
\textrm{, $\tilde{i}=1,..,3$}\\
(q, \psi_{q}) & \textrm{, $\tilde{i}=4$}
\end{array} \right.
\end{displaymath}
Then the mixing term in the superpotential has the form
\begin{equation}
\label{y-term}
{\cal W}_{mm} = H_{{\sm D}}L_{\tilde{i}}Y_{\tilde{i}j}E_{j},
\end{equation}
where $H_{\sm D} = (h_{{\sm D}},\chi_{{\sm D}})$ is the down Higgs
superfield and $E_{j} = (\tilde{e}_{{\sm R}j},e_{{\sm R}j})$ are
singlet lepton superfields. Hereafter $\tilde{i},\tilde{j}=1,..,4$ label
the three left lepton (and right quark) generations and the messenger
field, $i,j=1,..,3$ correspond to the three leptons (quarks) and
$Y^{(5)}_{\tilde{i}j}$ are the $4\times3$ matrices of
Yukawa couplings,
$$
Y_{\tilde{i}j} = 
\left(
\begin{array}{ccc}
Y_{e}& 0 & 0\\
0 & Y_{\mu} & 0\\
0 & 0 & Y_{\tau}\\
Y_{41}& Y_{42}& Y_{43}
\end{array}
\right).
$$
This mixing is natural in the sense that it does not break any MSSM
symmetry.  

Messenger masses are not a consequence of the Higgs effect in the observable
sector.  Consequently, overall mass matrices and matrices of Yukawa
couplings are not proportional to each other. It is convenient to
integrate out heavy messenger fields 
in order to analyse low energy processes with lepton flavour
violation.  One should first check, however, that these processes
are suppressed at the tree level in the model with
mixing (\ref{y-term}).

Let us diagonalize mass matrix of fermion fields (including left and
right messenger fermions)  in order to explore tree level mixing.

In principle, non-diagonal terms in the corresponding rotation
matrices can lead to lepton flavour violation through one loop
diagrams involving sleptons and gaugino in analogy to models with
gravity mediation~\cite{barbieri}. These terms in
the model with mixing are highly suppressed by the seesaw mechanism. Even at $Y_i\sim
1$ the corresponding rates of lepton decays are too small to be
observable in current experiments.

Mixing in scalar sector is also negligible at the tree level.  
After the scalar messengers are integrated out at the tree level,
the lepton flavour violating terms in the mass matrix of right
sleptons are of order 
\begin{equation}
\label{drev}
Y_iY_j(v_{{\sm D}}^2x^2+\frac{\mu ^2
v_{{\sm U}}^2x^2}{\Lambda ^2})
\end{equation}
for generic values\footnote{Hereafter, by generic values of $x$ we
  understand $x$ not very close to 1. Only at $x\approx 1$ some
  numerical coefficients slightly change in the following equations.} of
$x$.  These
terms are smaller than the one loop contributions (see below) at
$\Lambda \gtrsim 10$ TeV. Mixing terms in the left slepton sector
are even less significant.

 Mixing appears in the quadratic as well as in the interaction part of
the effective Lagrangian after messengers are integrated out. Let us
find the leading term in large parameter $\Lambda$.
The contribution to the lepton mixing matrix is proportional to the
vacuum expectation value of the corresponding Higgs field.
Contribution of the order $\Lambda$ due to the inertactions of SM fermions with
messengers is forbidden because of chirality~(see section~\ref{D1.3}).
So large mixing can appear only in the slepton sector. With
superpotential Eq.~(\ref{y-term}), corresponding terms appear in the mass
matrix of right sleptons due to their interactions with the Higgs and
messengers,

\begin{equation} 
\delta m^2_{ij}=
- \frac{1}{16\pi^2} \frac{\Lambda^2}{x^2} \Big\{ 
-\ln{(1-x^2)}-\frac{x}{2}
 \ln{\Big(\frac{1+x}{1-x}\Big)}\Big\}Y_{4i}^*Y_{4j}.
\label{delta-mixing}
\end{equation}

Let us note that the term~(\ref{delta-mixing}) is larger than the 
tree-level mixing of right sleptons. At small $x$,
Eq.~(\ref{delta-mixing}) has the form,
$$
\delta m^2_{ij}\approx -\frac{x^2}{6}
\frac{1}{16\pi^2}\Lambda^2Y_{4i}^*Y_{4j}.
$$
One-loop contribution exceeds the tree-level one provided the
following inequality holds (see Eq.~(\ref{delta-mixing})),
\begin{equation}
\label{***}
\upsilon_\sm{D}^2+\frac{\mu^2\upsilon_\sm{U}^2}{\Lambda^2}<\frac{1}{6}
\frac{1}{16\pi^2}\Lambda^2.
\end{equation}
At $\Lambda>10$~TeV and typical values of $\mu\approx 500$~GeV, the
relation ~(\ref{***}) is indeed satisfied. The contribution of Eq.~(\ref{delta-mixing})
to  $\tilde{m}_{\sm R}^2$ is negative so that the positivity of slepton
mass squared results in theoretical limits on Yukawa couplings 
$Y_{4i}$ \cite{dinem}. 

Let us consider the role of mixing in low-energy lepton physics.  It
was shown in Ref.\ \cite{we} that the rates of lepton flavour violating
processes are not too large in the natural range of parameters. For
instance, the current experimental limit on the branching ratio of
$\mu\to e\gamma$ decay $\mbox{Br}(\mu\to e\gamma)< 4.9\cdot
10^{-11}$~\cite{Particle_Data} gives for the mixing parameters,
\begin{equation}
\sqrt{Y_{41}Y_{42}}x\lesssim 3.0\times 10^{-3}
\label{y-me}
\end{equation}
It is worth noting that parameters $Y$ and $x$ appear in the
combination $Yx$ in this inequality. However, this fact does not imply
that $Y$ can be arbitrarily large provided $x$ is small enough. At the
values of $x$ of order gauge coupling constants, mixing in gauge
vertices due to the loops involving messengers and Higgs particles is
significant~\cite{we_new}. This sets a limit on
$Y\sqrt{\alpha_i}$.

Similar mechanisms are responsible for flavour violating $\tau$ decays, 
$\tau\to e\gamma$ and $\tau\to \mu\gamma$. 
Corresponding limits follow from the current bounds on the branching ratios of
these decays,
$$
\mbox{Br}(\tau\to e\gamma)<2.7\cdot 10^{-6}  ~~~ 
\makebox{and} ~~~ \mbox{Br}(\tau\to\mu\gamma)<3.0\cdot 10^{-6}.
$$

Messenger-matter mixing is possible in quark sector as well, because
some of the messengers have quantum numbers of $d$-quarks. This mixing
leads to additional contributions to FCNC processes such as
${K^0-\bar{K}^0}$ mixing, which happen in the Minimal Model due to CKM
matrix. Generally speaking, the corresponding Yukawa coupling is not
related directly to lepton couplings $Y_{\tilde{i}j}$ from~(\ref{y-term}).
However, the quark-messenger mixing constants also do not exceed
$10^{-3}$ which is compatible with possible unification.

Let us note that messenger-matter mixing leads to the negative
contributions of Higgs doublet masses squared~\cite{dinem}. It can
significantly alter electroweak breaking in the theory. For instance,
in the case of fundamental messenger representations, the 
limits on the Yukawa couplings $Y$ allow the contribution to
$m^2_{h_{\sm D}}$ to be large enough to make $\tg\beta\sim
1$. At these low values of $\tg\beta$, NLSP is neutralino so that
phenomenology can be rather different from the MM case.

Similar mixing is possible for other messenger representations as
well. For example, in the case of messengers in  antisymmetric
representations of $SU(5)$ group, the limits on the mixing parameters
coming from flavour physics are somewhat weaker than for fundamental
messenger representations at the same value of $\Lambda$.

Finally, let us note that messenger-matter mixing leads to 
the absence of cosmological problems related to 
stable charged (and coloured) particles (mesengers) (see section
\ref{dark-matter}). The solution of this problem due to tree level
mixing 
is one of the most natural approaches.

\section{Towards incorporation of the secluded sector}
\label{sec:secluded}
We have discussed so far the model-independent aspects of gauge
mediated supersymmetry breaking. All significant information about
the secluded sector was contained in parameters $\Lambda$, $x$ and
supersymmetric Higgs mass $\mu$. However, the combined analysis of
the effects both in the observable and in the secluded sector is one
of the essential advantages of gauge mediation models As compared to
gravity mediation scenario. In this section we
consider specific models incorporating secluded sector, where
dynamical supersymmetry breaking (DSB) takes place.

{\tolerance=1000
\subsection{Non-perturbative dynamics of supersymmetric gauge {theories}}
\label{dsb}
}
The recent revival of interest in construction of realistic
models with low energy supersymmetry is in great part due to
developments in the study of supersymmetric theories. On 
the one hand, these methods allow to enlarge significantly the
number of theories exhibiting DSB and, on the other hand, they provide
powerful tools for the detailed analysis of the consequences
of DSB.  We concentrate on the recent achievements in the
study of non-perturbative effects in supersymmetric theories before
the consideration of the specific models.  The earlier results are
discussed in detail in Refs.~\cite{ADS-84,ADS-85,instanton_contribution}.

\subsubsection{Flat directions: classical theory}
\label{class}
The presence of the classical flat directions --- submanifolds of the
space of fields which consist of, generally speaking, physically
different supersymmetric minima of the tree level potential --- is the
characteristic feature of supersymmetric theories. Eq.~\eq{FandD}
implies that the classical vacuum configuration satisfies the
following conditions,
\begin{equation} 
\label{vacua} 
D^a=0\;,\;\;\;F=0\;. 
\end{equation}
Usually, in order to solve Eqs. (\ref{vacua}), one first finds field
configurations with $D^a=0$, and study whether they satisfy the
condition of stationarity of the superpotential. The second stage is
purely technical and does not present fundamental difficulties usually.
That is why we will first consider the case $W=0$. One can find in
literature the explicit forms of vacuum configuration in a wide
number of specific models, but a universal method to obtain the
solutions of~(\ref{vacua}) is missing. As it has been pointed out
already in Ref. \cite{ADS-84}, the dynamics along the flat direction is
conveniently described in terms of holomorphic gauge
invariants composed of the matter fields. The following useful
theorem was formulated and proven in Ref.~\cite{luty}.

{\it Theorem.  \newline The classical space of vacua is isomorphic to
  the space of the constant field configurations, which cannot be
  transformed one to another by the action of the complexified
  gauge group. Holomorphic gauge invariants composed of the matter
  fields are the natural coordinates on this space.}

To prove this statement, gauge fixing conditions with additional
(in comparison to Wess-Zumino gauge)  symmetry were used in Ref.~\cite{luty}.
The field $V$ has the following form in this gauge,
$$ 
V=C-\theta\sigma^{\mu}\bar{\theta}v_{\mu}+ 
i\theta\theta\bar{\theta}\bar{\lambda}-i\bar{\theta}\bar{\theta}\theta\lambda 
+\frac{1}{2}\theta\theta\bar{\theta}\bar{\theta}D\;. 
$$
Then the theory is invariant under the group $G_c$, which is
complexification of the group $G$. One can show now that for every
constant field configuration it is possible to find the value of the
field $C$ in such a way that the vacuum energy is equal to zero. This
condition can be written in the following form in this gauge,
$$ 
\frac{\partial}{\partial C}\phi^{\dagger}_i\e^C\phi_i=0\;. 
$$
Every field configuration which is obtained in this way can be
uniquely (up to a usual gauge transformation) transformed to the
configuration with $C=0$ by the action of $G_c$. This means that flat
directions are in fact equivalent to the quotient space of constant fields
$\phi$ with respect to the group $G_c$.
The corresponding manifold is described in terms of the full set of
invariants of this action. Holomorphic gauge invariant polynomials
form this set.

This theorem does not provide us with the explicit form of the
classical flat directions in terms of elementary fields. However,
description of the space of vacua through gauge invariants is
convenient in the quantum theory, where these invariants become effective
low energy degrees of freedom (moduli). Generally speaking, taking into account
quantum corrections to the classical flat directions is necessary for the
complete analysis of the vacum structure. Let us consider this point now.
\subsubsection{Flat directions: quantum theory}
\label{flat-dir:quant-th}
Typically, quantum corrections change the vacuum structure of the
theory.  Non-renormalization theorems imply that quantum flat
directions coincide with the classical ones within the perturbation
theory (abelian theory with nonzero sum of charges of matter fields is
the only exception). But these theorems tell nothing about
non-perturbative corrections which can alter the classical
consideration. A classical direction can disappear completely or
in part because of the generation of non-perturbative superpotential,
can change its geometry, or remain the same but describe the theory
which is completely different from the classical one. Finally, even
the absence of classical flat directions can be very important in the
full quantum theory.

Practically all these possibilities are realized in the simplest
supersymmetric gauge theory --- Supersymmetric Quantum Chromodynamics
(SQCD).  It contains $N_F$ flav\-o\-urs of chiral superfields of
(anti-)quarks $(\bar{\Phi}_i), \Phi_i$ in the (anti-)fundamental
representation of the gauge group\footnote{In what follows all 
  formulas will be presented for the case $N_C>2$. The picture remains
  qualitatively the same for $N_C=2$, but concrete formulas can
  change due to the fact that the fundamental representation of $SU(2)$ 
  is isomorphic to its conjugate.} $SU(N_C)$. The Lagrangian is given by
Eq.~(\ref{matter_part}) with the superpotential
\begin{equation} 
\label{superpotential} 
W=\sum_1^{N_F}m_i\Phi_i\bar{\Phi}_i\;. 
\end{equation} 
Let us consider the dynamics of this theory at different values of $N_F$.
Supersymmetry implies holomorphy of superpotential as function of
chiral superfields and coupling constants~\cite{amati,seiberg92}.
This requirement combined with the invariance under non-anomalous
symmetries is a powerful tool allowing to find out the exact form
of the superpotential in many cases. Also, it  might occur useful to find out 
the superpotential in some limiting cases, when the theory gets simplified.

$\bf{N_F=0}$ \newline We are dealing with Yang-Mills theory with one
adjoint fermion (superpartner of gauge boson --- gaugino) in this
simplest case. There are no flat directions at the classical level,
but non-trivial quantum effects are present. Namely, $R$-symmetry is
broken by the gaugino condensate.\footnote{It is worth
mentioning, that in the absence of the matter fields the continuous
$R$-symmetry is broken by anomalies to its discrete subgroup
$Z_{2N_C}$. It is the latter one which gets broken down $Z_{2}$ by
gaugino condensate.} This condensation leads to the effective
superpotential
\begin{equation} 
\label{gauginosuperpotential} 
W_{eff}=c\Lambda^3\;. 
\end{equation}
As usual, $\Lambda=\mu\e^{\frac{2\pi i\tau(\mu)}{3N_C}}$ is the scale
where the gauge dynamics becomes non-perturba\-ti\-ve and $c$ is some
constant. One can obtain the expression (\ref{gauginosuperpotential})
from symmetry considerations if $\tau$ is treated as an external field
\cite{peskin} transforming as follows,
$$ 
\tau\to\tau+\frac{N_C}{\pi}\alpha 
$$
under $R$-symmetry (which becomes non-anomalous then).  The
meaning of the superpotential (\ref{gauginosuperpotential}) is that
the value of gaugino condensate is given by
\begin{equation} 
\label{gauginocondensate}
<\lambda\lambda>=16\pi i\frac{\partial}{\partial\tau}W_{eff}=- 
\frac{32\pi^2}{N_C}c\mu^3\e^{\frac{2\pi i\tau(\mu)}{N_C}}\;. 
\end{equation}
There are $N_C$ vacua with different phases of the gaugino condensate.
A surprising fact is that it is absolutely unclear which diagrams
contribute to the condensate (\ref{gauginocondensate}). Such a Green's
function corresponds to the transition with the change of the
topological charge at $\frac{1}{N_C}$ and cannot be obtained by means
of instanton calculations.

The parameter $\Lambda$ becomes a function of  dynamical fields in
more complicated theories and the superpotential
(\ref{gauginosuperpotential}) generates additional terms in the
scalar potential. An interesting application of this mechanism is described
below (see\ section \ref{heavy-messengers}).

$\bf{0<N_F<N_C}$ \newline In this case, at zero quark masses, there
are classical flat directions parametrised by the set of ``meson'' gauge
invariants,
\begin{equation} 
\label{mesons} 
M_{ij}=\Phi_i\bar\Phi_{j}\;. 
\end{equation}
The effective superpotential which is invariant under all non-anomalous
symmetries of the theory has the following form \cite{ADS-84},
\begin{equation} 
\label{mesonsuperpotential} 
W_{eff}=c\left(\frac{\Lambda^{3N_C-N_F}}{\det M}\right)^{\frac{1}{N_C-N_F}}\;. 
\end{equation} 
The superpotential (\ref{mesonsuperpotential}) lifts all classical
flat directions and there appear ``run-away'' vacua at the infinite
values of squark fields. At non-zero quark masses there are again
$N_C$ different vacua at finite values of the matter fields with the
gauge group broken to $SU(N_C-N_F)$. The chiral symmetry is broken by the
gaugino condensate, which is related to the squark condensate through
the Konishi anomaly (see, e. g., Ref.~\cite{amati}).

The mechanism of generating the superpotential
(\ref{mesonsuperpotential}) depends on the number of flavours
$N_F$. At $N_F=N_C-1$ one can explicitly obtain the value of the
constant $c$ in Eq.~(\ref{mesonsuperpotential}) by calculating the
one-instanton Green's function
\cite{ADS-84,instanton_contribution,amati}, but at smaller $N_F$
the superpotential is generated due to gaugino condensation in the low
energy theory. If one knows $W_{eff}$ at some value of $N_F$ then it
is straightforward to obtain $W_{eff}$ for $N_F-1$ quark flavours. One
should give the mass $m$ to one of the quarks and take the limit
$m\to\infty$. The value of $\Lambda$ in the low energy theory can be
obtained from matching of the coupling constants at the decoupling
scale $m$. This procedure serves as a non-trivial check of the
results.

$\bf{N_F=N_C}$ \newline In addition to meson gauge invariants
(\ref{mesons}), there are baryon $B$ and antibaryon $\bar{B}$
invariants in this case (we present here their general form for arbitrary
$N_F\geq N_C$; only one pair of baryon and antibaryon is present at
$N_F=N_C$),\begin{equation}
\label{baryons} 
B^{i_{N_C+1}...i_{N_F}}=\epsilon^{i_{N_1}...i_{N_F}}\Phi_{i_1}\cdot...\cdot
\Phi_{i_{N_C}}
\end{equation}
and analogously for antibaryons. One can check explicitly that at the
classical level these invariants satisfy the following constraint
\begin{equation}
\label{classicalconstraint}
B\bar{B}-\det M=0\;.
\end{equation}

The classical space of vacua exhibits singularities. They correspond
to the appearance of additional massless degrees of freedom at some
points of the moduli space. These are gluons of the unbroken gauge
group.  At the quantum level, all singularities turn out to be
smoothed out and the gauge group is broken in the entire moduli space
\cite{seiberg1}. The mechanism responsible for this phenomenon is 
illustrated in Fig.~\ref{qms}. 
\begin{figure}
\label{qms}
\begin{picture}(0,0)(0,250)%
 \epsfig{file=kartinka.pstex}%
 \end{picture}%
 \setlength{\unitlength}{0.00075000in}%
 \begingroup\makeatletter\ifx\SetFigFont\undefined
 \def\x#1#2#3#4#5#6#7\relax{\def\x{#1#2#3#4#5#6}}%
 \expandafter\x\fmtname xxxxxx\relax \def\y{splain}%
 \ifx\x\y   % LaTeX or SliTeX?
 \gdef\SetFigFont#1#2#3{%
   \ifnum #1<17\tiny\else \ifnum #1<20\small\else
   \ifnum #1<24\normalsize\else \ifnum #1<29\large\else
   \ifnum #1<34\Large\else \ifnum #1<41\LARGE\else
      \huge\fi\fi\fi\fi\fi\fi
   \csname #3\endcsname}%
 \else
 \gdef\SetFigFont#1#2#3{\begingroup
   \count@#1\relax \ifnum 25<\count@\count@25\fi
   \def\x{\endgroup\@setsize\SetFigFont{#2pt}}%
   \expandafter\x
     \csname \romannumeral\the\count@ pt\expandafter\endcsname
     \csname @\romannumeral\the\count@ pt\endcsname
   \csname #3\endcsname}%
 \fi
 \fi\endgroup
 \begin{picture}(7224,2020)(1189,-1850)
 \put(4051,-1261){\makebox(0,0)[lb]
{\smash{\SetFigFont{11}{13.2}{rm}cocorrections}}}
 \put(4051,-811){\makebox(0,0)[lb]{\smash{\SetFigFont{11}{13.2}{rm}ququantum}}}
 \end{picture}
\mycaption{Quantum deformation of the classical moduli space.}
 \end{figure}
 Formally, the classical constraint (\ref{classicalconstraint}) takes
 the following form in the quantum theory
\begin{equation}
\label{quantumconstraint}
\det M-B\bar{B}=\Lambda^{2N_C}
\end{equation}
(vacuum expectation values (vev's) of the corresponding fields are
understood on the left hand side of this equation). To check it one
can add the sources, $W_{tree}= \Tr mM+bB+\bar{b}\bar{B}$ in order to
find the vev's of the corresponding operators and take the limit of
zero sources. It is possible to obtain all points of the manifold
(\ref{quantumconstraint}) by varying the ways of taking this
limit. For instance, the Lagrangian of massive SQCD is obtained if
$b=\bar{b}=0$ from the very beginning.  Selection rules and exact
instanton calculations give the following vev's of the gauge
invariants~(\ref{mesons}) and (\ref{baryons}) in this theory,
$$
M_{ij}=\Lambda^2(\det{m})^{\frac{1}{N_C}}m^{-1}_{ij}\;,
$$
$$
B=\bar{B}=0\;.
$$
These vev's indeed satisfy Eq.~(\ref{quantumconstraint}).

This result can be also related to the description of SQCD at
smaller number of flavours. To do this, one should give mass to
one of the quark fields and exclude this field by making use of the
constraint  (\ref{quantumconstraint}).

The phenomenon just described is called deformation of the moduli
space and may lead to supersymmetry breaking.

$\bf{N_F=N_C+1}$ \newline Mesons and baryons (\ref{mesons}),
(\ref{baryons}) describe the classical flat directions in this case as
well, but they satisfy three different constraints,
\begin{equation}
\label{classicalconstraint+1}
B_i\bar{B}_j-\det M(M^{-1})_{ij}=0
\end{equation}
$$
M_{ij}B_i=M_{ij}\bar{B}_{j}=0\;.
$$
The main difference with the previous case is that the classical
constraints (\ref{classicalconstraint+1}) are still valid at the
quantum level. One can check this by introducing sources in a similar
as in the case $N_F=N_C$. So we are dealing with the quantum moduli
space which exactly coincides with the classical one.  Classical
singularities are not smoothed out here, but they are present at the
quantum level as well. However, their physical meaning is
different. In both cases their presence is related to the
additional massless particles in the spectrum.  At the classical
level, these particles are gauge bosons of the unbroken group. This
cannot be the case in quantum theory, because the dynamics is
non-perturbative and colourless degrees of freedom are expected.

Let us consider a singular point $M=B=\bar{B}=0$ for example. The fact
that all global symmetries of the theory are unbroken at this point
indicates that all mesons (\ref{mesons}) and baryons (\ref{baryons})
are massless there. A powerful tool to check the conjectures of this
type is

{\it the 't Hooft anomaly matching conditions \cite{'thoft}:  \newline
  Anomalies calculated with elementary and composite degrees of
  freedom should coincide for all global
  symmetries of the theory that are not broken spontaneously.}  

The proof of this statement deserves to be presented here because of
its simplicity and elegance. Let us consider a theory with the gauge
group $G_C$ and global symmetry group $G_F$. Let us extend the gauge
group to $G_C\times G_F$, in such a way that dynamics of $G_F$ will
not alter the dynamics of $G_C$. Then, besides additional gauge
bosons, one should add massless fermions charged only under $G_F$ in
order to obtain non-anomalous theory with weakly coupled group $G_F$
at low energies.  In the infrared region the theory is described by
the same composite fields as the original theory and by the new fields
which are neutral under $G_C$ and, consequently, are not involved in
the non-perturbative dynamics of this group. The group $G_F$ is
non-anomalous at low energies as well. Then the fact that the
additional fields give the same anomalies at high and low energies
implies that composite degrees of freedom contribute to the anomalies
in the same way as elementary fields of the theory.

The 't Hooft matching conditions are meaningful if the matrix elements of
the divergences of the corresponding currents are well defined
both at high and low energies. For instance, the abelian axial current in
ordinary QCD does not satisfy this criterion, because its divergence
is related to the field strength of gluons, which is not well
defined in the non-perturbative region. This implies that there is no
axial symmetry in the low energy theory and anomaly matching is meaningless.

Certainly, matching of anomalies is not the proof of validity of a
given effective description of the theory under consideration, but it
serves as a highly non-trivial check\footnote{Recently
~\cite{intrilligator} an example of the model has been presented where
the 't Hooft conditions are satisfied but the supposed effective
theory does not provide a correct description of the low energy
dynamics.}.

In the case of SQCD with $N_F=N_C+1$, the non-anomalous group of global
symmetries is $SU(N_F)_L\times SU(N_F)_R\times U(1)_B\times U(1) _R$. Quarks
have the following quantum numbers with respect to this group, 
$$
\Phi=(N_F,1,1,\frac{1}{N_F})\;,~~~
\bar{\Phi}=(1,\bar{N}_F,-1,\frac{1}{N_F})\;, 
$$
and composite degrees of freedom are
\bg
M=(N_F,\bar{N}_F,0,\frac{2}{N_F})\notag \\
B=(\bar{N}_F,1,N_F-1,\frac{N_F-1}{N_F})\;,~~~
\bar{B}=(1,\bar{N}_F,-N_F+1,\frac{N_F-1}{N_F})\notag\;.
\end{gather}
It is straightforward to check that contributions to anomalies from
fundamental and composite particles are equal. Let us consider, for
example, Green's function of $SU(N_F)_L^2\times U(1)_R$ currents. At high
energies the contribution to its anomaly is equal to $N_Cd^{(2)}(N_F)$
(we leave only group factor, and skip the factor which is identically
the same for all Green's functions), $d^{(2)}(N_F)$ is the Dynkin index of
quark representation. In the effective theory, the anomaly is saturated
by baryons and it is equal to
$$
R(B)d^{(2)}(\bar{N}_F)=
(N_F-1)d^{(2)}(N_F)=N_Cd^{(2)}(N_F)\;.
$$
In a similar way, one can check the anomaly matching for other Green's
functions.

The correct description of low energy excitations about the vacua~
(\ref{classicalconstraint+1}) is given by the effective superpotential
\begin{equation}
\label{effectivesuperpotential+1}
W_{eff}=\frac{1}{\Lambda^{2N_F-3}}(B_iM_{ij}\bar{B}_{j}-\det M)\;,
\end{equation}
which is invariant under all non-anomalous symmetries. The condition
of its stationarity implies the constraints
(\ref{classicalconstraint+1}); the superpotential
of Eq.~(\ref{effectivesuperpotential+1}) describes massless fields at the
origin, while at the large vev's some of the fields become
massive, which corresponds to their explicit elimination by making use
of constraints (\ref{classicalconstraint+1}). Finally,  integrating
out  one of the quarks leads to the correct results at
$N_F=N_C$.

$\bf{N_F>N_C+1}$.  

The theory becomes more and more weakly coupled in the infrared when
the number of flavours increases. At $N_F>3N_C$ asymptotic freedom
disappears and the low energy theory describes non-interacting quarks
and gluons. The situation is much more interesting in the intermediate
region $N_C+1<N_F<3N_C$. In similarity to the case $N_F=N_C+1$,
classical flat directions coincide with the quantum moduli space and
are described by the invariants (\ref{mesons}), (\ref{baryons})
satisfying the constraints, which are straightforward generalizations
of Eq.~(\ref{classicalconstraint+1}).  However, an attempt to describe
the behaviour of the theory in analogy to the previous case
fails.  This follows, for instance, from the failure of anomaly
matching. Another manifestation of the same problem is that
the superpotential,
\begin{equation}
W_{eff}=\frac{1}{\Lambda^{2N_F-3}}(B_{i_1...i_{N_C-N_F}}M_{i_1j_1}\cdot...\cdot
M_{i_{N_C-N_F}j_{N_C-N_F}}\bar{B}_{j_1...j_{N_C-N_F}}-\det M)\;,
\end{equation}
which is a generalization of Eq.~(\ref{effectivesuperpotential+1}) and
leads to the classical constraints between mesons and baryons as
consequences of the equations of motion, does not have $R$-charge
equal to two, as is required by $R$-symmetry.

On the other hand, it has been argued already in Ref.~\cite{banks} that
in the usual QCD with a certain relation between $N_F$ and $N_C$,
non-trivial infra-red fixed points can be present. $\beta$-function in the
supersymmetric case has the form,
\cite{betafunction}
$$
\beta(\alpha)=-\frac{\alpha^2}{2\pi^2}\frac{3N_C-N_F(1-\gamma(\alpha))}
{1-N_C\frac{\alpha}{2}}\;,
$$
where
$$
\gamma(\alpha)=-\frac{\alpha}{2}\frac{N_C^2-1}{N_C}+{\cal
O}(\alpha^2)\; 
$$
is the anomalous dimension of the quark field. It is straightforward
to check that in the limit of large $N_F$ and $N_C$, with
$N_C\alpha$ and $3-\frac{N_F}{N_C}\ll 1$ kept constant,
$\beta$-function indeed has a non-trivial zero in the perturbative
region.  Consequently, low energy dynamics is described by the
superconformal theory of interacting quarks and gluons. It was
suggested by Seiberg~\cite{duality} that this is the case in the entire
region $3N_C/2<N_F<3N_C$ at arbitrary (not necessary large) $N_C$
and $N_F$.

The dimension of the chiral operator $M_{ij}$ is smaller than one at
$N_F<3N_C/2$, that is $M_{ij}$ cannot correspond to any unitary representation of
the superconformal algebra. At the boundary value $N_F=3N_C/2$ mesons
have dimension one and describe free fields. The
conjecture~\cite{duality} is that from this value of $N_F$ down to
$N_F=N_C+2$ all low energy degrees of freedom are free and they are
described by another effective gauge theory, called ``magnetic'' (dual
to the original, electric) theory. Baryons of the electric theory have
$N_F-N_C$ indices and occur to be composite states of $N_F-N_C$
``magnetic'' quarks $\tilde{\Phi}$. The significant difference of
this case from all the previous ones is that low energy theory is the gauge
theory, but with new gauge group $SU(N_F-N_C)$. Magnetic quarks
$\tilde{\Phi}$ and antiquarks $\bar{\tilde{\Phi}}$ are charged under this ``magnetic''
gauge group and fall in the following representations of the global
symmetries group (which should be the same both in the electric and magnetic
theories),
$$
\tilde{\Phi}=(\overline{N_F},1,\frac{N_C}{N_F-N_C},\frac{N_C}{N_F})\;,~~~
\bar{\tilde{\Phi}}=(1,N_F,-\frac{N_C}{N_F-N_C},\frac{N_C}{N_F})\;.
$$
Magnetic theory is infrared free at $N_F<3N_C/2$. At
$3N_C/2<N_F<3N_C$, magnetic theory (as well as electric one) has
non-trivial fixed point; both magnetic and electric theories represent
two descriptions of the same superconformal theory in different terms.
Dual theory becomes non-perturbative at larger number of flavours
where electric theory gives the correct description of the
dynamics. Mesons (\ref{mesons}) appear in the dual theory as
independent neutral fields interacting with the dual quarks through
the superpotential
\begin{equation}
\label{dualsup}
W_d=\tilde{\Phi} M\bar{\tilde{\Phi}}\;.
\end{equation}
Its role is to break extra $U(1)$ symmetry acting on the meson fields
and to ensure the proper correspondence between flat directions in the
electric and magnetic theories.

In this description all 't Hooft anomaly matching conditions are
satisfied. Moreover, the conjecture satisfies other tests such as
simultaneous deformation of the dual theories and integrating out a
flavour. Furthermore one can dualize the theory twice and obtain the
original theory, as it is natural to expect (at first sight there
appear two additional singlets corresponding to meson fields;
but one can integrate out one of them by making use of
Eq.~(\ref{dualsup}) and find that the second one is expressed
through the quark fields according to Eq.~(\ref{mesons})).
Finally, this duality can be related to $N=2$ and string
dualities~\cite{N2}. To conclude this section, let us note that most
of the presented results can be generalized to the case of other
simple groups~\cite{intsei1,intsei2}.
\subsection{Models with mediation via singlet field}
\label{D2.4}

We discuss specific models incorporating secluded sector in the
remainder of this section. We analyze several typical examples
illustrating different mechanisms of generating non-zero $\la s\ra$
and $\la F_s\ra$ and of dynamical supersymmetry breaking.  Let us
consider first the theories where Yukawa-type interaction $SQ\bar{Q}$ is
the only interaction between messengers and secluded sector.
So called models of direct mediation, where messenger fields are charged
under gauge group of the secluded sector are discussed in
section~\ref{sec:direct}.

\subsubsection{Models with the additional $U(1)_m$ gauge group}
\label{DN.1}
We start with the discussion of the Dine-Nelson models~\cite{DN,DNS,DNNS}.
These are the models, which attracted considerable interest to low
energy supersymmetry breaking recently. As a rule,
supersymmetry breaks down spontaneously due to non-perturbative dynamics of
the gauge group of the secluded sector. Supersymmetry breaking dynamics
of the secluded sector and a mechanism that generates non-zero $\la s\ra $
and $\la F_s\ra \lesssim \la s\ra^2$ are two basic components of the
full theory. The characteristic feature of the Dine-Nelson models is 
the presence of a weakly coupled
$U(1)_m$ gauge group which does not affect supersymmetry breaking.

New intermediate sector is incorporated in order to generate non-zero
$\la s\ra $ and $\la F_s\ra \lesssim \la s\ra^2$. Fields of this
sector interact with the singlet and do not affect non-perturbative
dynamics.

There is no interaction of these fields with the secluded sector in
superpotential. On the other hand, these fields are charged under
$U(1)_m$ group and obtain information about supersymmetry breaking
through non-zero $U(1)_m$ $D$-terms. The latter appear because some
of the secluded sector fields are also charged under $U(1)_m$. If this
group is unbroken in the supersymmetry breaking vacuum, then the interaction
of the charged under $U(1)_m$ fields $Z_\pm$ with singlet $S$ of the
type
$$
W=\lambda_1Z_+Z_-S+\frac{\lambda_3}{3}S^3
$$ 
leads to the generation of non-zero $\la s\ra$ and $\la F_s\ra$.

Supersymmetry breaking sector considered in Ref.~\cite{DN} is rather
complicated (due to the so called R-axion problem). We will consider
the model of Ref.~\cite{DNS} where the secluded sector is much
simpler. It is based on the well-known ``3-2''
model~\cite{ADS-85}, which we will describe now.

\paragraph{``3-2'' model.}
The gauge group of this model is $SU(3)\times SU(2)$ with the
following set of the matter fields (numbers in parentheses denote
$SU(3)\times SU(2)$ representation of the corresponding field),
$$
\begin{array}{c} 
\mbox{one field}\;Q\;(3,2),\\ 
\mbox{two fields}\;\bar{L}_I\;(\bar{3},1),\;I=1,\;2,\;\\ 
\mbox{one field} \;\bar{R}\;(1,2)\;.\\ 
\end{array} 
$$
The following renormalizable superpotential, compatible with all
symmetries of the theory (global R-symmetry and gauge symmetries) can
be introduced,
\begin{equation} 
\label{super} 
W=kQ\bar{L}_1\bar{R}\;
\end{equation} 
($k$ is the corresponding Yukawa coupling). Let us discuss
dynamical supersymmetry breaking in
this model along the line of Ref.~\cite{IntrThomas}.

The complete set of holomorphic invariant polynomials in this model is
$$
X_I=Q\bar{L}_I\bar{R},~~ Y=(Q\bar{L}_1)(Q\bar{L}_2). 
$$
Consequently, classical flat directions form 3-dimensional complex
manifold with coordinates $X_I,~Y$ in the absence of the superpotential.
The situation changes when the superpotential (\ref{super}) is included.  Let
us first consider the condition of stationarity of $W(\Phi)$ along the
direction $\bar{L}_I$,
\begin{equation} 
\label{fterm1} 
\frac{\partial W}{\partial \bar{L}_1^a}=kQ^a\bar{R}=0. 
\end{equation} 
Upon multiplying this equation by $\bar{L}_I^a$ and contracting the group
index $a$, one obtains that $X_I=0$. In analogy, stationarity along the
direction $\bar{R}$, 
$$ 
\frac{\partial W}{\partial \bar{R}^i}=kQ^i\bar{L}_1=0 
$$
implies that $Y=0$. Hence all flat directions are lifted in the
presence of superpotential (\ref{super}), and the only vacuum in
``3-2'' model is the origin of the field space. This observation indicates
that in fact ({\it i.e.}\ with quantum non-perturbative corrections 
taken into account) supersymmetry might be broken. Indeed, the following
simple criterion of supersymmetry breaking was suggested in
Ref.~\cite{ADS-85}, 
\newline 
{\it Supersymmetry is broken if
  non-compact flat directions are absent and some global
  symmetry is spontaneously broken in the theory.}

This statement is related to the presence of massless {\it real}
scalar field (Goldstone boson) in the theory with spontaneously broken
global symmetry. Let us suppose that supersymmetry is unbroken. Then
this field should have one more massless scalar partner in order to
form a chiral multiplet which includes {\it complex} scalar
field. While the usual Goldstone boson corresponds to the action of
the broken group generator on the vacuum, its partner corresponds to
the action of complexified generator (the moduli space in
supersymmetric theory is invariant under the complexification of the
group of global symmetries due to holomorphy of the
superpotential). The orbits of the action of the complex group are not
compact, so that this field parametrizes a non-compact flat direction,
in contradiction to the assumption of the theorem.

Global R-symmetry is present in ``3-2'' model. This symmetry is
unbroken only if all vev's of the matter fields are equal to zero. The
above criterion implies that in order to check that
supersymmetry is broken in this case, it is sufficient to demonstrate that the
origin of the field space cannot be the vacuum of the theory after all
quantum corrections are taken into account. The specific mechanism 
of generating  non-zero vev's depends on the ratio of the dynamical
scales $\Lambda_2$ and $\Lambda_3$ of $SU(2)$ and $SU(3)$ groups,
respectively.

{\bf $\Lambda_2\ll\Lambda_3$.} In this case low energy dynamics is
completely determined by $SU(3)$ group. There are two flavours in the
fundamental ($SU(2)$-components of the field $Q$) and in
antifundamental ($\bar{L}_I$) representations of this group. So, we
are dealing with the specific case of SQCD with $N_F=N_C-1$. As 
discussed above, one-instanton superpotential 
\begin{equation}
\label{effect} W_{ins}=\frac{\Lambda_3^7}{Y}.  
\end{equation} 
is generated in this case. The superpotential (\ref{effect}) is
singular in the origin, so this point cannot represent the vacuum of
the theory and supersymmetry is indeed broken. So, the 
``3-2'' model in this limit is probably the simplest example of
supersymmetry breaking by means of the generation of non-perturbative
superpotential.

{\bf $\Lambda_2\gg\Lambda_3$.}  Non-perturbative dynamics of $SU(2)$
group is of importance in this case. Fundamental representation of this
group coincides with its conjugate, so that in low energy theory there
are four fields (three $SU(3)$-components of the field $Q$ and
$\bar{R}$) in this representation of $SU(2)$ group. This is the
simplest example of SQCD with $N_F=N_C$. Deformation of the moduli
space implies according to Eq.~(\ref{quantumconstraint})
that the origin of the field space cannot be the vacuum and the
criterion of supersymmetry breaking is satisfied again. This limiting
case of the ``3-2'' model illustrates dynamical supersymmetry breaking
due to the deformation of the moduli space.

{\bf $\Lambda_2\sim\Lambda_3$.}  In this case dynamics of both groups 
plays a role. The most general effective superpotential
compatible with all symmetries, has the form,
\begin{equation} 
\label{fulll} 
W_f=\frac{\Lambda_3^7}{Y}+kX_1+{\cal A}(Z-\Lambda_2^4). 
\end{equation} 
The first term in this equation is the effective superpotential,
generated, as discussed above, due to non-perturbative $SU(3)$
dynamics.  The second term is the tree-level superpotential
(\ref{super}). In the third term, a new gauge invariant combination of
fields appears,
$$ 
Z=Q^3\bar{R}. 
$$
A question arises, why this invariant had not been taken into
account earlier, in the study of classical flat directions? It is
sufficient to write down explicitly all group index contractions in the
expression for $Z$ in order to understand the answer,
$$ 
Z=\epsilon^{abc}\epsilon^{\alpha\beta}\epsilon^{\gamma\delta} 
Q_{\alpha a}Q_{\beta b}Q_{\gamma c}\bar{R}_{\delta}. 
$$ 
Greek letters denote $SU(2)$ indices here and Latin letters denote
$SU(3)$ indices, $\epsilon$ is completely antisymmetric tensor.
Making use of the identity for the product of $\epsilon$-symbols,
$$ 
\epsilon^{\alpha\beta}\epsilon^{\gamma\delta}=\delta^{\alpha\gamma}\delta^{ 
\beta\delta}-\delta^{\alpha\delta}\delta^{\beta\gamma}, 
$$
one obtains that the invariant $Z$ is identically zero at the
classical level. At the quantum level, vev's of the fields satisfy the
constraint,
$$
Z=\Lambda_2^4 , 
$$
which is similar to Eq.~(\ref{quantumconstraint}). It is this constraint
that leads to the deformation of the moduli space in the case
$\Lambda_2\gg\Lambda_3$. The last term in Eq.~(\ref{fulll}) is
nothing but expression of this constraint in terms of a Lagrange
multiplier.  It is straightforward to see from Eq.~(\ref{fulll})
that the origin never belongs to the quantum space of vacua in ``3-2''
model and, consequently, supersymmetry is always broken here.
\newline

Let us now discuss one of the possible mechanisms of generating non-zero
$\la s\ra$ and $\la F_s\ra$ in this model.

The theory exhibits global $U(1)$ symmetry with the following charges
of the matter fields $Q(1/3)$, $\bar{L}_1(2/3)$, $\bar{L}_2(-4/3)$,
$\bar{R}(-1)$.  If this symmetry is gauged it can play a role of
$U(1)_m$. It is worth noting, however, that this symmetry is 
anomalous in the original theory, so that it is necessary to include
additional fields (e.g., field $E$ with charge equal
to $+2$) which are singlets under non-abelian groups.
The key point of the mechanism of Ref.~~\cite{DNS} is to use  $D$-term of this 
group for generating non-zero $\la
s\ra$ and $\la F_s\ra$.

Let us consider the case $\Lambda_2\ll\Lambda_3$ for
definiteness. Then the
effective superpotential takes the form
$$
W=kX_1+\frac{\Lambda_3^7}{Y}\;.
$$
All dimensionful parameters in this theory are characterized by the unique
scale $\upsilon\sim\Lambda_3/k^{1/7}$. Namely, masses of the scalar fields are
of order $m_s\sim k\upsilon$, and masses of vector fields are $m_v\sim g\upsilon$.
Some of these fields are charged under $U(1)_m$ (e.g., the scalar component of
$X_2$). One can demonstrate~\cite{DNS} that supersymmetry breaking results in
non-zero vev of the $D$-term of this group,
 $$
\la D\ra =\frac{\alpha_m}{4\pi}m_s^2\ln\frac{m_v^2}{m_s^2}\;.
$$ 
Then one can add a pair of fields  $Z_\pm$ with charges  $\pm1$ under $U(1)_m$
and with the following superpotential,
\begin{equation}
W=\lambda_1Z_+Z_-S+\frac{\lambda_2}{2}EZ_-^2+\frac{\lambda_3}{3}S^3\;.
\label{DNS_5}
\end{equation}

As a result, rather cumbersome potential in the messenger sector arises,
$$
V=
\left|\frac{\d W}{\d\Phi}\right|^2+
2\pi\alpha_m \l\la D\ra+2|e|^2+|z_+|^2-|z_-|^2\r^2- 
m_{z_\pm}^2 \l 4|e|^2+|z_+|^2+|z_-|^2 \r\;,
$$
where the dominant contributions to parameters $m_{z_\pm}$, $m_e^2$
at small $\lambda_1\ll\alpha_m$ are given by two loop corrections due
to abelian gauge interaction,
$$
m_{z_\pm}^2=-\frac{2\alpha_m^2}{\pi^2}m_s^2\ln\frac{m_v^2}{m_s^2}\;,~~~
m_e^2=4m_{z_\pm}^2\;.
$$
There is a region of parameter space where minimization of this
potential results in non-zero vev's of the fields $z_+,z_-$ and $e$,
leading in turn to non-zero $\la s\ra $ and $\la F_s\ra $. The basic
difficulty of this model is fine tuning of parameters,
$\lambda_1\ll \alpha_m$. This difficulty can be avoided in more
complicated models \cite{DNNS}.

Phenomenology of the models with $U(1)_m$ group is typical for 
gauge mediation models and was discussed in sections \ref{singlet},
\ref{sec:extensions}. It is worth noting that 
supersymmetry breaking scale in the secluded sector $F_{\rm DSB}$ is rather
high there, because it is related to $\Lambda$ through small
coupling constant $\alpha_m\lesssim 1$ of $U(1)_m$ group,
$$
\Lambda^2\sim\left({\alpha_m\over 4\pi}\right)^2 F_{\sm{DSB}}.
$$
The corresponding gravitino mass at $\Lambda\sim$100~TeV is 
\begin{equation}
m_{3/2}={F_{\sm{DSB}}\over\sqrt{3}M^*_{Pl}}\simeq 3~{\rm eV}{F_{\sm{DSB}}
\over (100~{\rm TeV})^2}\gtrsim 0.5~\mbox{keV}\;.
\label{46*}
\end{equation}

\subsubsection{Secluded sector with vector-like matter}
\label{D4.2}
Recent rapid development of methods of investigation of
supersymmetric theories made it possible to demonstrate that
supersymmetry breaking may happen in models with only vector-like matter
representations as well.

The simplest theory of this type~\cite{IzawaYanagida,IntrThomas} is
based on the $SU(2)$ gauge group with four fundamental fields $\Phi_i$ and
six singlets which compose antisymmetric tensor $Z^{ij}$ under
$SU(4)_{\sm F}$ group of global symmetries. Classical flat directions
are parametrized by vev's of singlets $Z^{ij}$ and by usual meson
invariants $M_{ij}=\epsilon^{ab}\Phi_i^a\Phi_j^b$ subject to a
classical constraint
$$
\epsilon^{ijkl}M_{ij}M_{kl}=0\;.
$$
Let us add $SU(4)_{\sm F}$-symmetric superpotential
\begin{equation}
\label{HIY_1} 
W_{tree}=\frac{1}{2}k \Phi_i\Phi_jZ^{ij}\;.
\end{equation}
Fields $\Phi_i$ become heavy in the limit $kZ^{ij}\gg\Lambda_2$, where $\Lambda_2$
is the dynamical scale of the gauge group, and the low energy 
theory is described by supersymmetric $SU(2)$ gluodynamics with 
the effective dynamical scale
$$
\Lambda_{eff}^6=k^2\Lambda_2^4 Z_{ij}Z^{ij}\;.
$$
Gluino condensation takes place
in this model (see section~\ref{dsb}). As a result, the effective
superpotential~(\ref{gauginosuperpotential}) is generated. The
relevant effective scale $\Lambda_{\sm G}=\Lambda_{eff}$ is the
function of fields $Z$, so that O'Raifeartaigh
type potential arises which breaks supersymmetry. At small values of $Z$ and at $k=0$ quantum
deformation of moduli space takes place so that the origin does
not belong to the quantum moduli space. This fact combined with the
requirement $M_{ij}=0$ following from stationarity of superpotential
at $k\neq 0$ leads to supersymmetry breaking in this limit as well.

It was shown in Ref.~\cite{HIY} that it is possible to obtain non-zero
$\la s\ra$ and $\la F_s\ra$ without additional $U(1)_m$ gauge
interaction, but by making use of corrections to Kahler potential.

Let us rewrite the fields in terms of singlets  $(\hat{M},~Z)$ and five-plets
of the flavour group $SP(4)_{\sm F}\subset SU(4)_{\sm{F}}$. Effective potential
can be rewritten in the following form~\cite{HIY}
$$
W_{eff}={\cal
A}(\hat{M}^2+\hat{M}^a\hat{M}^a-\Lambda_2^4)+k_zZ\hat{M}+kZ^a\hat{M}^a\;,
$$
where ${\cal A}$ is a Lagrange multiplier and constants $k$ and $k_z$
can be different if global symmetry $SP(4)_{\sm F}$ is broken by some
mechanism.  We will consider a simple case when breaking of
this symmetry does not affect supersymmetry breaking and leads to small
values of $k_z$. Let us consider the following interaction of the
secluded sector field with singlet,
$$
W_{int}=\lambda_1S\hat{M}-\frac{\lambda_3}{3}S^3\;.
$$
Fields $Z^a$ and $\hat{M}^a$ are massive with the mass $k\Lambda$
and can be integrated out. The resulting low energy superpotential is 
$$
W_{low}=k_z\Lambda^2_2Z+\lambda_1\Lambda^2_2S-\frac{\lambda_3}{3}S^3\;.
$$

Kahler potential invariant under R-symmetry and discrete
$Z_2$-symmetry ($\hat{M}$, $S$, $Z$, $\lambda_3$ (as an external field) are odd
under this symmetry) has the form,
$$
K=ZZ^++SS^+-\frac{\eta}{\Lambda_2^2}| 
k_z Z+\lambda_1S|^4 
-\lambda_1^2\lambda_3^2\frac{\xi}{\Lambda_2^2}|k_z Z+\lambda_1S|^2SS^++\dots\;,
$$
where positive constant $\eta$ is of order one and loop factor
$\xi\sim 10^{-3}$. It is the latter one which is responsible for
mediation of supersymmetry breaking in the observable sector at
non-zero $\lambda_3$. Phenomenological parameters $\Lambda$ and $x$
(see Eq.~(\ref{Lambda_x})) are related to the parameters of the
secluded sector as follows, 
$$
\Lambda\simeq\frac{\xi}{2}k_z^4
\lambda_1^{3/2}\lambda_3^{3/2}
\Lambda_2,~~~~~~~
x\simeq\frac{\xi}{2}k_z^4\lambda_1\;.
$$
It was shown that with the proper choice of parameters one can satisfy the
limits on gravitino masses, and provide the stability of the
Higgs potential.

\subsubsection{Composite singlets}
Rather cumbersome intermediate sector and additional
$U(1)_m$ gauge group in the models  considered above are related to 
the necessity to generate non-zero vev's $\la s\ra $ and $\la F_s\ra $
for singlet which interacts with messengers through superpotential
$SQ\bar Q$.  In several cases it may be simpler to deal not with the
fundamental singlet $S$ but with composite field~\cite{R1--R2--Shadmi}
(interactions of the messengers with fundamental fields become
non-renormalizable in this case). Phenomenology of the specific models
can be rather different then because the secluded sector can affect low
energy theory not only through non-zero  $\la s\ra $ and $\la F_s\ra $.

\subsubsection{Vacuum metastability}
\label{D3.1}
There are a number of theoretical problems which are common in gauge
mediation models. These problems require new mechanisms to resolve them.
One such problem is that the global
minimum of the potential of the full theory including
secluded and observable sector is actually 
supersymmetric and breaks MSSM gauge group
in the majority of the models.
Supersymmetry breaking minimum turns out to be metastable in this case.

For instance, it was shown in Refs.~\cite{lisa,hiddenmur} that in
Dine-Nelson models considered in section \ref{DN.1}, the global
minimum of the potential is supersymmetric and is located at the point
$$
\la F_s\ra =\la s\ra =0,~~~~~~\lambda'\la e_+e_-\ra =-\lambda \la \bar{q}q\ra 
$$
when singlet-messenger interaction is taken into account.
Generally speaking, this implies breaking of $SU(3)\times U(1)$
symmetry and phenomenologically acceptable vacuum discussed in section
\ref{DN.1}
is metastable in this case. However,  its lifetime is  longer then the age
of the Universe at rather natural choice of parameters in these models.

Another way to solve this problem is to construct a model with the
global supersymmetry breaking and colour conserving minimum. The
simplest solution is to introduce explicit mass
terms for messengers. The corresponding mass should be in the range
$10^5-10^{15}$~GeV. Some new mechanisms are necessary then to
provide the naturalness of this new scale. Alternatively, one can make
use of additional singlets or new particles in the observable sector.
However, the latter solution is incompatible with perturbativity of MSSM up to
the Grand Unification scale \cite{lisa}. 
 
Additional chiral fields can be used when there is an abelian gauge group
rotating the fields $Z_\pm$ (see section \ref{DN.1})~\cite{lisa}.
However, a new problem arises in this case due to generation of the 
Fayet-Iliopoulos
terms in the potential of the secluded sector. These terms lead to 
so-called mass instability due to mixing of $U(1)_m$ group with the
MSSM hypercharge group~\cite{U(1)_mixing}.

It is worth to discuss this problem in more detail.
Mixing of the kinetic terms is possible for the abelian gauge
groups. In supersymmetric case it leads to mixing of
corresponding D-terms. The mixing term $F^{\mu\nu}_1F^{\mu\nu}_2$ is
renormalizable and can be present in the effective Lagrangian without
being suppressed by a dimensionful factor. Mixing of
the D-terms implies the generation of unacceptably large (of the order
of supersymmetry breaking scale in the secluded sector) scalar masses for MSSM
particles. Some additional symmetry can forbid this mixing. Alternatively, one
can use more complicated gauge group for fields  $Z_\pm$ instead of $U(1)_m$,
so that mixing would be impossible.

A successful attempt to obtain a theory with global supersymmetry
breaking vacuum due to variation of the secluded sector has been
presented in Ref.~\cite{NTY}. $SU(2)$ vector model of section
\ref{D4.2} was used to break supersymmetry, and $U(1)_m$ was one of the
abelian subgroups of $SP(4)_{\sm F}$ global symmetry group. Fields
$Z_\pm$ acquire soft positive masses due to threshold effects of
$U(1)_m$ gauge interaction. Non-zero $\la s\ra$ and $\la F_s\ra$
appear due to the Higgs mechanism after Yukawa interaction of singlet
with $Z_\pm$ is taken into account. In spite of the presence of
additional abelian group, the problem with mixing of D-terms does
not arise in this case because this mixing is forbidden by the charge
conjugation symmetry of the original theory.

\subsection{Models with direct mediation}
\label{sec:direct}
The approach described in section~\ref{D2.4} allows one to construct
phenomenologically viable models with gauge mediation. Some
shortcomings of this approach, in particular, those discussed at 
the end of the previous section,
call for modification of these models. One of the approaches 
is to allow messengers to carry quantum numbers not only of the
Standard Model but also of the secluded sector gauge group. Such
theories are called models with direct mediation. In this case
messengers themselves participate in the dynamics of supersymmetry
breaking (hence another name -- models with dynamical messengers). 

This approach certainly looks more natural and historically the
first attempts to construct realistic models with dynamical
supersymmetry breaking~\cite{ADS-85} exploited this
scheme. There are some problems, however, which lead to additional 
complications in model building, when the mechanism of direct
mediation is involved. The main difficulty is the following. Let
messengers transform as some representation of the secluded
sector gauge group. This group plays a role of a flavour group from
the point of view of the visible sector, i.e., several copies of
messenger fields appear in the spectrum. The number of copies 
coincides with the dimension of the corresponding representation of the 
secluded sector gauge group. Since messengers carry Standard Model
quantum numbers, they give contribution to the $\beta$ functions of
gauge coupling constants of QCD and electroweak theory, that may
result in the loss of asymptotic freedom of QCD. If the dimension of the 
representation of the secluded sector gauge group, i.e., the number of
messenger copies, is larger than four in the case of $SU(5)$ GUT 
representation $5+\bar5$ (or one for $10+\bar{10}$) 
and messenger threshold $M\lesssim
10^8$~GeV, then the coupling constants of the Standard Model 
groups become
large below $10^{16}$~GeV. This fact contradicts the attractive idea
of their perturbative unification. The majority of 
models with dynamical supersymmetry breaking known up to now do not 
lead to representations of  dimension less than
four\footnote{The recently proposed model~\cite{Agashe} of 
direct mediation with only two messenger copies 
is discussed in section~\ref{local-vacuum}.},
that had even provoked some authors to call this way 
``not the most clever''~\cite{DNS}. Nevertheless, models with
direct mediation of supersymmetry breaking are studied intensively and
several realistic and rather elegant theories have been
found already. 

The approaches to solve the problem with asymptotic freedom of
the Standard Model in theories with dynamical messengers are
conventionally separated into three types. The most developed approach
(see section~\ref{heavy-messengers}) deals with heavy messengers: if
the messenger scale is sufficiently high, then Landau pole
occurs above the GUT scale and perturbative unification of coupling
constants remains intact. Two other approaches look less
conventional. One of them is based on the refusal to
describe physics up to GUT scale in terms of the Standard Model. If
observed matter fields represent composite low energy degrees of freedom of
some effective theory whose coupling constant becomes strong at the
intermediate scale between $M_W$ and $M_{GUT}$, then the fundamental
matter content might differ from the content at low energies. The
observed particles may be composed of very small number of fundamental
constituents. The Standard Model gauge group would be
asymptotically free at high energies in this case despite of a large number of
charged degrees of freedom in the visible low energy spectrum. This
possibility has not been studied in great detail, we will discuss this scenario in
section~\ref{compositeness}. Finally, the last approach is to refuse 
to unify MSSM coupling constants in perturbative domain in favour of
the unification in the strong coupling regime. Such a possibility does
not contradict phenomenological requirements and the idea of
Grand Unification; moreover, this way is
even more preferable from several points of view. The general features of
this scenario have been discussed in section~\ref{strong-unif}.

\subsubsection{Heavy messengers}
\label{heavy-messengers}
We turn now to the discussion of the models with direct mediation, where the
perturbative unification of the Standard Model gauge coupling
constants is achieved due to the large values of thresholds pushing 
Landau pole beyond the GUT scale. First, note that the use of 
heavy messengers is consistent with sufficiently large soft masses of
superpartners of the Standard Model particles. The reason is that it
is the {\em ratio} $\Lambda=\la F_s\ra /\la s\ra$ that determines 
soft masses, see Eqs.~\eq{gaugin},~\eq{scalmas}, while the
messenger mass scale is determined by $\la s\ra$. Messengers do not
affect running of the Standard Model coupling constants below the 
scale $\la s\ra$. This value may be large at phenomenologically viable 
$\Lambda\sim$~(100~TeV)$/\sqrt{n}$, where $n$ 
is the effective number of messengers. As discussed in
section~\ref{D1.3}, the main constraints on the messenger mass come
from cosmology and give $\la s\ra\lesssim 10^{11}$~GeV. 

\paragraph{How to make messengers heavy?}
The standard method to get large masses of messengers is to provide 
large vacuum expectation value for the scalar component of the superfield $S$,
which interacts with messengers $Q$, $\bar Q$ through Yukawa
superpotential $SQ\bar Q$. The general scheme is the following. Let some flat
direction be parametrized by the field~\footnote{More exactly, by a 
holomorphic gauge invariant
composed of the components of this field: although $S$ is the
Standard Model singlet, it may transform nontrivially under the
secluded gauge group in direct mediation models.} $S$. If the
superpotential is generated along this direction due to quantum effects
(as in case of SQCD at $N_f<N_c$ considered in
section~\ref{flat-dir:quant-th}), then one has run-away vacuum,
i.e., the energy reaches its minimum only at $S=\infty$. To obtain finite
vacuum expectation value of $S$, one needs a competitive
contribution to the potential, which would lift the flat direction at
large $S$. The corresponding terms are either included in
the superpotential ``by hands'', or generated due to quantum effects. 

Let us consider a supersymmetric flat direction, along which the scalar
potential is zero. The potential for
singlet $S$ may appear due to nonperturbative effects in the
following way. Let $S$ interact, apart from $Q$, $\bar Q$, 
with fields $f$, $\bar f$ from the secluded sector, 
$$
W_f=Sf\bar f,
$$
where $n$ flavours ($\bar f$) $f$ transform as the
(anti)fundamental representation of some asymptotically free gauge
group $SU(m)$, whose gauge coupling becomes strong at the scale
$\Lambda$. The fields $f$, $\bar f$ effectively become heavy at the
values of scalar component $S\gg\Lambda$ and have to be integrated out
from the low energy theory. If the dynamics of gauge group $SU(m)$
dominates and there are no other fields charged under this
group, then the low energy theory is a supersymmetric Yang-Mills
theory where gaugino condensation provides the nonperturbative
superpotential 
$$
W_{\rm eff}=\Lambda_{\rm eff}^3,
$$
with the scale of effective theory $\Lambda_{\rm eff}$ determined from
the matching condition 
\begin{equation}
\Lambda_{\rm eff}^{3m}=S^n \Lambda^{3m-n}.
\label{matching}
\end{equation}
The powers of $\Lambda$ on the right and left hand sides of this
formula are the first coefficients of the $\beta$ function below and
above the matter field thresholds, respectively; 
in this case $S$ plays a role of
mass of heavy fields and the power of $S$ is determined by
dimensional arguments. Thus, the effective superpotential depends on $S$ and
is equal to 
$$
W_{\rm eff}=\Lambda^{3- n/m} S^{n/m}.
$$
The perturbative expansion for the Kahler potential of $S$ about the
canonical expression $K_S=SS^\dagger$ is valid far from $S=0$. The
corresponding contribution to the scalar potential has the form 
\begin{equation}
V_{\rm eff}=\left|F_s\right|^2=\left|{\d W\over \d s}\right|^2=
\Lambda^{(6-2n/m)}S^{2(n/m-1)},
\label{scalarPotential}
\end{equation}
so the potential is independent of $S$ at $n=m$, i.e., the flat
direction is uniformly lifted forming a plateau with constant scalar
potential. At $n<m$ the corresponding potential would push the vacuum
away to large $S$. In the latter case, a contribution of the different
origin is required to stabilize the flat direction and to obtain a
vacuum with
nonzero but finite $\la s\ra$ and $\la F_s\ra$.
The simplest way is to add the
terms increasing with $S$ directly into the tree-level superpotential. Such
terms usually have to be nonrenormalizable in order that the
nonsupersymmetric minimum appears at large $S$. Nonrenormalizable
operators in the tree superpotential have to be suppressed by
a dimensional parameter whose physical meaning is an energy cutoff,
where this description fails. One usually implies that this parameters
is the Plank mass, therefore the contribution of nonrenormalizable operators
is small. This leads to the required minimum of the scalar potential at very
large $S$, because the stabilizing effect is weak in comparison with
``repulsion'' to the region $S\to\infty$. Large $S$ automatically
provides large values of messenger masses. The other way to stabilize
the flat direction is to involve the dynamics of one more gauge group,
if it generates the increasing potential. Similar arguments are valid 
in more complicated models, in particular, when the field $S$ is not
a singlet. 

We turn now to the case $n=m$ in Eq.~\eq{matching}, when
nonperturbative effects provide  positive and independent of $S$ scalar 
potential~(\ref{scalarPotential}), i.e.,
there is a flat direction of inequivalent supersymmetry
breaking minima (recall this statement is valid only at large $S$, since near
$S=0$ the corrections to Kahler potential are uncalculable and
may lead to additional contributions to the scalar potential. 
In particular, these contributions may restore supersymmetry, therefore
the minima with nonzero vacua might be local). We have to take into account
the next order contributions, namely, perturbative corrections to the
Kahler potential, because the potential does not depend on $S$ along
flat direction in the leading order. Had we studied a supersymmetric flat
direction, these corrections would not alter the scalar
potential along it: they are multiplied by $F_S=0$ in the scalar potential.
In our case, $F_S=\Lambda^2\ne 0$,
therefore taking into account wave function renormalization gives (small)
corrections to the potential, which may provide isolated vacuum. This
mechanism to obtain vacuum at large values of $S$ is known as {\it
inverse hierarchy}~\cite{inverse_hierarchy}. As will be shown below, 
to exploit this mechanism one needs a nonsinglet field 
$S$ transforming as a nontrivial
representation of an asymptotically free group. 

If the field $S$ transforms under the gauge group $G$, then a nonzero
vacuum expectation value of the scalar component of this field breaks the
group down to subgroup $G'$. In the one-loop approximation, the Kahler
potential for $S$ gets contributions due to the loops generated by fields 
$Q$, $\bar Q$, $f$, $\bar f$ and heavy vector fields, whose mass is 
of order $S$ when $G$ is broken down to $G'$ (the gauge fields of $G'$
do not contribute to the Kahler potential at the one loop level,
since they do not
interact with $S$). If $g$ is the gauge coupling constant of $G$, and
$\lambda$ is the Yukawa constant of interaction $SQ\bar Q+Sf\bar f$, then
in one loop approximation 
\begin{equation}
K(S,S^\dagger)=SS^\dagger \left(1+
\left(c_g
g^2-c_\lambda\lambda^2\right)\ln\left(SS^\dagger/\Lambda_{\sm G}^2\right)
\right),
\label{Kahler1loop}
\end{equation}
where $c_{g,\lambda}$ are positive coefficients and $\Lambda_{\sm G}$
is a cutoff parameter. The leading logarithms may be summed at large
$S$ by making use of the renormalization group, then constants $g$, 
$\lambda$ will
be changed to their effective values at the scale $S$. With 
this contribution taken into account the potential takes the form 
\begin{equation}
V_{\rm eff}=
{
\displaystyle
 \left|
       {\d W\over \d S}
 \right|^2 
\over 
\displaystyle
  {\d^2 K\over \d S\d S^\dagger}
}
.
\label{correctScalarPotential}
\end{equation}
The behaviour of the potential is determined by running of the coupling
constants. The Yukawa constant often grows at large energies providing the
growth of scalar potential at infinity; the gauge coupling of the 
asymptotically free group $G'$ grows at low
energies. The combination of the two effects provides a maximum of the
coefficient in front of $SS^\dagger$ in (\ref{Kahler1loop}) at a 
value of $S$ which corresponds to the minimum of the effective
potential~(\ref{correctScalarPotential}). Because of the slow 
variation of logarithm, this minimum corresponds to large $S$, 
that is to heavy
messengers, as in the mechanism with nonrenormalizable terms. 

We have described above several ways to obtain the vacuum with large $\la
s\ra$ and nonzero $\la F_s\ra$. These approaches were exploited to 
construct concrete models, which
incorporate dynamical supersymmetry breaking and its direct mediation to
MSSM through gauge interactions. 

\paragraph{Models with nonrenormalizable terms in the superpotential.}
\label{NREN}
To make use of the direct mediation mechanism, one has to satisfy certain requirements
on the supersymmetry breaking sector. In fact, as it has already been 
mentioned, a given set of matter fields and their interactions should
have a global flavour symmetry group large enough to include the 
Standard Model group; the latter has to remain anomaly free 
after being gauged. Finally, the group has to remain unbroken despite
of supersymmetry breaking. To know whether this is the case, the exact
information about the supersymmetry breaking vacuum is required. 

The theories with the gauge group $SU(N)\times SU(M)$, whose infrared
 dynamics was discussed in Refs. \cite{PopShTr1--PopShTr2}, are
 exploited as a supersymmetry breaking sector in some of these 
models~\cite{A-HM-RM,PopTr-grav--PopTr-dyn-mess}. It is difficult to
 obtain an exact information about infrared dynamics at $M<N-2$. The
 global minimum of the potential is supersymmetric at $M=N$ and these
 models are incorporated in schemes where the supersymmetry breaking
 vacuum is local (see below). The theories with $M=N-1$ and $M=N-2$ break
 supersymmetry and their low energy dynamics is under control. 

Unfortunately, these theories seem to be phenomenologically
unacceptable due to the following reasons. As has been pointed out 
in Ref.~\cite{A-HM-RM} and discussed in detail in Ref.\ \cite{PopTr-Strace}, masses
of scalar superpartners of the Standard Model particles are not determined
by the formulae from section \ref{singlet} in this
case. The formulae~(\ref{gaugin}), (\ref{scalmas}) were derived under 
the assumption of quite specific supersymmetry breaking messenger
spectrum: masses of components of supermultiplet 
are splitted in such a way that 
$\str=0$. This condition is not always valid in
models with direct mediation and the main contribution to the masses
of scalar particles of MSSM is given by the logarithmically divergent
diagrams, which are proportional to $\str$ and the logarithm of an 
ultraviolet cutoff,
\begin{equation}
\delta m^2\sim-g^4\l\str\r \ln{\Lambda_M\over m}.
\label{STr-contribution}
\end{equation} 
Here $\Lambda_M$ is the scale of ultraviolet cutoff, $m$ is the mass
of light messengers. One can see that the MSSM soft terms 
depend significantly on the high energy physics; Eq.\ 
\eq{STr-contribution} gives larger contribution than Eq.~(\ref{scalmas}). 
Since $\str>0$ in these models, large negative
contribution to the masses squared of sleptons and squarks appears. This
contribution exceeds positive one coming from the usual diagrams 
(Fig.\ref{gc}) and leads to breaking of the gauge group of the Standard
Model, the fact that makes these theories phenomenologically unacceptable. 

To avoid breaking of the Standard Model group due to the 
contribution \eq{STr-contribution}, it is necessary to ensure that
there is no light particles with large soft masses charged under this
group. One may follow one of two ways. 
First, one can exclude all light charged fields from the secluded sector, for
instance, by lifting all flat directions parametrized by fields with
nontrivial quantum numbers of the Standard Model 
at the level of nonrenormalizable operators (see,
e.g., Ref.~\cite{Shirman-Nren}). 
Second, one can provide large soft masses for scalar
components of these light fields, so that the 
positive $\str$ does not appear; this is the case in models with local vacuum. 

\paragraph{Models with local vacuum.}
\label{local-vacuum}
It has already been pointed out that the basic mechanism providing a local
vacuum at large fields is the inverse hierarchy. The general scheme
of this mechanism have been formulated in Ref~\cite{Dimop-1} and
its essence is the following\footnote{We consider only models with direct
mediation, though this scheme can be applied with slight variations to
the models where supersymmetry breaking is mediated via 
singlet~\cite{Dimop-1}.}. The gauge group consists of three factors 
$G_S\times G_B\times G_w$. Roughly speaking, the strong group $G_S$
provides supersymmetry breaking, the weak $G_w$ contains the
Standard Model group, and asymptotically free ``balancing'' group $G_B$
is required for the inverse hierarchy. The field $S$ is charged
only under $G_B$ and there is one flat direction, 
parametrized by an invariant composed of the components of
$S$. $G_B$ is broken to some subgroup $H_B$ along this flat
direction. The fields $Q$, $\bar Q$ become heavy at large $S$ and
all arguments presented in the beginning of this section are correct. 

The supersymmetry breaking vacuum in all these models is false, so a
natural question arises, whether this vacuum is stable 
for cosmological intervals of time? We have
already concluded in section~\ref{singlet} that such a possibility is not excluded,
i.e., lifetime of a metastable vacuum can be much larger than the age 
of the Universe. This lifetime $\Gamma$ is given by the semiclassical
exponent in these models~\cite{Dimop-1},
$$
\Gamma\propto \e^{-S_B}, ~~~
S_B\sim 2 \pi^2\left(\frac{\la s\ra}{\Lambda}\right)^2.
$$
One obtains that the lifetime of supersymmetry breaking metastable state
exceeds the age of the Universe at $\la s\ra/\Lambda\gtrsim 10$.

\paragraph{Gauge messengers.} 
Unfortunately, there are specific contributions to the masses squared of
the MSSM scalars appear in specific models with 
the inverse hierarchy mechanism 
(see e.g., Ref.~\cite{Murayama-model,Dimop-1}). These contributions are
related to the fact that $G_B\times G_w$ is broken at 
the scale $M_{\rm mess}\ll M_{GUT}$ to a subgroup which includes the 
Standard Model group. The masses squared of squarks and
sleptons become negative due to these terms~\cite{GR:wave-function}, 
so that colour and
electromagnetic groups are broken. Hence, these models are not
realistic. 

The presence of other fields (besides ordinary messengers) that 
mediate supersymmetry breaking to the visible sector is
responsible for this phenomenon. These fields in adjoint
representation of the Standard Model group are known as gauge
messengers. These are heavy vector fields, which obtain masses 
as a result of breaking $G_B\times G_w\to \left(SU(3)\times
SU(2)\times U(1)\right)_{SM}$ by the vacuum expectation value of 
$\la s\ra$. 

The method proposed in Ref.~\cite{GR:wave-function} allows to
trace an appearance of the negative contribution to the
masses squared. Namely, the effective number of messengers $n$,
entering the expressions for soft masses, equals the value of a
jump of the first coefficient of $\beta$ function at the threshold 
which corresponds to messenger masses --- in case of fundamental
messengers it is merely the number of additional flavours $N_m$. If the
symmetry breaking $G\to H$ occurs at the same scale, then 
\begin{equation}
n=N_m-2(C_G-C_H),
\label{N-eff}
\end{equation}
where $C_G$ and $C_H$ are contributions to the first coefficient of 
$\beta$ function for the adjoint representations of $G$ and $H$,
respectively. The origin of the coefficient ($-2$) in Eq.~\eq{N-eff}
is the contribution from heavy vector bosons ($-3$)
and would-be Goldstone bosons ($+1$). Thus, the contribution of gauge
messengers is negative even at one loop level, and may lead to 
$n<0$. This statement is valid also in the case of breaking $G_B\times
H\to H'$ (where all matter fields of the Standard Model are 
assumed to be charged under $H$, $H'$ groups). 

\paragraph{Models without gauge messengers.}
There are various ways to overcome this difficulty. First, one can 
consider the same model but with additional matter charged under the
balancing group $G_B$ (or one can suppose that a part of the Standard
Model matter is charged under $H'$, while another part is charged
under $G_B$). Then $C_G$ becomes smaller and 
dangerous negative contribution to masses squared of superpartners 
proportional to $\sqrt{n}$ decreases. Unfortunately, 
``weakening'' of the balancing group may
destabilize the minimum obtained by making use of the inverse
hierarchy, or may shift this minimum to the unacceptable region of small
$S$. Phenomenologically viable models with gauge messengers are still
unknown. 

The other method is to modify the models in such a way that gauge
messengers are absent, that is symmetry breaking 
$G_B\times H'\to G_{SM}$ does not occur at the scale $M_{\rm mess}$. 
In other words,
the field $S$ providing the supersymmetry breaking masses to messengers $Q$,
$\bar Q$ is neutral under $G_B$. At first sight,
this idea contradicts the inverse hierarchy, because the latter
requires a charged field parameterizing a lifted flat direction. The
elegant solution was proposed in Ref.\cite{the_best_model}. Let
the field $S$ be a singlet. Let, however, the 
interaction in superpotential relates its vacuum expectation value
$\la s\ra$ to the vacuum expectation value of
another operator, which in turn breaks $G_B$. The flat direction, which the
inverse hierarchy mechanism works along, is parametrized by a {\em linear
combination} of the singlet interacting with messengers and 
another field charged under
the balancing group. Massive vector bosons appearing as a
result of breaking of $G_B$ are neutral under the Standard Model
group, so they do not serve as gauge messengers. 

Since the balancing group is separated from the Standard Model, it is 
possible to simplify the gauge sector of the theory. However, 
additional degrees of
freedom and new parameters are required, which 
describe interaction of fields which enter the 
linear combination mentioned above. An example of a model based on
this trick is a theory with $SU(5)$ strong group and balancing group
$SU(2)$ with one fundamental flavour $\psi$, $\bar\psi$ (any
asymptotically free theory with vector matter may serve as a balancing group)
\cite{the_best_model}. Five messenger generations $Q$, $\bar Q$
contribute to the masses of superpartners of the Standard Model particles. To
make this fact consistent with perturbativity of coupling constants, 
messengers have to be heavy that may contradict Eq.~\eq{mass-bound:nucleos}. 

To obtain correct hierarchy of the parameters in the Higgs sector of
MSSM, one might include two additional singlets with a specific
superpotential. The price for this solution of the $\mu$-problem is 
the introduction of new parameters without fine tuning their values. 

A completely different solution, associated with a significant
simplification of a model, has been proposed recently \cite{Agashe}
and deals with identifying $G_B$ and $G_S$. Namely, diagonal
$SU(2)_D$ subgroup of strong $SU(2)\times SU(2)$ group plays the role of
$G_B$. There are some fields ($\Sigma(2,2)$ and six flavours of
$Q(2,1)$ and $\bar Q(1,2)$) charged under the strong group and coupled
through the superpotential $\Sigma Q\bar Q$. The inverse hierarchy mechanism
works along the flat direction parametrized by $\det\Sigma$. The Standard
Model gauge group is embedded into $SU(6)$ global symmetry of the secluded
sector, so that this model does not contain gauge messengers but only two
generations of the ordinary messengers in the fundamental representation of 
$SU(5)$ GUT group. Thus, the problem with loss of asymptotic 
freedom of the Standard Model gauge group does not arise; however, the numerical
estimates demonstrate \cite{Agashe} that viable region of parameters of this
model corresponds to heavy messengers anyway, $M_{\rm mess}\gtrsim
10^{10}$~GeV (otherwise Landau pole for Yukawa coupling 
$\Sigma Q\bar Q$ would emerge at too low
energies). Nevertheless, a range of suitable parameters, 
$10^{10}$~GeV$\lesssim M_{\rm mess}\lesssim 10^{11}$~GeV, 
survives. Note that additional mechanism is required in this model to
solve the $\mu$-problem. 

Thus we have seen that, apparently, 
only a small number of theories with heavy messengers does not lead to
breaking of the Standard Model gauge symmetry and may be considered as
realistic. Unfortunately, these viable models have too heavy
messengers, the fact which conflicts with constraints from nucleosynthesis 
\eq{mass-bound:nucleos}. Note that these constraints may be
reconsidered (see above and Ref.\cite{Dimop-1,new-nucl}); moreover, there is an
example of a model \cite{Agashe} where a small region of
cosmologically acceptable messenger masses remains.

\subsubsection{Composite models}
\label{compositeness}
The idea that quarks and leptons are low energy degrees of
freedom of an effective theory that is strongly coupled at a certain energy
scale (similar to $\pi$-mesons, which are low energy degrees of
freedom of QCD), is rather popular because it allows both to solve
aesthetic problems of the Standard Model and to expect new
physics at energies significantly lower than the GUT scale. Since 
dynamical supersymmetry breaking also exploits strong dynamics
beyond the Standard Model, it might be natural to expect that one and
the same strongly coupled gauge group is responsible for the 
compositeness as well as for supersymmetry breaking. 

The way to understand that compositeness may help to solve
the problem of perturbative unification of
coupling constants even in models with a large number of
messengers is the following. The matter fields of MSSM and (or)
messengers are assumed to be low energy composite degrees of freedom 
of the strongly coupled secluded sector. Some fundamental degrees of freedom
(preons) are charged under the Standard Model gauge group, moreover, preons
belong to the complete multiplets of GUT group (e.g., $SU(5)$). Let
the fundamental gauge theory be strongly coupled at the scale
$\Lambda_S$. Then at higher energies the contributions to the
$\beta$ functions of the Standard Model gauge coupling constants come only
from preonic degrees of freedom, while at lower energies only
matter fields of the visible sector give such a contribution. There is a
threshold at the energy $\Lambda_S$ where one set of degrees of freedom
replaces the other. It is possible (combinatorically) to construct a
number of
low energy degrees of freedom from a few independent preons, therefore
the low energy theory may be asymptotically non-free (many
messengers). Coefficients of
$\beta$ functions may change, however, yet in the weak coupling domain
of the MSSM; at 
high energies the theory is asymptotically free and
gauge couplings remain small in this case. As both fundamental and composite
fields belong to complete $SU(5)$ multiplets, the perturbative
unification of the coupling constants remains intact (at least, at the
one loop level) -- the first coefficients of $\beta$ functions of 
all three Standard Model groups 
would be shifted by the same number. 

To construct composite models with perturbative unification of
coupling constants, one has to be able to analyze the low energy
description of strongly coupled gauge theory in order to learn which 
composite degrees of freedom enter the effective theory, to 
describe the vacuum and to confirm that the Standard Model group is not broken
in this vacuum. For our purposes, this vacuum has to break
supersymmetry and the fields mediating supersymmetry breaking to MSSM
fields have to appear. The methods of studying the supersymmetric gauge
theories described in section~\ref{dsb} allow one to
carry out this analysis in some cases 
(see, e.g., Ref.~\cite{we:compose--comp2}). 
In models with composite quarks and leptons originated from the
supersymmetry breaking sector, superpartners of matter fields get soft
masses due to the dynamics responsible for supersymmetry
breaking; global symmetries provide universality of soft terms. 
Gaugino masses may arise from the loop diagrams due to the interaction
either with messengers or with quarks and leptons. 
Messengers are not required at all in the latter case, 
but gaugino might be too light. 

To complete this section, it is worth pointing out that models with direct
mediation have been invented to simplify theory in comparison with
models with mediation via singlet. However, specific direct mediation
models are something even more contrived. 

\section{Gauge mediation and cosmology}
\label{D3.3}
Let us consider cosmological implications of gauge mediation theories. 
In particular, we discuss the following topics: 
\begin{enumerate}
\item Gravitino as a candidate for dark matter.
\item Moduli and dilaton in the early Universe. 
\item The role of new particles (messengers and fields from the secluded
sector) in the evolution of the Universe.
\end{enumerate}

\subsection{Low energy supersymmetry breaking and light particles}
As it has been mentioned above, the relatively low scale of supersymmetry
breaking in the secluded sector $F_{\sm {DSB}}\lesssim
(10^{10}~{\rm GeV})^2$ is typical for models with gauge mediation. This
leads to light gravitino and, in compactified string theories, to light
masses for some other fields. 

\subsubsection{Light gravitino} 
\label{LG.1}    

\paragraph{Lower bounds on $m_{3/2}$.} 
Various lower bounds on light gravitino mass are set by 
low energy measurements and 
astrophysics. The strongest bound, however, comes from physics at 
accelerators. Let us recall the conservative limit 
$\Lambda\gtrsim 3\times 10^4\frac{1}{\sqrt{n}}$~GeV, presented in 
section~\ref{D1.3} ($n$ is the 
effective number of messenger generations). This limit implies the
estimate 
$$
m_{3/2}=\frac{F_{\sm{DSB}}}{\sqrt{3}M^*_{Pl}}>\frac{\Lambda^2}{\sqrt{3}M^*_{Pl}}
=0.5\frac{1}{n}\mbox{eV}\;.
$$
The lower bound on gravitino mass may be significantly higher in
specific models~\cite{moduli_inflation}. For example, this is the
case, if the singlet $S$ gets its 
vacuum expectation value through interaction with $U(1)_{m}$ charged 
fields from the secluded sector (the mechanism discussed in section 
~\ref{DN.1}). If the gauge coupling constant $\alpha_m$ is assumed 
to be small up to $M_{\sm{GUT}}$ then 
$$
\frac{\alpha_{m}}{4\pi}(\Lambda)\lesssim\frac{1}{b_{m}
\ln\frac{M_{\sm{GUT}}}{\Lambda}},
$$
where $b_{m}$ is a sum of $U(1)_m$ charges squared of messenger fields 
(usually this value is of order 10). Hence we have $\alpha_m\sim 0.2$,
and 
$$
m_{3/2}\gtrsim
70\mbox{keV}\frac{1}{nk^2_m}
\frac{0.2}{\alpha_m}\l\frac{m_{\tilde{e}_{\sm R}}}{45\mbox{GeV}}\r^2\;,
$$
where $k_{m}$ is a coefficient in the relation 
$
\Lambda=k_{m}\frac{\alpha_{m}}{4\pi}\sqrt{F_{\sm{DSB}}}.
$
For $n=1\div 4$ we obtain $m_{3/2}\gtrsim 100$~keV, $\Lambda\simeq 3\times
10^4$~GeV.  

\paragraph{Constraints on $m_{3/2}$ related to 
the evolution of the Universe.} 
The contemporary picture of the evolution of the Universe is most
sensitive to the particle physics when 
primordial nucleosynthesis and energy density of the Universe are discussed. 

NLSP decoupled from the visible sector before the beginning of 
nucleosynthesis (i.e., before 1 s after Big Bang). Photons from their decay,
however, might affect significantly the abundance of chemical elements in the
Universe because the light nuclei would be destroyed. One may avoid
this problem if the reheating temperature after inflation 
was not too high, so that dangerous
concentration of superpartners was not reached. Another solution is
based on the requirement that NLSP had to decay during the first
second of the evolution of the Universe. 
In the latter case one obtains the following constraint on the parameters of
the theory~\cite{vacuum_relaxation},
$$
\sqrt{F_{\sm{DSB}}}
\lesssim 10^5\l\frac{m_{\sm{NLSP}}}{\mbox{100~GeV}}\r^{5/4}~\mbox{TeV},
$$
thus the models with heavy gravitino ($m_{3/2}\gtrsim 5$~GeV, that is 
$F_\sm{DSB}\sim m_{3/2} M^*_{Pl}\gtrsim (10^6~{\rm TeV})^2$) look
the least promising from the cosmological point of view. 

Another limit on the gravitino mass comes from the requirement that
the average energy density of the Universe, $\Omega\rho_c$, does not 
exceed the critical value $\rho_c$. For instance, assuming that
gravitino is thermalized in the early Universe, one obtains the bound 
$m_{3/2}<2h^2$keV~\cite{warm_gravitino}, where $h$ is Hubble constant
expressed in units $100{\mbox{km}}/{\mbox{s/ Mpc}}$. At $h\sim 0.7$
only gravitino lighter than $1$~keV is 
acceptable. In this case we obtain an upper bound on the scale of
supersymmetry breaking: $\sqrt{F_{\sm{DSB}}}<2\times
10^3$~TeV~\cite{gravitino_phenomenology}. Together with the lower
bound coming from physics at accelerators (see section~\ref{tg-beta}), we
find that gravitino would not be cosmologically dangerous in the model with
NLSP decaying inside a detector. 

It is known from observation of rotation curves of
galaxies that the significant part of an average density of the
Universe consists of invisible matter. In addition, consideration of
nucleosynthesis implies that the main part of the dark matter is
non-baryonic. {\it Light gravitino 
in thermal equilibrium may play the role of warm dark 
matter}~\cite{warm_gravitino}.

The typical gravitino masses are $1$~keV~$\lesssim
m_{3/2}\lesssim$~100~keV in Dine-Nelson type models. 
In this case NLSP would decay relatively late and gravitino might 
overclose the Universe. This does not occur in the cosmological models
with late inflation, where at the reheating
temperature $T_r<\tilde{m}_{\sm{NLSP}}\sim m_{\sm Z}$, 
multiple production of NLSP and increasing of number of gravitino~\cite{MMY} do
not occur. 
We note, that {\it since gravitino from NLSP decays are not thermalized,
it might be a candidate for cold dark matter}~\cite{cold_gravitino_decay}.

In the models with direct mediation 
(see section~\ref{sec:direct}), gravitino heavier than 100~keV are
typical, and the main contribution to their production rate
comes from scattering $A+B\to C+\psi_{3/2}$. 
\begin{figure}[htb]
\begin{picture}(0,0)%
\centerline{\epsfxsize=1.0\textwidth \epsfbox{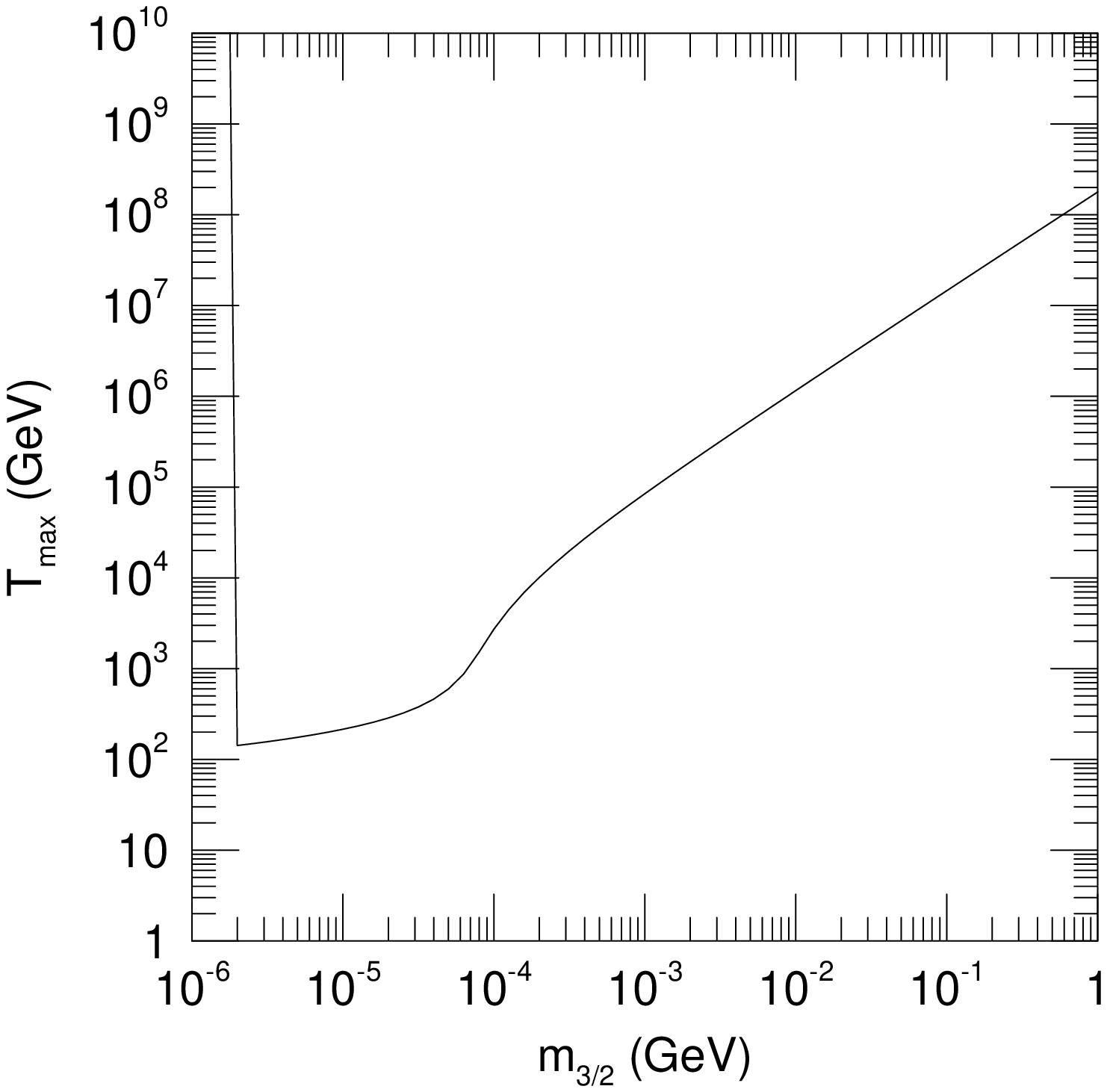}}
\end{picture}%
\setlength{\unitlength}{3947sp}%
\begingroup\makeatletter\ifx\SetFigFont\undefined%
\gdef\SetFigFont#1#2#3#4#5{%
  \reset@font\fontsize{#1}{#2pt}%
  \fontfamily{#3}\fontseries{#4}\fontshape{#5}%
  \selectfont}%
\fi\endgroup%
\begin{picture}(5971,5648)(1741,-7974)
\put(5931,-5536){\makebox(0,0)[lb]{\smash{\SetFigFont{12}{14.4}
{\familydefault}{\mddefault}{\updefault}\parbox{3cm}{\sl ACCEPTABLE REGION}}}}
\end{picture}
\mycaption{The upper bounds on the reheating temperature $T_{max}$ 
from the requirement $\Omega_{3/2}h^2<1$ at 
various gravitino masses.~\cite{moduli_inflation}
\label{Omega}
}
\end{figure}  

The compilation of bounds on the reheating temperature are shown on
Fig.~\ref{Omega} (see Ref. \cite{moduli_inflation}).
We see that $T_{max}\ll 10^8$~GeV, which is hardly possible in the
usual inflation scenario. Therefore, at 
$m_{3/2}\gtrsim 1$~keV some additional entropy production is required to reduce the
gravitino density. One of the mechanisms of entropy production 
deals with the relaxation of the fields corresponding to the flat
directions of MSSM potential. If the energy accumulated in the field 
parameterizing the flat direction dominates the energy of
the Universe, then the decay of the field increases the temperature of the
Universe up to $1\div 10$~TeV. In this case gravitino 
are not thermalized and required entropy production may occur via fine
tuning of parameters responsible for the decay; 
in addition, the baryon-photon ratio remains at 
the acceptable level. 

In the general case, the large entropy production may wash out the baryon
number. This is not the case if baryogenesis is provided by means of 
the Affleck-Dine
mechanism~\cite{Affleck_Dine_baryogenesis,Affleck-Dine-new}~\footnote{The
electroweak baryogenesis~\cite{RuSh}, taking place also at relatively low
temperature, is not suitable in this case because of large 
$m_{\tilde{t}}$~\cite{Quiros}.}.

\paragraph{Example of a model.}
\label{D4.10}
The model proposed in Ref.~\cite{INTY} and based on the 
$SU(2)$-model (see section~\ref{D4.2}) in the secluded sector is one 
of the most suitable from the cosmological point of view. 
Recall that in the low energy theory, the only
singlet $Z$ remains which acquires nonzero vacuum expectation value of the
auxiliary component due to the superpotential 
$W=k_z\Lambda^2_{\sm{DSB}}Z$ 
($\Lambda_{\sm{DSB}}=\Lambda_2$ is a scale of $SU(2)$ group) and
acquires nonzero $\la z\ra$ due to the Kahler potential.

One takes the superpotential as in the classical work by 
O`Raifeartaigh~\cite{LOR}:
\begin{equation}
W_1=Z\l k_z\Lambda^2_{\sm{DSB}}+\lambda Q\bar{Q}\r+mQ\bar{Q}'+m'Q'\bar{Q}
\label{INTY_1}
\end{equation}
where $(Q,\bar{Q})$, $(Q',\bar{Q}')$ are two vector messenger
generations. There is a non-supersym\-metric vacuum with 
$\la F_z\ra =k_z\Lambda^2_{\sm{DSB}}$ and $k_z\la z\ra
\sim\Lambda_{\sm{DSB}}$. This vacuum is stable provided 
$|mm'|>|\lambda \la F_z\ra |$. 

Masses of gauginos and scalars will be given again by
formulae similar to Eqs.~(\ref{gaugin}) and~(\ref{scalmas}), where 
$\Lambda=\frac{\lambda \la F_z\ra }{\sqrt{mm'}}$, and $x=\frac{\lambda
\la F_z\ra }{mm'}$. However, an additional suppression appears 
due to a function with a maximal value 0.1 in the prefactor. At small
$x$, the gaugino masses turn out to be
suppressed by additional factor $x^2$ due to an accidental
cancellation. As a result, the 
strongest limits on the parameters of the theory are taken from 
searches for gaugino~\cite{Particle_Data}: $M_2\gtrsim 50~\mbox{GeV}$, 
$M_3\gtrsim 220~\mbox{GeV}$. Taking into account 
the upper bounds on squark
masses we obtain for the parameter $x$ that 
$1\ge x^2\gtrsim 0.1$, and for the gravitino mass, $m_{3/2}=\frac{\la F_z\ra }{\sqrt{3}M^*_{Pl}}
\sim\frac{1}{\lambda}x^{-5}10^{-2}~\mbox{keV}$, one has 
$$
10~\mbox{eV}
\lesssim m_{3/2}
\lesssim 
3~\mbox{keV}
$$
at $\lambda\sim 1$ and $\Lambda_{\sm{DSB}}\simeq 10^{5\div 6}$~GeV.

In all arguments above we assume $m\sim
m'\sim\Lambda_{\sm{DSB}}$. Certainly, this relation may occur to be
correct in the strong coupling regime. One can avoid explicit
introduction of mass terms by including additional fields in the theory. 
Thus, the model discussed above 
is an example of the theory where gravitino may
constitute warm dark matter in the framework of the standard scenario of
inflation. 

\subsubsection{Cosmology in the context of string theories}
\label{string} 
\paragraph{Moduli.} 
It is often assumed that supersymmetric gauge theories are
spontaneously 
compactified string theories, an assumption which may lead to 
specific cosmological problems. 
A part of them is related to the presence of so-called moduli fields in
string theories. The expectation values of moduli fields parametrize 
the compactified dimensions in the coordinate space. The
flat directions of the potential correspond to moduli fields in the theory with
unbroken supersymmetry. Supersymmetry breaking generates moduli
mass $m_{\sm{\Phi}}\sim m_{3/2}$.  
The evolution of the moduli is important for the models with 
hidden sector~\cite{Polonyi1} (so-called Polonyi
problem) as well as for the models with low energy supersymmetry 
breaking~\cite{Polonyi2}. We consider only 
models of the latter type; the general analysis of
this problem was performed in Ref.\cite{B.de_Carlos}.

The essence of the moduli problem is the following. Since their 
potential is almost flat, moduli fields evolve slowly from their initial
values, $\Phi\sim M_{Pl}$, which may result in overclosing the 
Universe. Indeed, the critical density decreases during the
expansion of the Universe. Moduli interact weakly with other fields; 
at the early stage their energy density receives the main contribution
from the nonzero potential and remains almost constant. In
this case the energy density plays a role of vacuum energy (cosmological constant) and
does not lead to slowing down the expansion of the Universe. However, 
at Hubble constant $H\sim m_\Phi$ moduli fields begin to oscillate and
to produce particles, whose density may exceed the critical 
value at this moment. 

\begin{figure}[htb]
\begin{picture}(0,0)%
\epsfig{file=mod.pstex}%
\end{picture}%
\setlength{\unitlength}{3947sp}%
\begingroup\makeatletter\ifx\SetFigFont\undefined%
\gdef\SetFigFont#1#2#3#4#5{%
  \reset@font\fontsize{#1}{#2pt}%
  \fontfamily{#3}\fontseries{#4}\fontshape{#5}%
  \selectfont}%
\fi\endgroup%
\begin{picture}(7154,5150)(28,-12339)
\put(1736,-8148){\makebox(0,0)[lb]{\smash{\SetFigFont{14}{16.8}
{\rmdefault}{\mddefault}{\updefault}{\sl Critical density}}}}
\put(2990,-11031){\makebox(0,0)[lb]{\smash{\SetFigFont{14}{16.8}
{\rmdefault}{\mddefault}{\updefault}{\sl X($\gamma$)-ray background}}}}
\put(1736,-9069){\makebox(0,0)[lb]{\smash{\SetFigFont{14}{16.8}
{\rmdefault}{\mddefault}{\updefault}{\sl Moduli density}}}}
\end{picture}
\mycaption{The limits on the moduli mass~\cite{moduli2} in model with
modified thermal inflation; acceptable region is painted in grey.
\label{moduli-fig}
}
\end{figure} 

One of the possible solution to this 
problem~\cite{moduli_inflation,inflation_GMM} is an additional late
inflationary stage (for instance, a stage of the thermal 
inflation~\cite{Lyth}). Moduli fields are diluted at this stage, and
baryon asymmetry decreases.
However, the Affleck-Dine mechanism of
baryogenesis~\cite{Affleck_Dine_baryogenesis} may lead to a
significant baryonic asymmetry~\cite{moduli_inflation} in this case. 
In order that this mechanism works successfully, the relic moduli 
density has not to be less than 
\begin{equation}
\label{hhh}
\Omega_{\sm{\Phi}}h^2\gtrsim 10^{-6}\frac{m_{\sm{\Phi}}}{100\mbox{keV}}
\end{equation}
In specific models (see, e.g., Ref.~\cite{moduli2}) stronger lower
bound on the moduli density may arise due to their additional
production after thermal inflation epoch. 

The additional constraint comes from measuring the flux of cosmic rays
and is related to the fact that the main channel of moduli decay is  
pair production of $\gamma$-quanta. Combining this limit with
bound~(\ref{hhh}) one obtains an upper bound on the moduli mass, 
$m_{\sm{\Phi}}\lesssim 2$~MeV~\cite{moduli_rays_1}.

The set of limits on the parameters of moduli is presented in 
Fig.~\ref{moduli-fig} (see\ Ref.\ \cite{moduli2}). Although these
results are obtained in a specific model (modified thermal inflation)
many of them are valid in other cases. 

It was found that solution of all these problems without
introduction of additional energy scales is possible only in the
framework of rather complicated inflationary models 
(see, e.g., Ref.~\cite{moduli2}). In this case, the Affleck-Dine mechanism
provides sufficient baryon production 
($\frac{n_{\sm B}}{s}\gtrsim
10^{-10}$)~\cite{moduli_inflation}. 

It is worth noting that the hypothesis of low energy supersymmetry
breaking leads, as a rule, to new long-range forces originating
from the interaction via 
light moduli fields~\cite{Signatures,moduli_force}. This
prediction may be checked in experiments.

\paragraph{Dilaton.}
\label{dilaton}
Similar problems arise for the dilaton field $\phi$. One
expects that this field acquires mass of order 
$m_{3/2}$ due to nonperturbative dynamics~\cite{B.de_Carlos}. 
Dilaton lifetime is much longer than the age of the Universe. 
One can use~\cite{moduli_inflation} the mechanism of late 
thermal inflation~\cite{Lyth} in order to reduce the energy
accumulated by the dilaton field below thecritical value. 
It was found that for the inflaton mass exceeding 130~GeV the lower 
bound on the ratio of dilaton energy density to entropy is
reached at the minimal reheating temperature $T_r=10$~MeV, which is acceptable
for successful nucleosynthesis. The value of present critical density 
$\frac{\rho_c}{s}=3.6\cdot 10^{-9}h^2$~GeV leads 
to a limit~\cite{str-dilaton}
$$
\frac{\rho_\phi}{\rho_c}h^2\equiv\Omega_\phi h^2\gtrsim 1.5\cdot  
10^{-2}\l\frac{m_\phi}{\mbox{MeV}}
\r^{-3/4}\;.
$$
The dilaton energy density can be less than critical at 
$m_\phi\gtrsim 20$~keV in this case. 
If the inflaton mass is not larger than 130~GeV, then~\cite{str-dilaton}
$$
\Omega_\phi h^2\gtrsim 2.3\cdot  
10^{-3}\l\frac{m_\phi}{\mbox{MeV}}
\r^{-3/14}\;.
$$
It means that in the region $m_\phi\simeq$ 10~MeV-1~GeV the lower
bound on $\Omega_\phi h^2$ is lower than current 
critical density $\Omega h^2\simeq$0.25 by a factor of about 30.

For both dilaton and moduli, the strongest constraints
come from the analysis of the observed $\gamma$-ray flux, which dilaton
decays $\phi\to 2\gamma$ contribute to. In the case of the simplest
interaction Lagrangian 
$$
{\cal L}\sim
\frac{1}{16\pi\alpha}\frac{\Phi}{M^*_{Pl}}F_{\mu\nu}^2
$$
the dilaton lifetime is $\tau_\phi\simeq 7\cdot
10^{23}\l\frac{\mbox{MeV}}{m_\phi}\r^3$~s, and the dilaton masses from 500~keV to
1~GeV contradict the experiment~\cite{moduli_rays_1}. 
It is natural to expect $m_\phi\sim m_{3/2}$
that provides a strong upper bound on the value of supersymmetry
breaking F-term: $\sqrt{F_{\sm{DSB}}}<5\cdot 10^7$~GeV. 
A possibility to reconcile string compactification with rather
wide class of models under discussion is excluded by this limit. 

\subsection{Stable particles and cosmology} 
\label{dark-matter}
Let us consider the cosmological consequences of 
the presence of new fields (messengers and fields from the secluded
sector) in the model. 

First, we consider the secluded sector. The conservation of possible
global charges in the secluded sector guarantees the stability of the lightest
charged particle, in analogy to the baryon number in the Standard
Model which ensures proton stability. Following Ref.~\cite{messenger's_cosmology} we consider the
constraints on the parameters of the theory, assuming that new stable 
particles (``baryons'') form dark matter. 
One can estimate the density of the  relic
particles from their total annihilation cross section to light neutral
particles~\cite{unitarity_cosmology},  
$$
\Omega_{\sm B}\cdot h^2\gtrsim \l\frac{m_{\sm B}}{300 \mbox{TeV}}\r^2\;.
$$
Hence, {\it ``baryons'' from the secluded sector with} 
$m_{\sm B}\sim 100$~TeV {\it are realistic candidates for cold dark
matter}, the latter playing key role in the contemporary picture of
the 
structure formation in the Universe. 

We turn now to the study of messenger fields. In the case of indirect 
mediation and in the absence of mixing discussed in
section~\ref{D3.4} these fields are separated from
the secluded sector by singlet $S$ and from the visible sector
by gauge interactions. It is not difficult to see that in the case of indirect
interaction and in the absence of mixing discussed in
section~\ref{D3.4}, a global quantum
number carried by messengers is conserved. The
lightest of these fields will be also stable and may contribute to 
dark matter. If the lightest stable messenger were electrically charged, 
its presence would result in the distortion of the spectrum of
cosmic rays \cite{charged_dark_matter}. 
To find the lightest messenger, we consider mass splitting between
electrically charged and neutral components of messengers in
the fundamental representation. This splitting
arises at the tree level due to the contribution of $SU(2)$ $D$-term, 
as a result of electroweak symmetry breaking,
\be
\l m^2_+\!-\!m^2_0\r_{tree}\!=\!
\sqrt{
\l\lambda \la F_s\ra \r^2+
\frac{1}{4}M_{\sm Z}^4\cos^2 2\beta
}
-\sqrt{
\l\lambda \la F_s\ra \r^2\!+\!
\l\sin^2\theta_{\sm W}\!-\!\frac{1}{2}\r ^{\! 2}M_{\sm Z}^4\cos^2\! 2\beta
}.
\label{6}
\end{equation}
This splitting is of order $M_{\sm Z}^2\frac{M_{\sm Z}^2}{\lambda \la
F_s\ra }$. The one loop contribution to the masses is proportional to
$\alpha M_{\sm Z}^2$. Which contribution dominates, is determined by the
ratio of suppression factors $\alpha$ and $\frac{M_{\sm
Z}^2}{\lambda\la F_s\ra}$. Assuming that
messengers are heavier than electroweak bosons, i.e.
$\frac{\Lambda^2}{x^2}\l 1-x\r\gtrsim 1$~TeV$^2$, the one loop
contribution to mass splitting between electrically charged and
neutral components of messenger fields takes the
form~\cite{messenger's_cosmology}:
\be
\label{7}
\l m^2_+-m^2_0\r_{1-loop}=\frac{\alpha}{4\pi}M_{\sm Z}^2\l4\ln\frac{x}{1-x}-
\ln\frac{1+x}{1-x}+\frac{2x}{1-x}\ln\frac{2x}{1+x}-4\r\;.
\end{equation}
From the analysis of Eqs.\ (\ref{6}) and (\ref{7}) one infers whether
the lightest messenger is neutral. 

The condition $m_+>m_0$ does not
guarantee the absence of problems with charged particles. 
Decay width of the charged messenger component $q^+$
$$
\Gamma\l q^+\to q^0e\nu\r=\frac{G_{\sm F}^2}{15\pi^3}\l m_+-m_0\r^5
$$
is strongly suppressed by a small factor $(\delta m)^5$. 
Late decay of a charged particle may affect nucleosynthesis and result
in descripancy with observed matter abundances in the Universe. 
Nucleosynthesis is not spoiled if the lifetime of $q^+$ is of order 1~s. From
this requirement one obtains 
\begin{equation}
\l m_+-m_0\r\gtrsim 5\mbox{MeV}
\label{lll}
\end{equation}

One more constraint comes from the condition that the density of the stable
messengers should not exceed the critical value. From the requirement 
$\Omega h^2<1$, one obtains the limit on the mass of the lightest messenger, 
\begin{equation}
m^2_0\lesssim 25~\mbox{TeV}^2.
\label{25}
\end{equation}

By comparing the inequalities~(\ref{lll}), 
(\ref{25}) with Eqs.~(\ref{6}) and (\ref{7}), 
we find the small region in the entire parameter space $\l\Lambda, x\r$ 
where the lightest stable messengers may be phenomenologically
acceptable, namely, 
$x\gtrsim 0.95$. The dominant contribution here is Eq.~(\ref{7}).

Introducing several messenger generations with small weak mixing 
and different parameters $\l\Lambda_i,x_i\r$ may extend somewhat the 
range of parameters.  In this case a similar constraint should be imposed 
on $max\{\Lambda_i\}$ and does not affect the parameters of the 
lightest messenger with 
$\Lambda_{light}\ll max\{\Lambda_i\}$, which is cosmologically
acceptable under one of the following conditions: 
$\sqrt{\la F_s\ra_{light} }\lesssim 350~\mbox{GeV}$ (in this case
(\ref{6}) and (\ref{lll}) are compatible; (\ref{6}) dominates provided
$x$ is not very small; this region is
phenomenologically unacceptable in models with one singlet $S$) or
$x_{light}\gtrsim 0.95$.

The extension to the case of messengers belonging to the other
representations is straightforward. However, it does not result in
extending the allowed parameter region. 

The possibility that messengers constitute dark matter may be tested in
direct searches for dark matter. The lightest messenger scatters 
off nuclei through $Z$-boson exchange. Searches for 
dark matter imply that such particles with masses $5$~TeV compose not
more than a quarter of mass of the Galactic halo with density 0.3~GeV/cm$^3$. It
means, that {\it it is impossible to solve the dark matter 
problem by making use of messenger fields only, even invoking fine tuning}. 
 
To conclude, the stable lightest messenger is 
unacceptable for cosmology for most values of parameters. 
In analogy to gravity mediation models
this problem may be solved by invoking late stage of
inflation and low reheating temperature (of order of $M_{\sm Z}$).
The violation of global messenger quantum numbers due to 
mixing with MSSM fields considered in section~\ref{D3.4} seems to be more
natural. 

\section{Conclusions}
\label{sec:concl}
We considered various models which invoke a beautiful mechanism of
gauge mediation of supersymmetry breaking. We discussed phenomenology
of these models and their cosmological consequences. The basic
distinstive features of this approach are the following.
\begin{itemize}
\item
In models of gauge mediation, it is possible to describe mediation of
supersymmetry breaking to the observable sector and to calculate
parameters of this breaking just in the frameworks of field theory. It
is not necessary to investigate the effects of quantum gravity or
string theory. Models with gauge mediated supersymmetry breaking have
very few free parameters; theoretical predictions of these models are
not in conflict with experimental data.
\item
A necessary ingredient of these models is a set of messenger fields,
that is
relatively heavy ($\gtrsim 10$~TeV) matter fields
which transform as vector-like representations of the Standard Model 
gauge group.
\item
Supersymmetry breaking occurs at relatively low energies 
($\lesssim 10^{10}$~GeV) and mass spectrum of superpartners of
the Standard Model fields is determined by their quantum numbers and 
values of gauge coupling constants. As a consequence, (a)~flavour changing
processes are naturally suppressed without fine tuning;
(b)~the lightest superpartner is gravitino, this fact resuts in specific
phenomenology. 
\item
To obtain a correct pattern of the
electroweak symmetry breaking, fine tuning or 
introduction of additional parameters is required these models
($\mu$-problem). 
\end{itemize}
Several specific realizations of the gauge mediation mechanism were
constructed. The difference among them concerns the ways to include 
the supersymmetry breaking sector. We point out the following characteristic 
features.
\begin{itemize}
\item
The ``Minimal'' model does not include additional parameters in
the Higgs sector but requires fine tuning of couplings responsible for 
electroweak symmetry breaking. The way out is to introduce 
additional parameters. 
\item
Specific realistic models where supersymmetry breaking sector is
incorporated often
require complicated dynamics. In a series of models, 
supersymmetry is broken in a metastable vacuum. 
\item
Relatively simple models with direct mediation require a large number of
messengers. Many models with heavy messengers are unrealistic because
they break $SU(3)\times U(1)$ and are problematic for cosmology.
The large number of light messengers 
results in lost of perturbativity of coupling constants at the scale of Grand
Unification. 
\end{itemize}

Therefore, many theories of this kind are consistent with plethora 
of experimental limits. However, in our opinion, there is no
model which satisfies constraints coming from both physics
at accelerators and cosmology and which does not require fine tuning. 
This feature is undesirable since the natural origin of parameters 
is one of the cornerstones of the models with gauge mediation. 
We would like to list a few interesting problems in theories
with gauge
mediated supersymmetry breaking, problems which probably will be
solved in the nearest future: 
\begin{itemize}
\item 
It would be interesting to find new ways to constrain the parameters
in the frameworks of each specific model. As a rule, this requires to
consider both the visible and the secluded sectors of the theory. An
example of such a constraint is provided by boundary conditions at the
high energy scale according to the Grand unification or string
compactification 
(see e.\ g.\ \cite{we:couplings,q98}). 
\item
Relatively strong constraints on gauge mediation models 
come from cosmology, so it would be important to make the considered
mechanism consistent with models of
the early Universe without fine tuning.  
\item
Many models with gauge mediated supersymmetry breaking are not 
studied well enough from the phenomenological point of view. 
Detailed analysis may lead to unexpected
consequences and make several models phenomenologically inconsistent. 
\item
Challenging approaches to the standard problems of model
building, such as refusal of the 
perturbative unification of gauge coupling
constants, may simplify the theories without lost of viability. 
\end{itemize}

The authors thank V.A.Rubakov for the valuable discussions and 
attention to the work. We are indebted to F.L.Bezrukov,
M.V.Libanov, A.A.Penin, Yu.F.Pirogov, D.V.Semikoz, P.G.Tinyakov, 
participants of seminars in INR RAS, NPI MSU, ITEP and JINR who
took part in discussions of various topics covered by this 
review. This work is supported in part 
by Russian Foundation for Basic Research (grant 96-02-17449Á). 
The work of S.D. and D.G. is supported in part 
by grant INTAS 96-0457 and ISSEP fellowships. The work of S.T. is
partially supported by grant CRDF RP1-187. 

{\bf Note added.}
The Russian version of this review has been submitted to {\em Uspekhi}
in 1998. Since then, a number of papers appeared where 
phenomenological \cite{p1,p2,p3,p4,p5,p6,p7,p8,p9,p10,p11,p12}, model
building \cite{m1,m2,m3,m4,m5,m6,m7,m8,m9,m10,m11}, and
cosmological \cite{c1,c2,c3,c4,c5} aspects of gauge
mediated supersymmetry breaking are discussed. 

%%%%%%%%%%%%%%%%%%%%%%%%%%%%%%%%%%%%%%%%%%%%%%%%%%%%%%%%%
\def\ijmp#1#2#3{{\it Int.\ Jour.\ Mod.\ Phys.\ }{\bf #1} #3 (19#2)}
\def\pl#1#2#3{{\it Phys.\ Lett.\ }{\bf B#1} #3 (19#2)}
\def\zp#1#2#3{{\it Z.\ Phys.\ }{\bf C#1} #3 (19#2)}
\def\prl#1#2#3{{\it Phys.\ Rev.\ Lett.\ }{\bf #1~} #3 (19#2)}
\def\rmp#1#2#3{{\it Rev.\ Mod.\ Phys.\ }{\bf #1~} #3 (19#2)}
\def\prep#1#2#3{{\it Phys.\ Rep.\ }{\bf #1~} #3 (19#2)}
\def\pr#1#2#3{{\it Phys.\ Rev.\ }{\bf D#1~} #3 (19#2)}
\def\np#1#2#3{{\it Nucl.\ Phys.\ }{\bf B#1~} #3 (19#2)}
\def\mpl#1#2#3{{\it Mod.\ Phys.\ Lett.\ }{\bf #1~} #3 (19#2)}
\def\arnps#1#2#3{{\it Ann.\ Rev.\ Nucl.\ Part.\ Sci.\ }{\bf #1~} #3 (19#2)}
\def\sjnp#1#2#3{{\it Sov.\ J.\ Nucl.\ Phys.\ }{\bf #1~} #3 (19#2)}
\def\jetp#1#2#3{{\it JETP Lett.\ }{\bf #1~} #3 (19#2)}
\def\app#1#2#3{{\it Acta Phys.\ Polon.\ }{\bf #1~} #3 (19#2)}
\def\rnc#1#2#3{{\it Riv.\ Nuovo Cim.\ }{\bf #1~} #3 (19#2)}
\def\ap#1#2#3{{\it Ann.\ Phys.\ }{\bf #1~} #3 (19#2)}
\def\ptp#1#2#3{{\it Prog.\ Theor.\ Phys.\ }{\bf #1~} #3 (19#2)}
\def\spu#1#2#3{{\it Sov.\ Phys.\ Usp.\ }{\bf #1~} #3 (19#2)}
\def\pu#1#2#3{{\it Phys.\ Usp.\ }{\bf #1~} #3 (19#2)}
\def\ufn#1#2#3{{\it Usp.\ Phys.\ Nauk}{\bf #1~} #3 (19#2)}
\def\apj#1#2#3{{\it Ap.\ J.\ }{\bf #1~} #3 (19#2)}
\def\epj#1#2#3{{\it Eur.\ Phys.\ J.\ }{\bf C#1~} #3 (19#2)}
\def\ppn#1#2#3{{\it Phys.\ Part.\ Nucl.\ }{\bf #1~} #3 (19#2)}
\def\jhep#1#2#3{{\it JHEP }{\bf #1~} #3 (19#2)}
%%%%%%%%%%%%%%%%%%%%%%%%%%%%%%%%%%%%%%%%%%%%%%%%%%%%%%%%%


\begin{thebibliography}{999}	
\bibitem{MatveevKrasnikov} 
Krasnikov N V, Matveev V A \ppn{28}{97}{441}, hep-ph/9703204
\bibitem{GolfandLikhtman} 
Gol'fand Yu A, Likhtman E P \jetp{13}{71}{323}
\bibitem{VolkovAkulov} Volkov D V, Akulov V P \pl{46}{73}{109}
\bibitem{WessZum} Wess J, Zumino B \np{70}{74}{39}
\bibitem{WessBagger} Wess J, Bagger J 
{\em Supersymmetry and Supergravity}
(Princeton Univ.\ Press, 1982)
\bibitem{West} West P {\em Introduction to Superymmetry and
Supergravity} (Singapore: World Scientific, 1986)
\bibitem{Ogiev} Ogievetskii V I, Mezincescu L \spu{18}{76}{960} 
\bibitem{nilles} Nilles H P {\em Phys. Rep.} {\bf 110} 1 (1984)
\bibitem{Gir} Girardello L, Grisaru M \np{194}{82}{65}
\bibitem{witten_ier} Witten E \np{188}{81}{513}
\bibitem{LOR} O'Raifeartaigh L {\em Nucl. Phys.} {\bf B96} 331 (1975)
\bibitem{FI} Fayet P, Iliopoulos J {\em Phys. Lett.} {\bf B51} 461 (1974)
\bibitem{broad_brush} Dine M,  hep-ph/9612389 
\bibitem{20questions} Dienes K R, Kolda C,
in {\em Perspectives on Supersymmetry} (Ed.\ G Kane) (World
Scientific, to appear), hep-ph/9712322
\bibitem{NelsonDSB} Nelson A E {\em Nucl. Phys. Proc. Suppl.} {\bf 62} 261
(1998), hep-ph/9707442
\bibitem{PopTrDSB} Poppitz E, Trivedi S P,
hep-th/9803107
\bibitem{Colda} Kolda C {\em Nucl. Phys. Proc. Suppl.} {\bf 62} 266 (1998),
hep-ph/9707450
\bibitem{GiudiceRattazziReview} Giudice G F, Rattazzi R, 
hep-ph/9801271
\bibitem{visotsky} Vysotsky M I \spu{28}{85}{667} 
\bibitem{kazakov} Kazakov D I
{\em Surveys
in High Energy Physics} {\bf 10} 153 (1997)
\bibitem{berezin} Berezin F A {\em Method of Second Quantization} (Moscow:
Nauka, 1986) (in Russian)
\bibitem{boer} Gladyshev A {\em et al.} \np{498}{97}{3}
\bibitem{FCNC} F.Gabbiani \emph{et al.,} \np{477}{96}{321}, hep-ph/9604387
\bibitem{vacuum_relaxation} Dvali G, Giudice G F, Pomarol A
\np{478}{96}{31}, hep-ph/9603238
\bibitem{DimH} 
Dimopoulos S, Georgi H 
\np{193}{81}{150};\\
Dimopoulos S, Georgi H,
in {\it Second Workshop on Grand Unification,
Ann Arbor, 
1981}
(Eds J Leveille, L Sulak, D Unger) (Birkhauser, 1981) p.285 
\bibitem{dipole} Fischler W, Paban S, Thomas S \pl{289}{92}{373}
\bibitem{dipole_exp} Smith K {\em et al.} \pl{234}{90}{191};\\
Altarer I {\em et al.} \pl{276}{92}{242}
\bibitem{dipole_review} Grossman Y, Nir Y, Rattazzi R, 
in 
{\em Heavy Flavours II} (Eds A J Buras, M Lindner)
(Singapore: World Scientific, to appear); 
hep-ph/9701231
\bibitem{spectrum_old} Dine M, Fischler W \pl{110}{82}{227};\\ 
Alvarez-Gaum\'e L, Claudson M, Wise M \np{207}{82}{96}
\bibitem{parents}
Dine M, Fischler W, Srednicki M \np{189}{81}{575};\\
Dimopoulos S, Raby S \np{192}{81}{353};\\
Dine M, Srednicki M \np{202}{82}{238};\\
Dine M, Fischler W \np{204}{82}{346};\\
Nappi C R, Ovrut B A \pl{113}{82}{175};\\
Dimopoulos S, Raby S \np{219}{83}{479}       
\bibitem{DN} Dine M, Nelson A \pr{48}{93}{1277}, hep-ph/9303230
\bibitem{DNS} Dine M, Nelson A E, Shirman Y \pr{51}{95}{1362},  hep-ph/9408384 
\bibitem{DNNS} Dine M {\em et al.} \pr{53}{96}{2658} hep-ph/9507378
\bibitem{Martin} Martin S \pr{55}{97}{3177}, hep-ph/9608224
\bibitem{messenger's_cosmology} Dimopoulos S, Giudice G, Pomarol A 
\pl{389}{96}{37}, hep-ph/9607225
\bibitem{renorm} Grisaru M, Ro\v cek M, Siegel W {\em Nucl. Phys.} 
{\bf B159} 429 (1979)
\bibitem{Dimopoulos_Thomas_Wells} Dimopoulos S, Thomas S, Wells J
\pr{54}{96}{3283}, hep-ph/9604452
\bibitem{kolda} Babu K S, Kolda C, Wilczek F {\em Phys. Rev. Lett.}
{\bf 77} 3070 (1996) hep-ph/9605408
\bibitem{borzumati} Borzumati F,  
hep-ph/9702307
\bibitem{Dimop-1} Dimopoulos S {\em et al.} \np{510}{98}{12}, hep-ph/9705307
\bibitem{new-nucl} Gherghetta T, Giudice G F, Riotto A, hep-ph/9808401 
\bibitem{generalization_GT} Fayet P \pl{70}{77}{4611}
\bibitem{kabak} Langacker P, Polonsky N \pr{52}{95}{3081}, hep-ph/9503214
\bibitem{GUT_fall_1} Carone C, Murayama H \pr{53}{96}{1658}, hep-ph/9510219
\bibitem{GUT_fall_3} Hamidian H {\em et al.} \pl{428}{98}{310}, hep-ph/9803228
\bibitem{Faraggi} Faraggi A \pl{387}{96}{775}, hep-ph/9607296
\bibitem{string_high_scale} Derendinger J, Ib\'a\=nez L, Nilles H
\pl{155}{85}{65};\\
Dine M {\em et al.} \pl{156}{85}{55};\\
Ferrara S {\em et al.} \pl{245}{90}{409};\\ 
Casas J {\em et al.} \np{347}{90}{243};\\ 
Cveti\v c M {\em et al.} \np{361}{91}{194}
\bibitem{Particle_Data} Caso C {\em et.\ al.} (Particle Data Group), 
\epj{3}{98}{1}
\bibitem{Non-Unified} Dimopoulos S, Pomarol A, \pl{353}{95}{222},
hep-ph/9502397 
\bibitem{mu_1--B_mu_2} Barbieri R, Giudice G \np{306}{88}{63}
\bibitem{Giudice-Masiero} Giudice G, Masiero A \pl{206}{88}{480}
\bibitem{Gherghetta} Gherghetta T, Jungman G, Poppitz E, hep-ph/9511317 
\bibitem{weak_scale_phenomenology} Bagger J {\em et al.}
\pr{55}{97}{3188}, hep-ph/9609444 
\bibitem{RS_2} Rattazzi R, Sarid U \np{501}{97}{297} hep-ph/9612464
\bibitem{Sparticle-Spectroscopy} Dimopoulos S, Thomas S, Wells J
\np{488}{97}{39}, hep-ph/9609434
\bibitem{GUT_fall_2} Blok B, Lu C, Zhang D \pl{386}{96}{146}, hep-ph/9602310
\bibitem{gravitino_decay} Gaillard M, Hall L, Hinchliffe I
\pl{116}{82}{279};\\ 
Ellis J, Hagelin J \pl{122}{83}{303}
\bibitem{MMY} Moroi T, Murayama H, Yamaguchi M \pl{303}{93}{289}
\bibitem{Signatures} Dimopoulos S {\em et al.} \prl{76}{96}{3494}, hep-ph/9601367
\bibitem{gravitino_phenomenology} Ambrosanio S {\em et
al.} \pr{54}{96}{5395}, hep-ph/9605398
\bibitem{CDF_predictions} Ambrosanio S {\em et al.}
\prl{76}{96}{3498}, hep-ph/9602239
\bibitem{gravitino_collider} Stump D, Wiest M, Yuan C-P
\pr{54}{96}{1936}, hep-ph/9601362
\bibitem{ep} Kiers K, Ng J, Wu G \pl{381}{96}{177},
hep-ph/9604338
\bibitem{CDF} Park S (for CDF Coll.) in {\it Proc.\ 10th Topical Workshop
on Proton-Antiproton Collider Physics} (Eds R Raha, J Yoh) (New York:
AIP, 1995)
\bibitem{Ambrosanio_Mele} S.Ambrosanio and B.Mele, \pr{55}{97}{1399},
{\it Erratum}, \pr{56}{97}{3157}, hep-ph/9609212.
\bibitem{B_to_ll} Goto T {\em et al.} \pr{55}{97}{4273},
hep-ph/9609512;\\  
Huang C-S, Liao W, Yan Q-S, hep-ph/9803460
\bibitem{Maiani--Cabibbo-Farrar} Maiani L, Parisi G, Petronzio R
{\em Nucl. Phys.} {\bf B136} 115 (1978);\\
Cabibbo N, Farrar G R
{\em Phys. Lett.} {\bf 110B} 107 (1982);\\ 
Maiani L,  Petronzio R {\em Phys. Lett.} {\bf 176B} 120 (1986);
Erratum -- ibid. {\bf 178B} 457 (1986)
\bibitem{nonpert-unif} 
Brahmachari B, Sarkar U, Sridhar K {\em Mod. Phys. Lett.} {\bf A8} 3349
(1993);\\ 
Hempfling R \pl{351}{95}{206}, hep-ph/9502201;\\
Babu K S, Pati J C \pl{384}{96}{140}, hep-ph/9606215;\\
Kolda C, March-Russell J \pr{55}{97}{4252}, hep-ph/9609480 
\bibitem{Ross}  Ghilencea D, Lanzagorta M, Ross G G \pl{415}{97}{253}, hep-ph/9707462
\bibitem{to-appear} Gorbunov D S, Troitsky S V, to appear
\bibitem{q98} Dubovsky S L, Gorbunov D S, Troitsky S V, 
in {\em Proc. 10th Int.Seminar 'Quarks-98', Suzdal, 1998}, to appear
(Eds V A Matveev {\em et al.}) (Moscow: INR, 1998), hep-ph/9809484
\bibitem{NewRoss} Amelino-Camelia G, Ghilencea D, Ross G G
\np{528}{98}{35}, hep-ph/9804437 
\bibitem{dinem} Dine M, Nir Y, Shirman Y \pr{55}{97}{1501}, hep-ph/9607397 
\bibitem{we} Dubovsky S, Gorbunov D \pl{419}{98}{223}, hep-ph/9706272
\bibitem{we_new} Dubovsky S, Gorbunov D, hep-ph/9807347 
\bibitem{Zhang} Han T, Zhang R-J, \pl{428}{98}{120}, hep-ph/9802422
\bibitem{Dvali-Shifman} Dvali G, Shifman M \pl{399}{97}{60}, hep-ph/9612490
\bibitem{zurab} Tavartkiladze Z \pl{427}{98}{65}, hep-ph/9706332
\bibitem{cc} Mohapatra R N, Nandi S, \prl{79}{97}{181},
hep-ph/9702291; Chacko Z {\em et al.}, \pr{56}{97}{5466}, hep-ph/9704307
\bibitem{barbieri} Barbieri R, Hall L {\em Phys. Lett.}
{\bf B338} 212 (1994), hep-ph/9408406;\\ 
Barbieri R, Hall L, Strumia A 
\np{445}{95}{219}, hep-ph/9501334
\bibitem{ADS-84} Affleck I, Dine M, Seiberg N \np{241}{84}{493}
\bibitem{ADS-85} Affleck I, Dine M, Seiberg N \np{256}{85}{557}
\bibitem{instanton_contribution} Shifman M, Vainshtein A, Zakharov V 
\spu{28}{85}{709}
\bibitem{luty} Luty M A, Washington Taylor IV \pr{53}{96}{3399}, hep-th/9506098
\bibitem{amati}  Amati D {\em et al.} \prep{162}{88}{169}
\bibitem{seiberg92} Seiberg N \pl{318}{93}{469}, hep-th/9309335
\bibitem{peskin} Peskin M, hep-th/9704217
\bibitem{seiberg1} Seiberg N \pr{49}{94}{6857}, hep-th/9402044
\bibitem{'thoft} 't Hooft G, in {\it Recent developments in gauge
theories} (Eds G 't Hooft {\em et al.}) (New York: Plenum, 1980) 
\bibitem{intrilligator} Brodie J, Cho P, Intriligator K \pl{429}{98}{319}, hep-th/9802092
\bibitem{banks} Banks T, Zaks A \np{196}{82}{189}
\bibitem{betafunction} Novikov V {\em et al.} \np{229}{83}{381} 
\bibitem{duality} Seiberg N \np{435}{95}{129}, hep-th/9411149
\bibitem{N2} See, e.g., Argyres P, Plesser M, Seiberg N
\np{471}{96}{159}, hep-th/9603042; 
Elitzur S {\em et al.} \np{505}{97}{202}, hep-th/9704104; 
Schmaltz M, Sundrum R \pr{57}{98}{6455}, hep-th/9708015 
\bibitem{intsei1} Intriligator K, Seiberg N \np{444}{95}{125},
hep-th/9503179
\bibitem{intsei2} Intriligator K, Pouliot P \pl{353}{95}{471}, hep-th/9505006 
\bibitem{IntrThomas} Intriligator K, Thomas S \np{473}{96}{121}, hep-th/9603158
\bibitem{IzawaYanagida} Izawa K-I, Yanagida T {\em Prog. Theor. Phys.} {\bf 95}
829 (1996), hep-th/9602180
\bibitem{HIY} Hotta T, Izawa K, Yanagida T \pr{55}{97}{415}, hep-ph/9606203
\bibitem{R1--R2--Shadmi} Randall L \np{495}{97}{37}, hep-ph/9612426;\\ 
Shadmi Y \pl{405}{97}{99}, hep-ph/9703312;\\
Cs\'aki C, Randall L, Skiba W \pr{57}{98}{383},
hep-ph/9707386
\bibitem{lisa} Dasgupta I, Dobrescu B, Randall L 
{\em Nucl. Phys.} {\bf B483} 95 (1997), hep-ph/9607487
\bibitem{hiddenmur} Arkani-Hamed N {\em et al.} \pr{54}{96}{7032}, hep-ph/9607298
\bibitem{U(1)_mixing} Dienes K, Kolda C, March-Russell J 
\np{492}{97}{104} hep-ph/9610479
\bibitem{NTY} Nomura Y, Tobe K, Yanagida T \pl{425}{98}{107}, hep-ph/9711220
\bibitem{Agashe} Agashe K \pl{435}{98}{83}, hep-ph/9804450
\bibitem{inverse_hierarchy} Witten E {\em Phys. Lett.} {\bf B105} 267 (1981)
\bibitem{PopShTr1--PopShTr2} 
Poppitz E, Shadmi Y, Trivedi S \pl{388}{96}{561}, hep-th/9606184;\\
Poppitz E, Shadmi Y, Trivedi S \np{480}{96}{125}, hep-th/9605113
\bibitem{A-HM-RM}  Arkani-Hamed N, March-Russell J,
Murayama H {\em Nucl. Phys.} {\bf B509} 3 (1998),  hep-ph/9701286
\bibitem{PopTr-grav--PopTr-dyn-mess} 
Poppitz E, Trivedi S P \pr{55}{97}{5508}, hep-ph/9609529;\\
Poppitz E, Trivedi S P
{\em Nucl. Phys. Proc. Suppl.} {\bf 62} 281 (1998), hep-ph/9707439 
\bibitem{PopTr-Strace} Poppitz E, Trivedi S P 
{\em Phys. Lett.} {\bf B}, to appear, hep-ph/9703246  
\bibitem{Shirman-Nren} Shirman Y \pl{417}{98}{281}, hep-ph/9709383
\bibitem{Murayama-model} Murayama H \prl{79}{97}{18}, hep-ph/9705271
\bibitem{GR:wave-function} Giudice G, Rattazi R \np{511}{98}{25}, hep-ph/9706540.
\bibitem{the_best_model} Dimopoulos S, Dvali G,
Rattazzi R \pl{413}{97}{336}, hep-ph/9707537
\bibitem{we:compose--comp2}  Arkani-Hamed N, Luty M A, Terning J \pr{58}{98}{015004} 
hep-ph/9712389;\\
Dubovsky S L, Gorbunov D S, Troitsky S V 
{\em Phys. Lett.} {\bf B423} 301 (1998), hep-ph/9712397
\bibitem{moduli_inflation} de Gouv\^ea A, Moroi T, Murayama H 
\pr{56}{97}{1281}, hep-ph/9701244
\bibitem{warm_gravitino} Pagels H, Primack J \prl{48}{82}{223}
\bibitem{cold_gravitino_decay} Borgani S, Masiero A, Yamaguchi M 
\pl{386}{96}{189}, hep-ph/9605222 
\bibitem{Affleck_Dine_baryogenesis} Affleck I, Dine M \np{249}{85}{1} 
\bibitem{Affleck-Dine-new} Dine M, Randall L, Thomas S
\np{458}{96}{291}; hep-ph/9507453
\bibitem{RuSh} Rubakov V, Shaposhnikov M \pu{39}{96}{461}, 
hep-ph/9603208.
\bibitem{Quiros} Carena M, Quiros M, Wagner C E M, \pl{380}{96}{81}
\bibitem{INTY} Izawa K, Nomura Y, Tobe K, Yanagida T
\pr{56}{97}{2886}, hep-ph/9705228
\bibitem{Polonyi1} Coughlan G {\em et al.} \pl{131}{83}{59}
\bibitem{Polonyi2} Banks T, Kaplan D, Nelson A \pr{49}{94}{779}, hep-ph/9308292
\bibitem{B.de_Carlos} de Carlos B {\em et al.} \pl{318}{93}{447}, hep-ph/9308325
\bibitem{moduli2} Asaka T {\em et al.} \pr{58}{98}{083509}, hep-ph/9711501
\bibitem{inflation_GMM} Dine M, Riotto A \prl{79}{97}{2632}, hep-ph/9705386
\bibitem{Lyth} Lyth D, Stewart E \prl{75}{95}{201}, hep-ph/9502417;
\pr{53}{96}{1784}, hep-ph/9510204
\bibitem{moduli_rays_1} Kawasaki M, Yanagida T \pl{399}{97}{45}, hep-ph/9701346
\bibitem{moduli_force} Dimopoulos S, Giudice G \pl{379}{96}{105}, hep-ph/9602350  
\bibitem{str-dilaton} Hashiba J, Kawasaki M, Yanagida T, \prl{79}{97}{4525}, hep-ph/9708226
\bibitem{unitarity_cosmology} Griest K, Kamionkowski M
\prl{64}{90}{615} 
\bibitem{charged_dark_matter} De Rujula A, Glashow S, Sarid U 
\np{333}{90}{173};\\ 
Basdevant J {\em et al.} \pl{234}{90}{395};\\ 
Spiro M {\it Nucl. Phys. Proc. Suppl.} {\bf B19} 234 (1990);\\
Dimopoulos S {\em et al.} {\it Phys. Rev.} {\bf D41} 2388 (1990);\\ 
Gould A {\em et al.} \pl{238}{90}{337} 
\bibitem{we:couplings}  Dubovsky S L, Gorbunov D S,
Troitsky S V \pl{413}{97}{89}, hep-ph/9707357
\bibitem{p1}Chun E J, hep-ph/9901220
\bibitem{p2}Hinchliffe I, Paige F E, hep-ph/9812233
\bibitem{p3}Moroi T \pl{447}{99}{75}, 
hep-ph/9811257 
\bibitem{p4}Muller D J, Nandi S,
hep-ph/9811248
\bibitem{p5}Liu C, Song H S,
hep-ph/9811203 
\bibitem{p6}Hamidian H, Huitu K, Puolamaki K, Zhang D-X,
hep-ph/9808341 
\bibitem{p7}Joshipura A S, Vempati S K,
hep-ph/9808232 
\bibitem{p8}Dutta B, Muller D J, Nandi S,
hep-ph/9807390
\bibitem{p9}Baer H, Mercadante P G, Paige F, Tata X, Wang Y \pl{435}{98}{109},
hep-ph/9806290 
\bibitem{p10}Martin S P, Wells J D \pr{59}{99}{035008},
hep-ph/9805289 
\bibitem{p11}Gabrielli E, Sarid U \pr{58}{98}{115003},
hep-ph/9803451
\bibitem{p12} Wagner C E M, \np{528}{98}{3},
hep-ph/9801376
\bibitem{m1}Luty M A, Terning J, hep-ph/9812290
\bibitem{m2}Dvali G, Nir Y \jhep{9810}{98}{014},
hep-ph/9810257 
\bibitem{m3}Mahanthappa K T, Oh S \pl{441}{98}{178},
hep-ph/9807231
\bibitem{m4}Kaplan D E, Lepeintre F, Masiero A, Nelson A E, Riotto A,
hep-ph/9806430 
\bibitem{m5}Chacko Z, Luty M A, Ponton E \pr{59}{99}{035004},
hep-ph/9806398 
\bibitem{m6}Nelson A, Strassler M J,
hep-ph/9806346
\bibitem{m7}Maru N \pl{436}{98}{311},
hep-ph/9806273
\bibitem{m8}Hirayama T, Ishimura N, Maekawa N,
hep-ph/9805457
\bibitem{m9}Asaka T, Yamaguchi M \pl{437}{98}{51},
hep-ph/9805449 
\bibitem{m10}Chikira Y, Haba N, Mimura N,
hep-ph/9804279
\bibitem{m11}Kong O C W, Wright B D \pr{58}{98}{015002},
hep-ph/9802221
\bibitem{c1}Choi K, Hwang K, Kim H B, Lee T, hep-ph/9902291 
\bibitem{c2}Izawa K-I, Nomura Y, Yanagida T, hep-ph/9901345
\bibitem{c3}Asaka T, Yamaguchi M, hep-ph/9811451 
\bibitem{c4}Asaka T, Kawasaki M, Yamaguchi M,
hep-ph/9810334
\bibitem{c5}Moroi T \pr{58}{98}{124008},
hep-ph/9807265 
\end{thebibliography}
\end{document}